\documentclass[12pt,titlepage,letterpaper]{utarticle}

\usepackage{amsmath}
\usepackage{mathrsfs}
\usepackage{amsfonts}
\usepackage{amssymb}
\usepackage{amsthm}
\usepackage{mathtools}
\usepackage{mathbbol}
\usepackage{graphicx}
\usepackage{color}
\usepackage{ucs}
\usepackage[utf8x]{inputenc}
\usepackage{xparse}
\usepackage{tocloft}
\usepackage[titletoc,title]{appendix}
\usepackage{cite}
\usepackage{hyperref}
\usepackage{longtable}

\definecolor{aqua}{rgb}{0, 1.0, 1.0}
\definecolor{fuschia}{rgb}{1.0, 0, 1.0}
\definecolor{gray}{rgb}{0.502, 0.502, 0.502}
\definecolor{lime}{rgb}{0, 1.0, 0}
\definecolor{maroon}{rgb}{0.502, 0, 0}
\definecolor{navy}{rgb}{0, 0, 0.502}
\definecolor{olive}{rgb}{0.502, 0.502, 0}
\definecolor{purple}{rgb}{0.502, 0, 0.502}
\definecolor{silver}{rgb}{0.753, 0.753, 0.753}
\definecolor{teal}{rgb}{0, 0.502, 0.502}

\makeatletter
\newdimen\itex@wd%
\newdimen\itex@dp%
\newdimen\itex@thd%
\def\itexspace#1#2#3{\itex@wd=#3em%
\itex@wd=0.1\itex@wd%
\itex@dp=#2ex%
\itex@dp=0.1\itex@dp%
\itex@thd=#1ex%
\itex@thd=0.1\itex@thd%
\advance\itex@thd\the\itex@dp%
\makebox[\the\itex@wd]{\rule[-\the\itex@dp]{0cm}{\the\itex@thd}}}
\makeatother

\makeatletter
\newif\if@sup
\newtoks\@sups
\def\append@sup#1{\edef\act{\noexpand\@sups={\the\@sups #1}}\act}%
\def\reset@sup{\@supfalse\@sups={}}%
\def\mk@scripts#1#2{\if #2/ \if@sup ^{\the\@sups}\fi \else%
  \ifx #1_ \if@sup ^{\the\@sups}\reset@sup \fi {}_{#2}%
  \else \append@sup#2 \@suptrue \fi%
  \expandafter\mk@scripts\fi}
\def\tensor#1#2{\reset@sup#1\mk@scripts#2_/}
\def\multiscripts#1#2#3{\reset@sup{}\mk@scripts#1_/#2%
  \reset@sup\mk@scripts#3_/}
\makeatother

\makeatletter
\newbox\slashbox \setbox\slashbox=\hbox{$/$}
\def\itex@pslash#1{\setbox\@tempboxa=\hbox{$#1$}
  \@tempdima=0.5\wd\slashbox \advance\@tempdima 0.5\wd\@tempboxa
  \copy\slashbox \kern-\@tempdima \box\@tempboxa}
\def\slash{\protect\itex@pslash}
\makeatother

\def\clap#1{\hbox to 0pt{\hss#1\hss}}

\let\oldroot\root
\def\root#1#2{\oldroot #1 \of{#2}}
\renewcommand{\sqrt}[2][]{\oldroot #1 \of{#2}}

\DeclareSymbolFont{symbolsC}{U}{txsyc}{m}{n}
\SetSymbolFont{symbolsC}{bold}{U}{txsyc}{bx}{n}
\DeclareFontSubstitution{U}{txsyc}{m}{n}

\DeclareSymbolFont{stmry}{U}{stmry}{m}{n}
\SetSymbolFont{stmry}{bold}{U}{stmry}{b}{n}

\DeclareFontFamily{OMX}{MnSymbolE}{}
\DeclareSymbolFont{mnomx}{OMX}{MnSymbolE}{m}{n}
\SetSymbolFont{mnomx}{bold}{OMX}{MnSymbolE}{b}{n}
\DeclareFontShape{OMX}{MnSymbolE}{m}{n}{
    <-6>  MnSymbolE5
   <6-7>  MnSymbolE6
   <7-8>  MnSymbolE7
   <8-9>  MnSymbolE8
   <9-10> MnSymbolE9
  <10-12> MnSymbolE10
  <12->   MnSymbolE12}{}

\makeatletter
\def\re@DeclareMathSymbol#1#2#3#4{%
    \let#1=\undefined
    \DeclareMathSymbol{#1}{#2}{#3}{#4}}
\re@DeclareMathSymbol{\neArrow}{\mathrel}{symbolsC}{116}
\re@DeclareMathSymbol{\neArr}{\mathrel}{symbolsC}{116}
\re@DeclareMathSymbol{\seArrow}{\mathrel}{symbolsC}{117}
\re@DeclareMathSymbol{\seArr}{\mathrel}{symbolsC}{117}
\re@DeclareMathSymbol{\nwArrow}{\mathrel}{symbolsC}{118}
\re@DeclareMathSymbol{\nwArr}{\mathrel}{symbolsC}{118}
\re@DeclareMathSymbol{\swArrow}{\mathrel}{symbolsC}{119}
\re@DeclareMathSymbol{\swArr}{\mathrel}{symbolsC}{119}
\re@DeclareMathSymbol{\nequiv}{\mathrel}{symbolsC}{46}
\re@DeclareMathSymbol{\Perp}{\mathrel}{symbolsC}{121}
\re@DeclareMathSymbol{\Vbar}{\mathrel}{symbolsC}{121}
\re@DeclareMathSymbol{\sslash}{\mathrel}{stmry}{12}
\re@DeclareMathSymbol{\bigsqcap}{\mathop}{stmry}{"64}
\re@DeclareMathSymbol{\biginterleave}{\mathop}{stmry}{"6}
\re@DeclareMathSymbol{\invamp}{\mathrel}{symbolsC}{77}
\re@DeclareMathSymbol{\parr}{\mathrel}{symbolsC}{77}
\makeatother

\makeatletter
\def\Decl@Mn@Delim#1#2#3#4{%
  \if\relax\noexpand#1%
    \let#1\undefined
  \fi
  \DeclareMathDelimiter{#1}{#2}{#3}{#4}{#3}{#4}}
\def\Decl@Mn@Open#1#2#3{\Decl@Mn@Delim{#1}{\mathopen}{#2}{#3}}
\def\Decl@Mn@Close#1#2#3{\Decl@Mn@Delim{#1}{\mathclose}{#2}{#3}}
\Decl@Mn@Open{\llangle}{mnomx}{'164}
\Decl@Mn@Close{\rrangle}{mnomx}{'171}
\Decl@Mn@Open{\lmoustache}{mnomx}{'245}
\Decl@Mn@Close{\rmoustache}{mnomx}{'244}
\makeatother

\makeatletter
\DeclareRobustCommand\widecheck[1]{{\mathpalette\@widecheck{#1}}}
\def\@widecheck#1#2{%
    \setbox\z@\hbox{\m@th$#1#2$}%
    \setbox\tw@\hbox{\m@th$#1%
       \widehat{%
          \vrule\@width\z@\@height\ht\z@
          \vrule\@height\z@\@width\wd\z@}$}%
    \dp\tw@-\ht\z@
    \@tempdima\ht\z@ \advance\@tempdima2\ht\tw@ \divide\@tempdima\thr@@
    \setbox\tw@\hbox{%
       \raise\@tempdima\hbox{\scalebox{1}[-1]{\lower\@tempdima\box
\tw@}}}%
    {\ooalign{\box\tw@ \cr \box\z@}}}
\makeatother

\makeatletter
\NewDocumentCommand\mathraisebox{moom}{%
\IfNoValueTF{#2}{\def\@temp##1##2{\raisebox{#1}{$\m@th##1##2$}}}{%
\IfNoValueTF{#3}{\def\@temp##1##2{\raisebox{#1}[#2]{$\m@th##1##2$}}%
}{\def\@temp##1##2{\raisebox{#1}[#2][#3]{$\m@th##1##2$}}}}%
\mathpalette\@temp{#4}}
\makeatletter

\makeatletter
\def\udots{\mathinner{\mkern2mu\raise\p@\hbox{.}
\mkern2mu\raise4\p@\hbox{.}\mkern1mu
\raise7\p@\vbox{\kern7\p@\hbox{.}}\mkern1mu}}
\makeatother

\newcommand{\gt}{>}

\theoremstyle{plain}

\theoremstyle{definition}

\theoremstyle{remark}

\begin{document}

\preprint{
UTTG--03--17\\
}

\title{Tinkertoys for the $E_7$ Theory}

\author{Oscar Chacaltana
    \address{
    ICTP South American Institute for\\ Fundamental Research,\\
    Instituto de F\'isica Te\'orica,\\Universidade Estadual Paulista,\\
    01140-070 S\~{a}o Paulo, SP, Brazil\\
    {~}\\
    \qquad and\\
    {~}\\
    Howard Community College\\
    10901 Little Patuxent Pkwy\\
    Columbia, MD 21044, USA\\
    {~}\\
    \email{chacaltana@ift.unesp.br}\\
    },
    Jacques Distler
     \address{
     Theory Group\\
     Department of Physics,\\
     University of Texas at Austin,\\
     Austin, TX 78712, USA \\
     {~}\\
      \email{distler@golem.ph.utexas.edu}\\
      \email{zhuyinan@physics.utexas.edu}\\
      {~}\\
     \hskip-2ex${}^{\mathrm{c}}$ Department of Mathematics, \\
     Gwinnett School of Mathematics,\\
     Science and Technology,\\
     970 McElvaney Lane,\\
     Lawrenceville, GA 30044, USA\\
     {~}\\
     \email{anderson.trimm@gsmst.org}
     },
     Anderson Trimm ${}^\mathrm{c}$     
     and Yinan Zhu ${}^\mathrm{b}$
}
\date{\today}

\Abstract{We classify the class $S$ theories of type $E_7$. These are four-dimensional $\mathcal{N}=2$ superconformal field theories arising from the compactification of the $E_7$ $(2,0)$ theory on a punctured Riemann surface, $C$. The classification is given by listing all 3-punctured spheres (``fixtures"), and connecting cylinders, which can arise in a pants-decomposition of $C$. We find exactly 11,000 fixtures with three regular punctures, and an additional 48 with one ``irregular puncture" (in the sense used in our previous works). To organize this large number of theories, we have created a web application at \href{https://golem.ph.utexas.edu/class-S/E7/}{https://golem.ph.utexas.edu/class-S/E7/}~. Among these theories, we find 10 new ones with a simple exceptional global symmetry group, as well as a new rank-2 SCFT and several new rank-3 SCFTs. As an application, we study the strong-coupling limit of the $E_7$ gauge theory with 3 hypermultiplets in the $56$. Using our results, we also verify recent conjectures that the $T^2$ compactification of certain $6d$ $(1,0)$ theories can alternatively be realized in class $S$ as fixtures in the $E_7$ or $E_8$ theories.
}

\maketitle

\tocloftpagestyle{empty}
\tableofcontents
\vfill
\newpage
\setcounter{page}{1}

\section{Introduction}\label{introduction}
Class $S$ theories are a large class of four-dimensional $\mathcal{N}=2$ superconformal field theories arising from the partially-twisted compactification of a six-dimensional $(2,0)$ theory on a punctured Riemann surface \cite{Gaiotto:2009hg, Gaiotto:2009we}. Along with Lagrangian $\mathcal{N}=2$ SCFTs of vector and hypermultiplets, class $S$ contains many strongly-interacting SCFTs which have no known Lagrangian description \cite{Minahan:1996fg, Minahan:1996cj}. Nevertheless, the six-dimensional construction gives rise to  powerful tools to study their properties (for an extensive review of recent progress see, e.g., the collection \cite{Teschner:2014oja}).

As the $6d$ $(2,0)$ theories have an ADE classification, the corresponding four-dimensional theories resulting from their compactification also come in ADE type. A program to classify these theories was initiated in \cite{Chacaltana:2010ks, Chacaltana:2011ze}, where we  provided a method for classifying the $A$ and $D$ series, carrying out this classification explicitly for low ranks, before moving on to the $E_6$ theory in \cite{Chacaltana:2014jba}\footnote{This class of theories can be enlarged for types $A,D,$ and $E_6$ by twisting the $(2,0)$ theory by an outer-automorphism when traversing a nontrivial cycle on the punctured Riemann surface. This construction gives rise to a sector of twisted punctures, leading to many new SCFTs. A classification for the twisted theories of type $A_{2N-1}, D_N,$ and $E_6$ was given in \cite{Chacaltana:2012ch, Chacaltana:2013oka, Chacaltana:2014ica, Chacaltana:2015bna, Chacaltana:2016shw}. Though a complete classification of the theories of type $A_{2N}$ is still lacking, twists of this type were utilized \cite{Chacaltana:2014nya} to construct the $R_{2,2N}$ family of SCFTs. Similarly, a full classification for the $S_3$-twisted $D_4$ theory has not yet been carried out, but these twists were used to construct additional $4d$ theories and study their S-duality frames in \cite{Tachikawa:2010vg}.}. In this work, we classify the $E_7$-type class $S$ theories. We find 11,000 fixtures with three regular punctures. Of these, 962 have enhanced global symmetries or additional free hypermultiplets. It would be formidable to  list all of these here; instead, we have created a web application at \href{https://golem.ph.utexas.edu/class-S/E7/}{https://golem.ph.utexas.edu/class-S/E7/} where the interested reader can explore them. A description and instructions are given in \S\ref{interacting_and_mixed_fixtures}. Among the theories on our list, we find a new rank-2 and several new rank-3 interacting SCFTs \footnote{In \cite{Argyres:2015ffa, Argyres:2015gha, Argyres:2016xua, Argyres:2016xmc, Argyres:2016yzz} a proposed classification of four-dimensional rank-1 $\mathcal{N}=2$ SCFTs was given by constructing the rigid special K\"ahler geometries consistent with the interpretation as the Coulomb branch of an $\mathcal{N}=2$ SCFT. A natural follow up would be to extend these works to rank-2 and higher.}. Additionally, we find several new SCFTs with a simple exceptional global symmetry group.

Using our results, we construct the $E_7$ gauge theory with matter in the $3(56)$. We determine its S-duality frames and provide the $k$-differentials specifying its Seiberg-Witten solution. Additionally, we confirm predictions in \cite{Ohmori:2015pua,Ohmori:2015pia,DelZotto:2015rca} that the $T^2$ compactifications of the worldvolume theories on $M5$ branes probing ALE singularities of type $E$ have class $S$ realizations.

\section{The $E_7$ theory}\label{the__theory}
\subsection{Coulomb branch geometry}\label{coulomb_branch_geometry}

The Coulomb branch geometry for our theories can be realized either by studying parabolic Hitchin systems on the punctured Riemann surface, $C$, or by studying the Calabi-Yau integrable system for a certain family of non-compact Calabi-Yaus fibered over $C$.

In the former description, the Seiberg-Witten curve $\Sigma\to C$ is spectral curve

\begin{displaymath}
\Sigma = \{\text{Det}(\lambda \mathbb{1}- \Phi(z))=0\}\subset \text{Tot}(K_C)
\end{displaymath}
where (for definiteness) the determinant is taken in the adjoint representation and $\lambda$ is the Seiberg-Witten differential.

In the latter description, the noncompact Calabi-Yau is the hypersurface

\begin{displaymath}
\begin{aligned} X_{\vec{u}} =& \bigl\{ 0 = -w^2 - x^3+16x y^{3} + \phi_2(z) y^4 + \phi_6(z) y^3 +\phi_8(z) x y + \phi_{10}(z) y^2 +\phi_{12}(z) x +\phi_{14}(z)y+\phi_{18}(z)\bigr\}\\ & \subset \text{Tot}(K_C^9\oplus K_C^6\oplus K_C^4) \end{aligned}
\end{displaymath}
In both cases, the Seiberg-Witten geometry is expressed in terms of meromorphic $k$-differentials, $\phi_k(z)$, on $C$, which have poles of various orders at the punctures \cite{Tachikawa:2011yr}. It is most convenient to work in the Katz-Morrison basis \cite{1992alg.geom..2002K}, where the $\phi_k(z)$ are related to the invariant traces, $P_k = Tr(\Phi^k)$, by

\begin{displaymath}
\begin{aligned}
\phi_2 =& \tfrac{1}{18}P_2\\
\phi_6 =& -\tfrac{2}{3}P_6 + \tfrac{1}{2916}P_2^3\\
\phi_8 =& -\tfrac{4}{25}P_8 - \tfrac{22}{675}P_6 P_2 - \tfrac{7}{524880}P_2^4\\
\phi_{10}  =& - \tfrac{32}{315}P_{10} - \tfrac{1}{175}P_8 P_2 + \tfrac{17}{36450}P_6 P_2^2 - \tfrac{1}{9447840}P_2^5\\
\phi_{12} =& \tfrac{128}{225}P_{12} - \tfrac{4096}{42525}P_{10}P_2 + \tfrac{737}{127575}P_8 P_2^2 - \tfrac{992}{2025}P_6^2 + \tfrac{167}{492075}P_6 P_2^3 - \tfrac{149}{1530550080P_2^6}\\
\phi_{14} =& \tfrac{1024}{20867}P_{14} - \tfrac{140864}{18109575}P_{12}P_2 + \tfrac{132848}{311155425}P_{10}P_2^2 - \tfrac{992}{60975}P_8P_6 - \tfrac{1289}{1866932550}P_8 P_2^3\\ & + \tfrac{5648}{2963385}P_6^2 P_2 - \tfrac{11609}{7201025550}P_6 P_2^4 + \tfrac{11083}{31357297819008}P_2^7\\
\phi_{18} =& - \tfrac{8192}{167487} P_{18} + \tfrac{78810880}{94363683183}P_{14}P_2^2 + \tfrac{308224}{12561525}P_{12}P_6 - \tfrac{871487200}{9827303577201}P_{12}P_2^3\\ & + \tfrac{7553024}{439653375}P_{10}P_8 - \tfrac{72249472}{11870641125}P_{10}P_6P_2 + \tfrac{24365269174}{4221273582024975}P_{10}P_2^4 - \tfrac{619144}{732755625}P_8^2 P_2\\& + \tfrac{18510930376}{48254156173125}P_8P_6P_2^2 - \tfrac{14715122551}{63319103730374625}P_8 P_2^5 - \tfrac{1921408}{339161175}P_6^3\\
& - \tfrac{4632094024}{5025325692886875}P_6^2P_2^3 - \tfrac{886993691}{8508142378495752491520}P_2^9
\end{aligned}
\end{displaymath}
At a puncture, $\Phi(z)$ has a simple pole with nilpotent residue,

\begin{displaymath}
\Phi(z) = \frac{N}{z} + \text{regular}
\end{displaymath}
where $N$ is a representative of the ``Hitchin'' Nilpotent orbit which is Spaltenstein-dual \cite{Chacaltana:2012zy} to the Nahm orbit (which we use to label our punctures)

\begin{displaymath}
O_H = d (O_N)
\end{displaymath}
Taking traces, one finds an elaborate set of constraints on the coefficients of the polar parts of the $\phi_k(z)=\sum_j \frac{c^{(k)}_j}{z^j}+\text{regular}$. When the special piece of $O_N$ has more than one element, we have an additional quotient by a finite group (the ``Sommers-Achar group'') acting on the coefficients. \cite{Chacaltana:2012zy}

\subsection{Puncture properties}\label{puncture_properties}

Here we review the puncture properties listed in our table below, leaving most of the details to \cite{Chacaltana:2014jba}.

 As in our previous works, we use Bala-Carter notation \cite{BalaCarter1, BalaCarter2} to label the nilpotent orbits, where $O_N=0$ is the full puncture and $O_N=E_7(a_1)$ is the simple puncture. The flavour symmetry algebra, $\mathfrak{f}$, associated to a puncture is the centralizer of $\rho_N(\mathfrak{su}(2))$ inside $\mathfrak{e}_7$. For the distinguished orbits ($E_7(a_i)$, $i=1,\dots,5$), $\mathfrak{f}$ is trivial, whereas for the $0$ orbit $\mathfrak{f}$ is all of $\mathfrak{e}_7$. The level of each factor $\mathfrak{f}_i\subset\mathfrak{f}$ is determined from the decomposition of the adjoint representation under the embedding $\mathfrak{e}_7\supset \mathfrak{su}(2)\times \mathfrak{f}_i$ as

\begin{equation}
\mathfrak{e}_7 = \bigoplus_n V_n\otimes R_{n,i}
\label{decomp}\end{equation}
where $V_n$ denotes the $n$-dimensional irreducible representation of $\mathfrak{su}(2)$ and $R_{n,i}$ the corresponding representation of $\mathfrak{f}_i$, which is in general reducible. \footnote{We list this decomposition for each puncture in Appendix \ref{appendix_embeddings_of__in_}.} The level of $\mathfrak{f}_i$ is then given by \cite{Chacaltana:2012zy, Tachikawa:2015bga}

\begin{equation}
k_i = \sum_n l_{n,i}
\end{equation}
where $l_{n,i}$ is the Dynkin index of the representation $R_{n,i}$. For $\mathfrak{f}_i = \mathfrak{u}(1)$, $l_{n,i}$ is the $\mathfrak{u}(1)$ charge squared. In the table below, we normalize the $\mathfrak{u}(1)$ generators so that the free hypermultiplets in the mixed fixtures have charge 1.

The $\phi_k(z)$ have poles at the punctures of order at most $p_k$:

\begin{displaymath}
\phi_k(z) = \sum_{j=1}^{p_k} \frac{c^{(k)}_j}{z^j} + \text{regular}
\end{displaymath}
where the set $\{p_k\}$ is called the \emph{pole structure} of the puncture.
The coefficients, $c^{(k)}_j$, typically are not all independent, but instead obey certain polynomial relations, which we list below.

Finally, for each puncture we also list its contribution to the effective number of vector and hypermultiplets $(n_h,n_v)$, which are given in terms of the conformal central charges $a$ and $c$ by $n_v=4(2a-c)$ and $n_h=4(5c-4a)$.

\subsection{Regular punctures}\label{regular_punctures}

The pole structure of an $E_7$ puncture at $z=0$ is denoted $\{p_2,p_6,p_8,p_{10},p_{12},p_{14},p_{18}\}$, and is defined to be the set of \emph{leading} pole orders in $z$ of the differentials $\phi_k$, for $k=2,6,8,10,12,14,18$. As discussed above, for certain punctures, there are constraints among \emph{leading} coefficients, and sometimes even for \emph{subleading} ones, in the expansion of the $\phi_k$ in $z$.

{\scriptsize
\setlength\LTleft{-.5in}
\renewcommand{\arraystretch}{2.25}

\begin{longtable}{|c|c|c|c|c|c|}
\hline
\mbox{\shortstack{Nahm\\B-C label}}&\mbox{\shortstack{Hitchin\\B-C label}}&Pole structure&Constraints&Flavour group&$(\delta n_h,\delta n_v)$\\
\hline 
\endhead
0&$E_{7}$&$\{1,5,7,9,11,13,17\}$&$-$&$(E_7)_{36}$&$(1596,1533)$\\
\hline
$A_1$&$E_7(a_1)$&$\{1,5,7,9,11,13,16\}$&$-$&$Spin(12)_{28}$&$(1544, 1498)$\\
\hline
$2A_1$&$E_7(a_2)$&$\{1,5,7,9,11,12,16\}$&$-$&$Spin(9)_{24}\times SU(2)_{20}$&$(1508, 1471)$\\
\hline
$(3A_1)''$&$E_6$&$\{1,5,7,8,11,12,16\}$&$-$&$(F_4)_{24}$&$(1488, 1452)$\\
\hline
$(3A_1)' \,\,(\underline{ns})$&$(E_7(a_3),\mathbb{Z}_2)$&$\{1,5,7,9,10,12,16\}$&$-$&$Sp(3)_{20}\times SU(2)_{19}$&$(1479, 1448)$\\
\hline
$A_2$&$E_7(a_3)$&$\{1,5,7,9,10,12,16\}$&$c^{(18)}_{16}={\bigl(a^{(9)}_{8}\bigr)}^2$&$SU(6)_{20}$&$(1460, 1430)$\\
\hline
$4A_1 \,\,(\underline{ns})$&$(E_6(a_1),\mathbb{Z}_2)$&$\{1,5,7,8,10,12,16\}$&$-$&$Sp(3)_{19}$&$(1457, 1429)$\\
\hline
$A_2+A_1$&$E_6(a_1)$&$\{1,5,7,8,10,12,16\}$&$c^{(18)}_{16}={\bigl(a^{(9)}_{8}\bigr)}^2$&$SU(4)_{18}\times U(1)_{42}$&$(1436, 1411)$\\
\hline
$A_2+2A_1$&$E_7(a_4)$&$\{1,5,7,8,10,12,15\}$&$-$&$\begin{gathered}SU(2)_{16} \times SU(2)_{28}\\ \times SU(2)_{84}\end{gathered}$&$(1416, 1394)$\\
\hline
$A_3$&$D_6(a_1)$&$\{1,5,7,8,10,12,15\}$&$\begin{gathered}\begin{aligned}c^{(18)}_{15}=&\left(\tfrac{1}{3}c^{(12)}_{10}\right.-3\,c^{(6)}_5a^{(6)}_5\\&\left.-9\,{\bigl(a^{(6)}_5\bigr)}^2\right)a^{(6)}_5\end{aligned}\\c^{(14)}_{12}=c^{(8)}_7 a^{(6)}_5\end{gathered}$&$Spin(7)_{16}\times SU(2)_{12}$&$(1364, 1343)$\\
\hline
$2A_2$&$D_5+A_1$&$\{1,5,7,8,10,11,15\}$&$-$&$(G_2)_{16}\times SU(2)_{36}$&$(1388, 1367)$\\
\hline
$A_2+3A_1$&$A_6$&$\{1,5,6,8,10,12,15\}$&$-$&$(G_2)_{28}$&$(1400, 1379)$\\
\hline
$(A_3+A_1)''$&$D_5$&$\{1,5,7,8,10,11,14\}$&$-$&$Spin(7)_{16}$&$(1352, 1332)$\\
\hline
$2A_2+A_1 \,\,(\underline{ns})$&$(E_7(a_5),S_3)$&$\{1,5,6,8,10,11,15\}$&$-$&$SU(2)_{36}\times SU(2)_{38}$&$(1370, 1352)$\\
\hline
$(A_3+A_1)' \,\,(\underline{ns})$&$(E_7(a_5),\mathbb{Z}_2)$&$\{1,5,6,8,10,11,15\}$&$\begin{gathered}\begin{aligned}c^{(18)}_{15}= &-\tfrac{2}{81}\bigl(c^{(6)}_{5}\bigr)^3 - \tfrac{1}{27}c^{(12)}_{10}c^{(6)}_{5}\\& + \tfrac{2}{81}a^{(6)}_{5}\left(4\bigl(a^{(6)}_{5}\bigr)^2\right.\\&\left. - 3\left(c^{(12)}_{10} + \bigl(c^{(6)}_{5}\bigr)^2\right)\right)\end{aligned}\end{gathered}$&$\begin{gathered}SU(2)_{13}\times SU(2)_{24}\\ \times SU(2)_{12}\end{gathered}$&$(1345, 1328)$\\
\hline
$D_4(a_1)$&$E_7(a_5)$&$\{1,5,6,8,10,11,15\}$&$\begin{gathered}\begin{aligned}c^{(12)}_{10}=& - \bigl(c^{(6)}_{5}\bigr)^2 + \bigl(a^{(6)}_{5}\bigr)^2\\& - \bigl({a'}^{(6)}_{5}\bigr)^2\end{aligned}\\ \begin{aligned}c^{(18)}_{15}= &-\tfrac{2}{81}\bigl(c^{(6)}_{5}\bigr)^3\\& - \tfrac{1}{27}c^{(12)}_{10}c^{(6)}_{5}\\& + \tfrac{2}{81}a^{(6)}_{5}\left(\bigl(a^{(6)}_{5}\bigr)^2 \right.\\&\left.+ 3\bigl({a'}^{(6)}_{5}\bigr)^2\right)\end{aligned}\end{gathered}$&${SU(2)}_{12}^3$&$(1332, 1316)$\\
\hline
$A_3+2A_1 \,\,(\underline{ns})$&$(E_6(a_3),\mathbb{Z}_2)$&$\{1,5,6,8,10,11,14\}$&$-$&$SU(2)_{13}\times SU(2)_{24}$&$(1333, 1317)$\\
\hline
$D_4(a_1)+A_1$&$E_6(a_3)$&$\{1,5,6,8,10,11,14\}$&$c^{(12)}_{10}=-\tfrac{3}{4}{\bigl(c^{(6)}_5\bigr)}^2 + 3{\bigl(a^{(6)}_5\bigr)}^2$&${SU(2)}_{12}^2$&$(1320, 1305)$\\
\hline
$D_4$&$(A_5)''$&$\{1,5,6,8,10,11,15\}$&$\begin{gathered}c^{(12)}_{10}=-{\bigl(c^{(6)}_5\bigr)}^2\\ \begin{aligned}c^{(12)}_{9}=&-2c^{(6)}_5 c^{(6)}_4 -3 c^{(8)}_6 a^{(4)}_3\\&-2 c^{(6)}_5 c^{(2)}_1 a^{(4)}_3- {\bigl(a^{(4)}_3\bigr)}^3\end{aligned}\\ \begin{aligned}c^{(14)}_{11}=&-2c^{(10)}_8 a^{(4)}_3-\tfrac{1}{27}\left(3c^{(8)}_6 c^{(6)}_5\right.\\&\left.+{\bigl(c^{(6)}_5\bigr)}^2c^{(2)}_1+3 c^{(6)}_5 {\bigl(a^{(4)}_3\bigr)}^2 \right)\end{aligned}\\ c^{(18)}_{15}=\tfrac{1}{81}{\bigl(c^{(6)}_5\bigr)}^3\\ 
\begin{aligned}c^{(18)}_{14}=&c^{(10)}_8{\bigl(a^{(4)}_3\bigr)}^2\\&+\tfrac{1}{27}\left(3c^{(8)}_6c^{(6)}_5a^{(4)}_3\right.\\&+{\bigl(c^{(6)}_5\bigr)}^2c^{(6)}_4\\&+2{\bigl(c^{(6)}_5\bigr)}^2c^{(2)}_1a^{(4)}_3\\&\left.+3c^{(6)}_5{\bigl(a^{(4)}_3\bigr)}^3\right)\end{aligned}\\ 
\begin{aligned}c^{(18)}_{13}= &-c^{(14)}_{10} a^{(4)}_{3} - c^{(10)}_{7} {\bigl(a^{(4)}_{3}\bigr)}^2\\& - \tfrac{1}{27}\left(c^{(12)}_{8}c^{(6)}_{5}\right.\\&\left.+{\bigl(c^{(6)}_{5}\bigr)}^2 c^{(2)}_{0} a^{(4)}_{3}\right.\\& \left.+ {\bigl(c^{(6)}_{5}\bigr)}^2 c^{(6)}_{3}\right.\\&\left. + 3c^{(8)}_{5}c^{(6)}_{5}a^{(4)}_{3}\right)\end{aligned}\end{gathered}$&$Sp(3)_{12}$&$(1196, 1181)$\\
\hline
$A_3+A_2$&$D_5(a_1)+A_1$&$\{1,5,6,8,9,11,14\}$&$-$&$SU(2)_{12}\times U(1)_{112}$&$(1308, 1294)$\\
\hline
$A_4$&$D_5(a_1)$&$\{1,5,6,8,9,11,14\}$&$\begin{gathered}c^{(10)}_{8}={\left(a^{(5)}_4\right)}^2\\c^{(14)}_{11}=-2\, a^{(5)}_4 a^{(9)}_7\\c^{(18)}_{14}={\left(a^{(9)}_7\right)}^2\end{gathered}$&$SU(3)_{12}\times U(1)_{24}$&$(1252, 1239)$\\
\hline
$A_3+A_2+A_1$&$A_4+A_2$&$\{1,4,6,8,9,11,14\}$&$-$&$SU(2)_{224}$&$(1296, 1283)$\\
\hline
$(A_5)''$&$D_4$&$\{1,5,6,6,9,9,12\}$&$-$&$(G_2)_{12}$&$(1144, 1132)$\\
\hline
$A_4+A_1$&$A_4+A_1$&$\{1,4,6,8,9,11,14\}$&$\begin{gathered}c^{(10)}_{8}={\left(a^{(5)}_4\right)}^2\\c^{(14)}_{11}=-2\, a^{(5)}_4 a^{(9)}_7\\c^{(18)}_{14}={\left(a^{(9)}_7\right)}^2\end{gathered}$&$U(1)_{54} \times U(1)_{24}$&$(1239, 1228)$\\
\hline
$D_4+A_1 \,\,(\underline{ns})$&$(A_4,\mathbb{Z}_2)$&$\{1,4,6,8,9,11,14\}$&$\begin{gathered}c^{(12)}_9 = {\bigl(a^{(4)}_3\bigr)}^3 +3 c^{(8)}_6 a^{(4)}_3\\c^{(14)}_{11}=2\, c^{(10)}_8 a^{(4)}_3\\c^{(18)}_{14}= c^{(10)}_8 {\bigl( a^{(4)}_3\bigr)}^2\\c^{(18)}_{13}=c^{(14)}_{10} a^{(4)}_3-c^{(10)}_{7} {\bigl(a^{(4)}_3\bigr)}^2\end{gathered}$&$Sp(2)_{11}$&$(1182, 1170)$\\
\hline
$D_5(a_1)$&$A_4$&$\{1,4,6,8,9,11,14\}$&$\begin{gathered}c^{(10)}_8 = {\bigl(a^{(5)}_4\bigr)}^2\\c^{(12)}_9 = {\bigl(a^{(4)}_3\bigr)}^3 +3 c^{(8)}_6 a^{(4)}_3\\c^{(14)}_{11}=2{\bigl(a^{(5)}_4\bigr)}^2 a^{(4)}_3\\c^{(18)}_{14}= {\bigl(a^{(5)}_4 a^{(4)}_3\bigr)}^2\\c^{(18)}_{13}=c^{(14)}_{10} a^{(4)}_3-c^{(10)}_{7} {\bigl(a^{(4)}_3\bigr)}^2\end{gathered}$&$SU(2)_{10}\times U(1)_{28}$&$(1170, 1160)$\\
\hline
$A_4+A_2$&$A_3+A_2+A_1$&$\{1,4,6,7,9,10,13\}$&$-$&$SU(2)_{108}$&$(1212, 1202)$\\
\hline
$D_5(a_1)+A_1$&$A_3+A_2$&$\{1,4,6,7,9,10,13\}$&$\begin{gathered}c^{(12)}_9 = 3 c^{(8)}_{6} a^{(4)}_{3} + \bigl(a^{(4)}_3\bigr)^3\\c^{(18)}_{13} = c^{(14)}_{10} a^{(4)}_{3} - c^{(10)}_{7} \bigl(a^{(4)}_{3}\bigr)^2\end{gathered}$&$SU(2)_{56}$&$(1160, 1151)$\\
\hline
$(A_5)' \,\,(\underline{ns})$&$(D_4(a_1)+A_1,\mathbb{Z}_2)$&$\{1,4,6,7,9,10,13\}$&$\begin{gathered}c^{(12)}_{9} = 2 a^{(4)}_{3} \left( 4\bigl(a^{(4)}_{3}\bigr)^2 + 3 c^{(8)}_{6} \right)\\c^{(14)}_{10} = -2 c^{(10)}_{7} a^{(4)}_{3}\\c^{(18)}_{13} = c^{(10)}_{7} \left( 4\bigl(a^{(4)}_{3}\bigr)^2 + 3 c^{(8)}_{6} \right)\end{gathered}$&$SU(2)_{9}\times SU(2)_{20}$&$(1133, 1124)$\\
\hline
$E_6(a_3)$&$D_4(a_1)+A_1$&$\{1,4,6,7,9,10,13\}$&$\begin{gathered}c^{(8)}_6 = - \left( \bigl(a^{(4)}_{3}\bigr)^2 + 3 \bigl({a'}^{(4)}_{3}\bigr)^2 \right)\\c^{(12)}_{9} = 2 a^{(4)}_{3} \left( \bigl(a^{(4)}_{3}\bigr)^2 - 9 \bigl({a'}^{(4)}_{3}\bigr)^2 \right)\\c^{(14)}_{10} = -2 c^{(10)}_{7} a^{(4)}_{3}\\c^{(18)}_{13} = c^{(10)}_{7} \left( \bigl(a^{(4)}_{3}\bigr)^2 - 9 \bigl({a'}^{(4)}_{3}\bigr)^2 \right)\end{gathered}$&$SU(2)_{20}$&$(1124, 1116)$\\
\hline
$A_5+A_1 \,\,(\underline{ns})$&$(D_4(a_1),S_3)$&$\{1,4,6,6,9,9,12\}$&$-$&$SU(2)_{26}$&$(1130, 1121)$\\
\hline
$D_6(a_2) \,\,(\underline{ns})$&$(D_4(a_1),\mathbb{Z}_2)$&$\{1,4,6,6,9,9,12\}$&$c^{(12)}_{9} = 2 a^{(4)}_{3} \left( 4\bigl(a^{(4)}_{3}\bigr)^2 + 3 c^{(8)}_{6} \right)$&$SU(2)_9$&$(1113, 1105)$\\
\hline
$E_7(a_5)$&$D_4(a_1)$&$\{1,4,6,6,9,9,12\}$&$\begin{gathered}c^{(8)}_{6}=-\left(\bigl(a^{(4)}_3\bigr)^2+3\bigl({a'}^{(4)}_3\bigr)^2\right)\\c^{(12)}_{9}=2a^{(4)}_3\left(\bigl(a^{(4)}_3)\bigr)^2 -9\bigl({a'}^{(4)}_3\bigr)^2\right)\end{gathered}$&none&$(1104, 1097)$\\
\hline
$D_5$&$(A_3+A_1)''$&$\{1,4,6,7,9,10,13\}$&$\begin{gathered}c^{(8)}_6=-{\bigl(a^{(4)}_3\bigr)}^2\\c^{(10)}_7=a^{(6)}_{4} a^{(4)}_{3}\\c^{(12)}_9=2{\bigl(a^{(4)}_3\bigr)}^3\\ \begin{aligned}c^{(12)}_8= &-3\left(c^{(8)}_{5} a^{(4)}_{3} - 6 c^{(6)}_{4} a^{(6)}_{4} \right.\\ &\left. - 6 a^{(6)}_{4} a^{(4)}_{3} c^{(2)}_{1} - 27 \bigl(a^{(6)}_{4}\bigr)^2\right) \end{aligned}\\c^{(14)}_{10}= - 2 a^{(6)}_{4} \bigl(a^{(4)}_{3}\bigr)^2\\ \begin{aligned}c^{(14)}_{9} = &-2 c^{(10)}_{6} a^{(4)}_{3} + c^{(8)}_{5} a^{(6)}_{4}\\ & - 3\bigl(a^{(6)}_{4}\bigr)^2 c^{(2)}_{1} \end{aligned}\\c^{(18)}_{13}= a^{(6)}_{4} \bigl(a^{(4)}_{3}\bigr)^3\\ 
\begin{aligned}c^{(18)}_{12}= & c^{(10)}_{6} \bigl(a^{(4)}_{3}\bigr)^2 - c^{(8)}_{5} a^{(6)}_{4} a^{(4)}_{3} \\ & + 3 \bigl(a^{(6)}_{4}\bigr)^2 c^{(6)}_{4}\\& + 6 \bigl(a^{(6)}_{4}\bigr)^2 a^{(4)}_{3} c^{(2)}_{1} \\ & + 18 \bigl(a^{(6)}_{4}\bigr)^3 \end{aligned}\\ 
\begin{aligned}c^{(18)}_{11}= & - c^{(14)}_{8} a^{(4)}_{3} + \tfrac{1}{3}c^{(12)}_{7} a^{(6)}_{4} \\&- c^{(10)}_{5} \bigl(a^{(4)}_{3}\bigr)^2\\& + c^{(8)}_{4} a^{(6)}_{4} a^{(4)}_{3} \\& - 3 \bigl(a^{(6)}_{4}\bigr)^2 c^{(6)}_{3}\\& - 3 \bigl(a^{(6)}_{4}\bigr)^2 a^{(4)}_{3} c^{(2)}_{0}\end{aligned}\\ \end{gathered}$&$SU(2)_{8}\times SU(2)_{12}$&$(988, 981)$\\
\hline
$A_6$&$A_2+3A_1$&$\{1,4,5,6,7,8,11\}$&$-$&$SU(2)_{36}$&$(1004, 998)$\\
\hline
$D_5+A_1$&$2A_2$&$\{1,4,5,6,8,9,12\}$&$\begin{gathered}c^{(12)}_{8} = -4 a^{(6)}_{4}\left( c^{(6)}_{4} - a^{(6)}_{4}\right)\\c^{(14)}_{9} = - \tfrac{2}{9} a^{(6)}_{4} \left( c^{(8)}_{5} + \frac{2}{3}c^{(2)}_{1}a^{(6)}_{4} \right)\\c^{(18)}_{12} = \tfrac{4}{27}\bigl(a^{(6)}_{4}\bigr)^2\left(c^{(6)}_{4} - \tfrac{4}{3}a^{(6)}_{4}\right)\\c^{(18)}_{11} = - \tfrac{2}{27}a^{(6)}_{4}\left(c^{(12)}_{7} + 2a^{(6)}_{4}c^{(6)}_{3}\right)\end{gathered}$&$SU(2)_{12}$&$(980, 974)$\\
\hline
$D_6(a_1)$&$A_3$&$\{1,4,6,6,9,9,12\}$&$\begin{gathered}c^{(8)}_{6}=-{\left(a^{(4)}_3\right)}^2\\c^{(12)}_{9}=-2{\left(a^{(4)}_3\right)}^3\\c^{(12)}_{8}=3\, a^{(4)}_3 c^{(8)}_{5}\\c^{(14)}_{9}=2\, a^{(4)}_3 c^{(10)}_{6}\\c^{(18)}_{12}={\left(a^{(4)}_3\right)}^2 c^{(10)}_{6}\\c^{(18)}_{11}=c^{(14)}_{8}a^{(4)}_3-c^{(10)}_5 {\bigl(a^{(4)}_3\bigr)}^2\end{gathered}$&$SU(2)_8$&$(976, 970)$\\
\hline
$E_7(a_4)$&$A_2+2A_1$&$\{1,4,5,6,7,8,10\}$&$-$&none&$(968, 963)$\\
\hline
$E_6(a_1)$&$A_2+A_1$&$\{1,4,5,6,7,8,10\}$&$\begin{gathered}c^{(6)}_4=3{\left(a^{(3)}_2\right)}^2\\c^{(8)}_5=-6\, a^{(5)}_3 a^{(3)}_2\\c^{(10)}_6={\left(a^{(5)}_3\right)}^2\\ c^{(12)}_7=-18\, a^{(9)}_5 a^{(3)}_2\\c^{(14)}_8=2\, a^{(9)}_5 a^{(5)}_3\\c^{(18)}_{10}={\left(a^{(9)}_5\right)}^2\end{gathered}$&$U(1)_{24}$&$(868, 864)$\\
\hline
$D_6 \,\,(\underline{ns})$&$(A_2,\mathbb{Z}_2)$&$\{1,4,4,4,6,6,8\}$&$-$&$SU(2)_7$&$(767, 763)$\\
\hline
$E_7(a_3)$&$A_2$&$\{1,4,4,4,6,6,8\}$&$c^{(6)}_4=3\left(a^{(3)}_2\right)^2$&none&$(760, 757)$\\
\hline
$E_6$&$(3A_1)''$&$\{1,3,4,5,6,7,9\}$&$\begin{gathered}\begin{aligned}c^{(8)}_4=&-\tfrac{1}{3}\left(2c^{(6)}_3 c^{(2)}_1\right.\\&\left.+2{\bigl(c^{(2)}_1\bigr)}^2a^{(4)}_2+3{\bigl(a^{(4)}_2\bigr)}^2\right)\end{aligned}\\c^{(10)}_5=-\tfrac{1}{9}\left(c^{(6)}_3+c^{(2)}_1a^{(4)}_2\right)a^{(4)}_2\\
\begin{aligned}c^{(12)}_6=&-{\bigl(c^{(6)}_3\bigr)}^2\\& +\left({\bigl(c^{(2)}_1\bigr)}^2+\right.\\&\left.2a^{(4)}_2\right){\bigl(a^{(4)}_2\bigr)}^2\end{aligned}\\ 
\begin{aligned}c^{(12)}_5= &-2 c^{(6)}_{3} c^{(6)}_{2}\\& - a^{(4)}_2 \left(3c^{(8)}_{3} + 2c^{(6)}_3 c^{(2)}_{0}\right.\\&\left.+ 2c^{(6)}_{2} c^{(2)}_{1}\right.\\&\left. + 2 a^{(4)}_{2} c^{(2)}_{1} c^{(2)}_{0}\right)\\ \end{aligned}\\ 
\begin{aligned}c^{(14)}_7= &\tfrac{1}{27}\left(\bigl(c^{(6)}_{3}\bigr)^2 c^{(2)}_{1} \right.\\&\left.+ a^{(4)}_{2} \left(2c^{(6)}_{3}\bigl(c^{(2)}_{1}\bigr)^2\right.\right.\\&\left.\left.+ 6c^{(6)}_{3}a^{(4)}_{2}+a^{(4)}_{2}\bigl(c^{(2)}_{1}\bigr)^3 \right.\right.\\&\left.\left.+ 6 \bigl(a^{(4)}_{2}\bigr)^2c^{(2)}_{1}\right)\right)\\ \end{aligned}\\  
\begin{aligned}c^{(14)}_6 = &- \tfrac{1}{27}\left(3c^{(8)}_{3} c^{(6)}_{3}\right.\\&\left. + \bigl(c^{(6)}_{3}\bigr)^2 c^{(2)}_{0}\right.\\&+ a^{(4)}_{2} \left( 54 c^{(10)}_{4} + 3c^{(8)}_{3} \right.\\&\left.c^{(2)}_{1} + 2c^{(6)}_{3} c^{(2)}_{1} c^{(2)}_{0}\right.\\&\left.\left.+ a^{(4)}_{2} \bigl(c^{(2)}_{1}\bigr)^2 c^{(2)}_{0} \right)\right)\\ \end{aligned}\\ \begin{aligned}c^{(18)}_{9} =& \tfrac{1}{81}\left( \bigl(c^{(6)}_{3}\bigr)^3 \right.\\&\left.- \bigl(a^{(4)}_{2}\bigr)^2 \left( 3c^{(6)}_{3} \bigl(c^{(2)}_{1}\bigr)^2\right.\right.\\&\left.\left.+ a^{(4)}_{2} \left(9 c^{(6)}_{3} + 2\bigl(c^{(2)}_{1}\bigr)^3\right) \right.\right.\\&\left.\left.+ 9\bigl(a^{(4)}_{2}\bigr)^2 c^{(2)}_{1} \right)\right)\\ \end{aligned}\\ 
\begin{aligned}c^{(18)}_8 =& \tfrac{1}{27}\left( \bigl(c^{(6)}_{3}\bigr)^2c^{(6)}_{2}\right.\\&\left. + a^{(4)}_{2} \left(3c^{(8)}_{3}c^{(6)}_{3}\right.\right.\\&\left.\left.+ 2c^{(6)}_{3} c^{(6)}_{2} c^{(2)}_{1}\right.\right.\\&+ 2 \bigl(c^{(6}_{3}\bigr)^2c^{(2)}_{0}\\& + a^{(4)}_{2} \left(27c^{(10)}_{4}+ 3c^{(8)}_{3} c^{(2)}_{1}\right.\\&\left. + c^{(6)}_{2} \bigl(c^{(2)}_{1}\bigr)^2\right.\\&\left. + 4 c^{(6)}_{3} c^{(2)}_{1} c^{(2)}_{0}\right)\\&\left.\left.+ 2\bigl(a^{(4)}_{2}\bigr)^2 \bigl(c^{(2)}_{1}\bigr)^2 c^{(2)}_{0}\right)\right)\\ \end{aligned}\\ 
\begin{aligned}c^{(18)}_{7} =& - \tfrac{1}{27}\left(c^{(12)}_{4} c^{(6)}_{3}\right.\\&\left. + \bigl(c^{(6)}_{3}\bigr)^2 c^{(6)}_{1} \right.\\&+ a^{(4)}_{2} \left( 27c^{(14)}_{5} + c^{(12)}_{4} c^{(2)}_{1}\right.\\&\left. + 3c^{(8)}_{2} c^{(6)}_{3}\right.\\&\left.+ 2 c^{(6)}_{3} c^{(6)}_{1} c^{(2)}_{1}\right)\\& + \bigl(a^{(4)}_{2}\bigr)^2 \left(27 c^{(10)}_{3}\right.\\&\left.\left.+ 3c^{(8)}_{2} c^{(2)}_{1} \right.\right.\\&\left.\left.+ c^{(6)}_{1} \bigl(c^{(2)}_{1}\bigr)^2\right)\right)\end{aligned}\end{gathered}$&$SU(2)_{12}$&$(604, 601)$\\
\hline
$E_7(a_2)$&$2A_1$&$\{1,3,4,4,6,6,8\}$&$\begin{gathered}c^{(6)}_3= a^{(4)}_2 c^{(2)}_1\\c^{(8)}_4=-\left(a^{(4)}_2\right)^2\\c^{(12)}_6=-2\left(a^{(4)}_2\right)^3\\c^{(12)}_5=3 c^{(8)}_3a^{(4)}_2\\c^{(14)}_6=2 c^{(10)}_4 a^{(4)}_2\\c^{(18)}_8= c^{(10)}_4 \, \left(a^{(4)}_2\right)^2\\c^{(18)}_7=c^{(14)}_5\, a^{(4)}_2-c^{(10)}_3\left(a^{(4)}_2\right)^2\end{gathered}$&none&$(592, 590)$\\
\hline
$E_7(a_1)$&$A_1$&$\{1,2,2,2,3,3,4\}$&$-$&none&$(384, 383)$\\
\hline
\end{longtable}

}

\subsection{Cataloging fixtures using the superconformal index}\label{cataloging_fixtures_using_the_superconformal_index}

There are 45 nilpotent orbits in $\mathfrak{e}_7$. Excluding the regular orbit (which corresponds to the trivial defect), this yields 44 codimension-2 defects (``punctures''). A 3-punctured sphere is specified by choosing a triple of such defects. There are 15,180 such triples, but 4,180 of them are ``bad'' (do not lead to well-defined 4D SCFTs\footnote{The simplest diagnostic for when an $n$-punctured sphere is ``bad" is that the Riemann-Roch index predicts a negative number for one or more of the graded Coulomb branch dimensions. Equivalently, the Hall-Littlewood index \eqref{SCI} diverges.}). Of the remaining\footnote{There are, in addition, 48 fixtures with an ``irregular'' puncture. These arise when the collision of two punctures \emph{would} have resulted in bubbling-off one of the 4,180 bad 3-punctured spheres. Of the 48, 36 are free-field fixtures, 10 are interacting fixtures and 2 are mixed. They are listed in the tables below.}  11,000, one is a free-field fixture (corresponding to three half-hypermultiplets in the $56$ of $E_7$), 262 are ``mixed'' fixtures (consisting of some number of hypermultiplets plus an interacting SCFT),  and the remaining 10,737 are isolated interacting SCFTs. Of these, 654 have ``enhanced'' global symmetry groups: the global symmetry group of the SCFT is larger than the ``manifest'' global symmetry associated to the three punctures.

Of the ``good'' fixtures, we will need to determine which are ``mixed'' (i.e., include free hypermultiplets) and which have enhanced global symmetries. To carry out this classification, we make recourse to the Hall-Littlewood limit of the superconformal index as we did in \cite{Chacaltana:2014jba} for the $E_6$ theory. This method is a generalization of the work of \cite{Kinney:2005ej,Gadde:2009kb,Gadde:2011ik,Gadde:2011uv,Lemos:2012ph} to type $E$ theories. Here, we briefly summarize our procedure in \cite{Chacaltana:2014jba}.

We assume the Hall-Littlewood index for a fixture in the $E_7$ theory takes the form

\begin{equation}
\mathcal{I}=\sum_{\lambda}\frac{\prod_{i=1}^3\mathcal{K}(\mathbf{a}_i)P^\lambda(\mathbf{a}_i|\tau)}{\mathcal{K}(\{\tau\})P^\lambda(\{\tau\}|\tau)}
\label{SCI}\end{equation}
where

\begin{itemize}%
\item The sum is over partitions $\lambda$ labeling the highest weights of finite-dimensional irreducible representations of $\mathfrak{e}_7$.

\item The $P^\lambda(\mathbf{a}_i|\tau)$ are Hall-Littlewood polynomials, defined for general $\mathfrak{g}$ by

\end{itemize}
\begin{displaymath}
\begin{split} P^\lambda&=W^{-1}(\tau) \sum_{w \in W}w\left(e^\lambda \prod_{\alpha \in R^+}\frac{1-\tau^2e^{-\alpha}}{1-e^{-\alpha}}\right) \\ W(\tau)& = \sqrt{\sum_{\stackrel{w \in W}{w\lambda=\lambda}}\tau^{2\ell(w)}} \end{split}
\end{displaymath}
where $R^+$ is the set of positive roots, $W$ the Weyl group, and $\ell(w)$ the length of the Weyl group element $w$.

\begin{itemize}%
\item $\mathbf{a}_i \equiv \{e^\alpha\}_{\alpha \in R^+}$ denotes a set of flavor fugacities for the flavor symmetry of the $i^\text{th}$ puncture. The set $\{\tau\}$ is the set of fugacities for the trivial puncture.

\item To compute the $\mathcal{K}$ factors, first decompose the adjoint representation of $\mathfrak{g}$ as in \eqref{decomp}. The $\mathcal{K}$ factors are then given by

\end{itemize}
\begin{displaymath}
\mathcal{K}(\mathbf{a})=\text{PE}[\sum_n \tau^{n+1}\chi^{R_n}_{\mathfrak{f}}(\mathbf{a})].
\end{displaymath}
We classify each fixture using the Hall-Littlewood superconformal index following \cite{Gaiotto:2012uq}. For a ``good" fixture, expanding the index in the superconformal fugacity $\tau$ gives

\begin{equation}
\mathcal{I} = 1 + \chi^{R}_\text{F} \tau + \chi^{adj}_{G_\text{fixt}}\tau^2+\cdots
\label{schematic}\end{equation}
The coefficient of $\tau$ signals the presence of free hypermultiplets transforming in the representation $R$ of flavor symmetry $F$, while the coefficient of $\tau^2$ is the character of the adjoint representation of the global symmetry of the fixture, which is a product $G_\text{fixt}=G_\text{SCFT} \times F$ of the global symmetry of the SCFT and the global symmetry of the free hypers.

Expanding the index $\mathcal{I}_\text{free}=PE[\tau \chi^{R}_\text{F}]$ of the free hypers and removing their contribution from \eqref{schematic}, we arrive at

\begin{displaymath}
\begin{split}
\mathcal{I}_\text{SCFT}&=\mathcal{I}/\mathcal{I}_\text{free} \\
&=1+ \chi^{adj}_{G_\text{SCFT}}\tau^2+\cdots
\end{split}
\end{displaymath}
from which we read off the global symmetry of the SCFT.

To determine when a fixture has an enhanced global symmetry, we note that in \eqref{SCI} the first term in the sum over representations (coming from the trivial representation of $\mathfrak{e}_7$) gives, to second order in $\tau$ \cite{Gaiotto:2012uq},

\begin{displaymath}
\mathcal{I}=1+\chi^{adj}_{G_\text{manifest}}\tau^2+\cdots
\end{displaymath}
encoding the manifest global symmetry group. The fixture has an enhanced global symmetry if there are terms contributing at order $\tau^2$ coming from the sum over $\lambda \gt 0$.

As explained in \cite{Chacaltana:2014jba}, to order $\tau^2$ \eqref{SCI} simplifies to  

\begin{equation}
\mathcal{I}=1+\chi^{adj}_{G_\text{manifest}}\tau^2 + {\left[\sum_{\lambda \gt 0}\frac{\prod_{i=1}^3\chi^\lambda(\mathbf{a_i}|\tau)}{\chi^\lambda(\{\tau\}|\tau)}\right]}_{\mathcal{O}(\tau^2)}
\label{expansion}\end{equation}
To compute \eqref{expansion}, we consider each $\mathfrak{e}_7$ representation in the sum to be a reducible representation of $\mathfrak{su}(2) \times \mathfrak{f}$ and plug in the corresponding character expansion, where the embedded $\mathfrak{su}(2)$ has fugacity $\tau$. The decomposition of any $\mathfrak{e}_7$ representation in terms of $\mathfrak{su}(2) \times \mathfrak{f}$ representations can be obtained using the projection matrices listed in Appendix \ref{projection_matrices}.

As an example of such a calculation, the fixture

\begin{displaymath}
 \includegraphics[width=92pt]{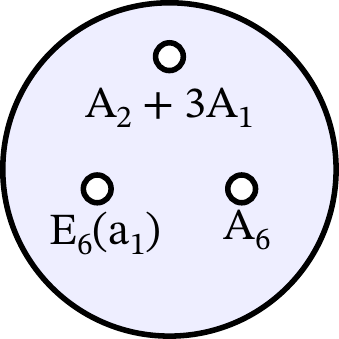}
\end{displaymath}
has manifest global symmetry $(G_2)_{28}\times SU(2)_{36}\times U(1)$. Its superconformal index has the expansion:

\begin{displaymath}
\begin{split} I=1+&\bigl[{(14,1)}_0+{(1,3)}_0+{(1,1)}_0\\ &+\underset{56}{\underbrace{{(1,2)}_1+{(1,2)}_{-1}}}+\underset{133}{\underbrace{{(7,3)}_0}}+\underset{912}{\underbrace{{(7,2)}_1+{(7,2)}_{-1}}}+\underset{1539}{\underbrace{{(14,1)}_0+{(7,1)}_0}}\bigr]\tau^2+\dots \end{split}
\end{displaymath}
where we've noted the representations in the sum in \eqref{SCI} which make additional contributions to the index at this order. Putting together these contributions, the global symmetry is enhanced to ${(E_6)}_{18}\times {(G_2)}_{10}$.

In computing the expansion of \eqref{SCI} to order $\tau^2$ we truncate the sum over representations. Knowing exactly at which representation we should truncate the sum for each fixture is tedious to determine due to the complicated Weyl group of $\mathfrak{e}_7$, so in practice we truncate the sum at a very large dimensional representation and check that our results are consistent with various S-dualities. Here, we summed over all irreducible representations of $\mathfrak{e}_7$ up to the $980,343$ dimensional irrep.

The largest representation of $\mathfrak{e}_7$ contributing at order $\tau^2$ was the $253,935$. This occurred for two fixtures:

\begin{displaymath}
 \includegraphics[width=92pt]{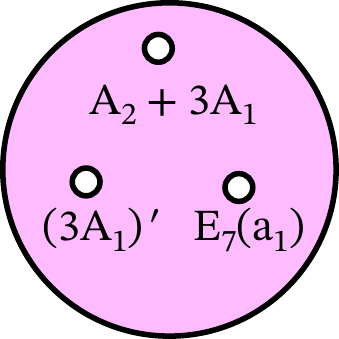}
\end{displaymath}
has manifest global symmetry ${Sp(3)}_{20}\times{SU(2)}_{19}\times{(G_2)}_{28}$. Its superconformal index picks up contributions at $O(\tau)$ from the $56$, $133$ and $912$ representations, indicating hypermultiplets transforming as the $\tfrac{1}{2}(6,1,7)+\tfrac{1}{2}(1,2,7)+\tfrac{1}{2}(6,1,1)$ of the manifest global symmetry. As shown below, the full index receives contributions from representations up to the $253,935$:

\[
\begin{split}
I=1&+\bigl[\underset{56}{\underbrace{(6,1,7)}}+\underset{133}{\underbrace{(1,2,7)}}+\underset{912}{\underbrace{(6,1,1)}}\bigr]\tau\\&+\bigl[\cdots+(1,3,1)+(1,1,14)+\underset{912}{\underbrace{(14',2,1)}}+\underset{40,755}{\underbrace{(14,1,7)}}+\underset{86,184}{\underbrace{(6,2,7)}}+\underset{253,935}{\underbrace{(21,1,1)}}\bigr]\tau^2
\end{split}
\]
where the $\dots$ indicate the contribution to $O(\tau^2)$ from the free hypermultiplets. This last contribution completes the enhancement of the global symmetry of the interacting SCFT to $(E_8)_{12}$, and this mixed fixture is the $(E_8)_{12}$ SCFT with 31 hypermultiplets.

The fixture

\begin{displaymath}
 \includegraphics[width=92pt]{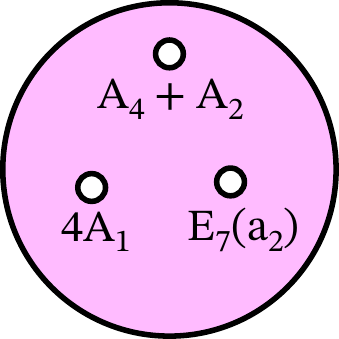}
\end{displaymath}
has hypermultiplets in the $\tfrac{1}{2}(6,3)+\tfrac{1}{2}(1,2)+\tfrac{1}{2}(1,4)$ and has manifest global symmetry ${Sp(3)}_{19}\times {SU(2)}_{108}$ enhanced to ${(E_7)}_{16}\times {SU(2)}_9$ with, again, the final enhancement coming from the $253,935$.

\section{Tinkertoys}\label{tinkertoys}
\subsection{Free-field fixtures}\label{freefield_fixtures}

We indicate a 3-punctured sphere, in the tables below, by listing the Bala-Carter labels of the three punctures. For all but one of the free-field fixtures, one of the punctures is an irregular puncture (in the sense used in our previous papers), which we denote by the pair $(\mathcal{O}, G)$, where $\mathcal{O}$ is the regular puncture obtained as the OPE of the two regular punctures which collide. This fixture is attached to the rest of the surface via a cylinder

\begin{displaymath}
(\mathcal{O}, G) \xleftrightarrow{\quad G\quad} \mathcal{O}
\end{displaymath}
with gauge group $G\subset E_7$. The exception is \#22, which consists of three regular punctures, and was first discussed in \cite{Chacaltana:2012zy}.

For each of the free-field fixtures, we indicate how the hypermultiplets transform under the manifest global symmetry of the fixture.

\newpage

\begin{longtable}{|c|c|c|c|}
\hline
\#&Fixture&$n_h$&Representation
\endhead
\hline
1&$\begin{matrix} \includegraphics[width=93px]{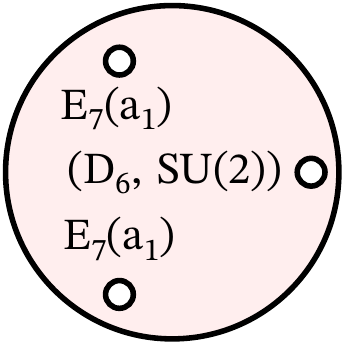}\end{matrix}$&1&$\tfrac{1}{2}(2)$\\
\hline
2&$\begin{matrix} \includegraphics[width=93px]{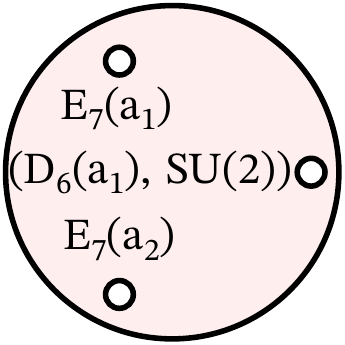}\end{matrix}$&0&empty\\
\hline
3&$\begin{matrix} \includegraphics[width=93px]{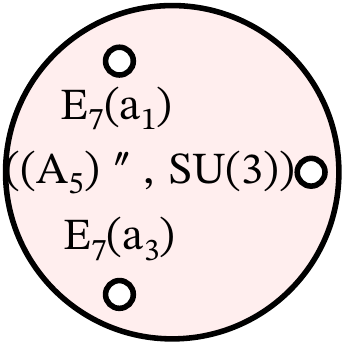}\end{matrix}$&0&empty\\
\hline
4&$\begin{matrix} \includegraphics[width=93px]{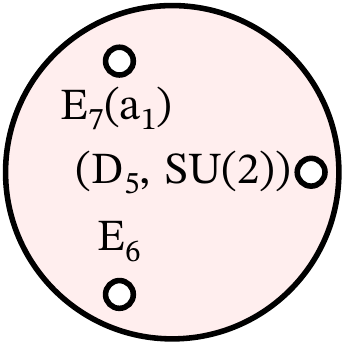}\end{matrix}$&0&empty\\
\hline
5&$\begin{matrix} \includegraphics[width=93px]{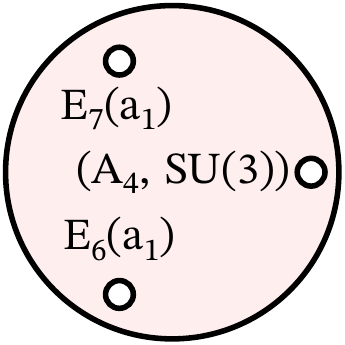}\end{matrix}$&0&empty\\
\hline
6&$\begin{matrix} \includegraphics[width=93px]{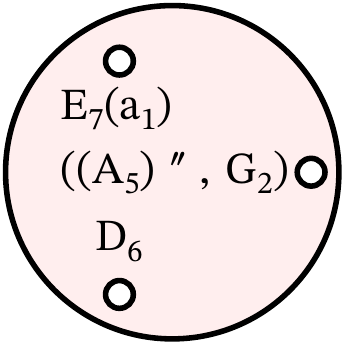}\end{matrix}$&7&$\tfrac{1}{2}(2,7)$\\
\hline
7&$\begin{matrix} \includegraphics[width=93px]{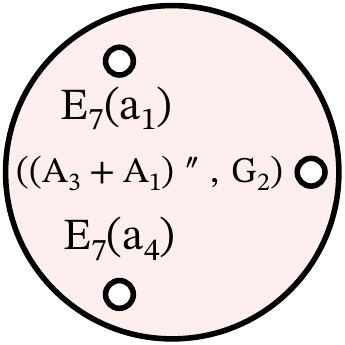}\end{matrix}$&0&empty\\
\hline
8&$\begin{matrix} \includegraphics[width=123px]{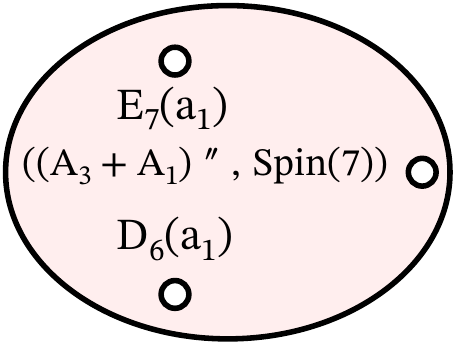}\end{matrix}$&8&$\tfrac{1}{2}(2,8)$\\
\hline
9&$\begin{matrix} \includegraphics[width=93px]{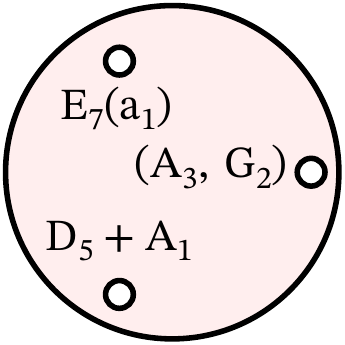}\end{matrix}$&0&empty\\
\hline
10&$\begin{matrix} \includegraphics[width=93px]{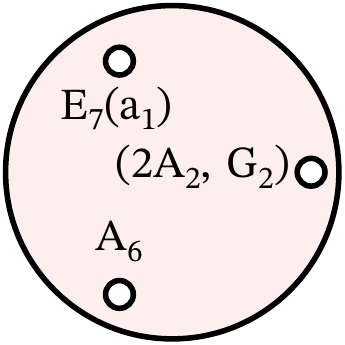}\end{matrix}$&0&empty\\
\hline
11&$\begin{matrix} \includegraphics[width=101px]{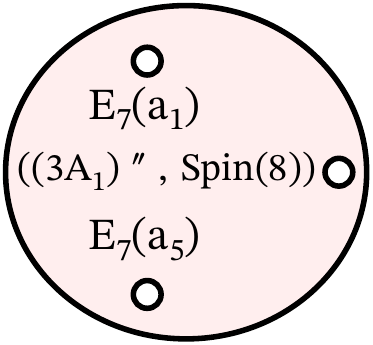}\end{matrix}$&0&empty\\
\hline
12&$\begin{matrix} \includegraphics[width=93px]{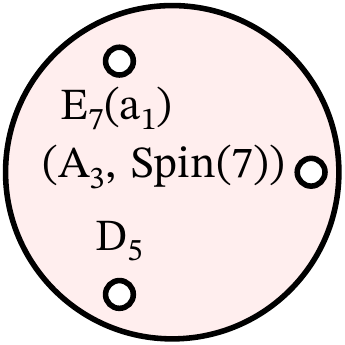}\end{matrix}$&8&$\tfrac{1}{2}(2,8)$\\
\hline
13&$\begin{matrix} \includegraphics[width=93px]{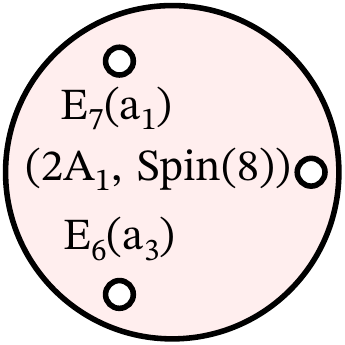}\end{matrix}$&0&empty\\
\hline
14&$\begin{matrix} \includegraphics[width=101px]{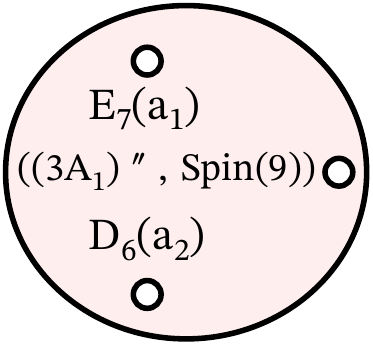}\end{matrix}$&9&$\tfrac{1}{2}(2,9)$\\
\hline
15&$\begin{matrix} \includegraphics[width=93px]{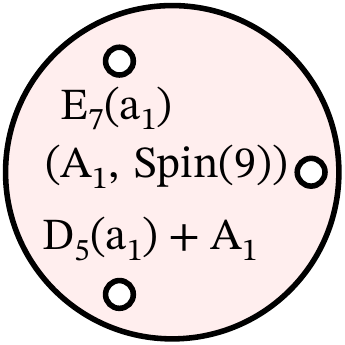}\end{matrix}$&0&empty\\
\hline
16&$\begin{matrix} \includegraphics[width=93px]{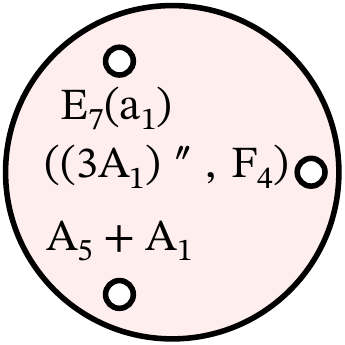}\end{matrix}$&26&$\tfrac{1}{2}(2,26)$\\
\hline
17&$\begin{matrix} \includegraphics[width=93px]{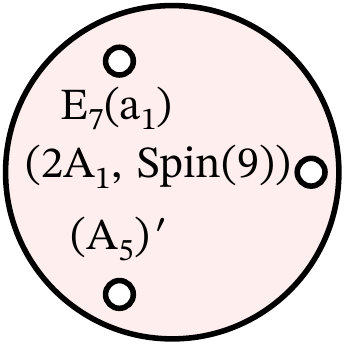}\end{matrix}$&9&$\tfrac{1}{2}(2,9)$\\
\hline
18&$\begin{matrix} \includegraphics[width=93px]{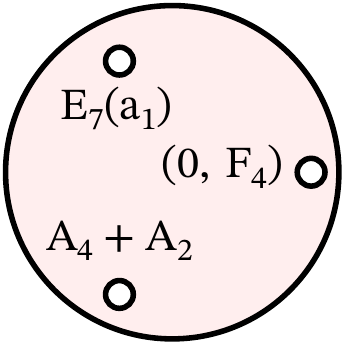}\end{matrix}$&0&empty\\
\hline
19&$\begin{matrix} \includegraphics[width=93px]{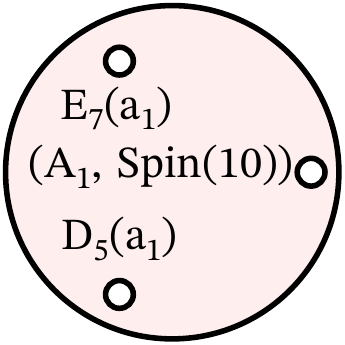}\end{matrix}$&10&$\tfrac{1}{2}(2,10)$\\
\hline
20&$\begin{matrix} \includegraphics[width=93px]{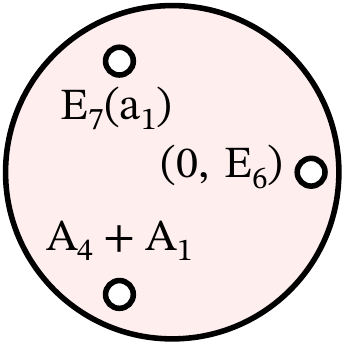}\end{matrix}$&27&$(27)$\\
\hline
21&$\begin{matrix} \includegraphics[width=93px]{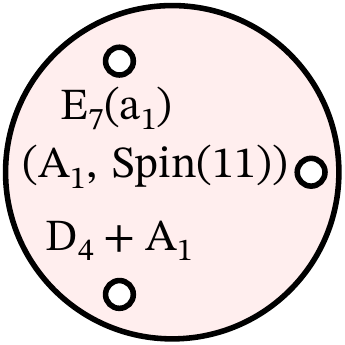}\end{matrix}$&22&$\tfrac{1}{2}(4,11)$\\
\hline
22&$\begin{matrix} \includegraphics[width=93px]{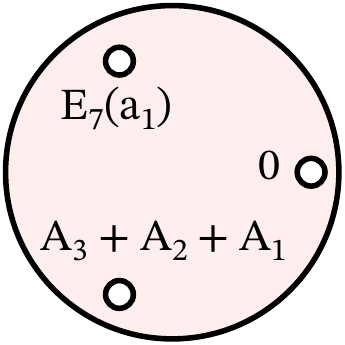}\end{matrix}$&84&$\tfrac{1}{2}(3,56)$\\
\hline
23&$\begin{matrix} \includegraphics[width=93px]{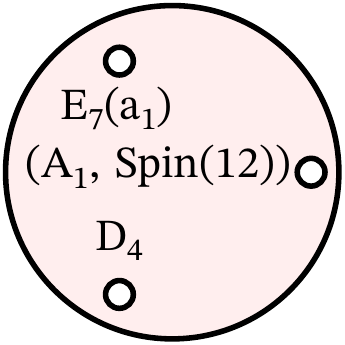}\end{matrix}$&36&$\tfrac{1}{2}(6,12)$\\
\hline
24&$\begin{matrix} \includegraphics[width=93px]{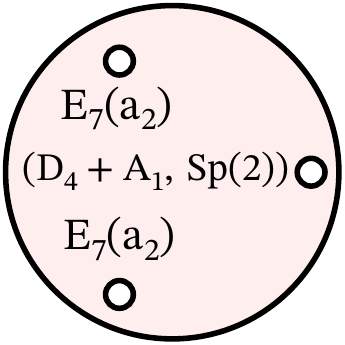}\end{matrix}$&2&$\tfrac{1}{2}(4)$\\
\hline
25&$\begin{matrix} \includegraphics[width=116px]{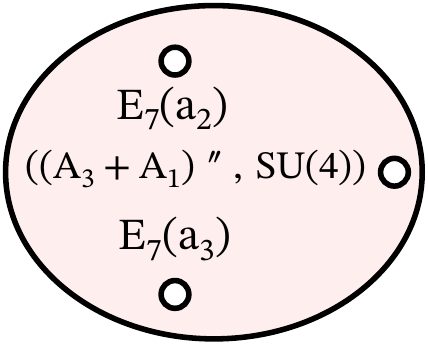}\end{matrix}$&0&empty\\
\hline
26&$\begin{matrix} \includegraphics[width=93px]{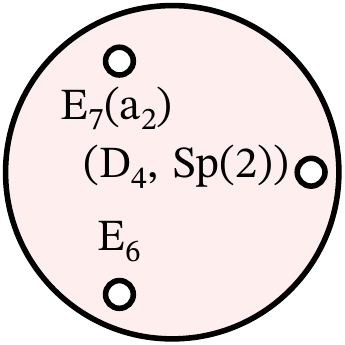}\end{matrix}$&0&empty\\
\hline
27&$\begin{matrix} \includegraphics[width=93px]{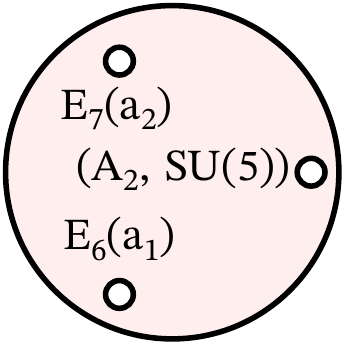}\end{matrix}$&0&empty\\
\hline
28&$\begin{matrix} \includegraphics[width=123px]{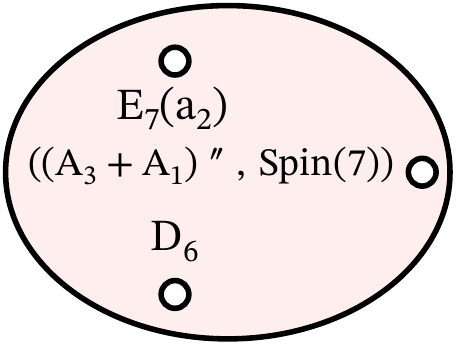}\end{matrix}$&7&$\tfrac{1}{2}(2,7)$\\
\hline
29&$\begin{matrix} \includegraphics[width=93px]{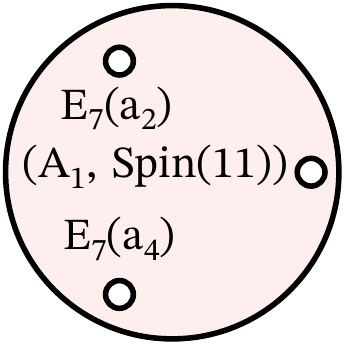}\end{matrix}$&16&$\tfrac{1}{2}(32)$\\
\hline
30&$\begin{matrix} \includegraphics[width=93px]{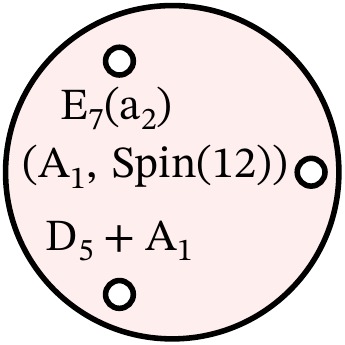}\end{matrix}$&28&$\tfrac{1}{2}(2,12)+\tfrac{1}{2}(1,32)$\\
\hline
31&$\begin{matrix} \includegraphics[width=93px]{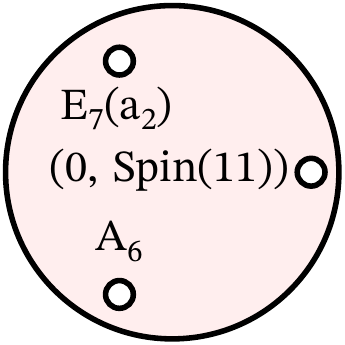}\end{matrix}$&0&empty\\
\hline
32&$\begin{matrix} \includegraphics[width=93px]{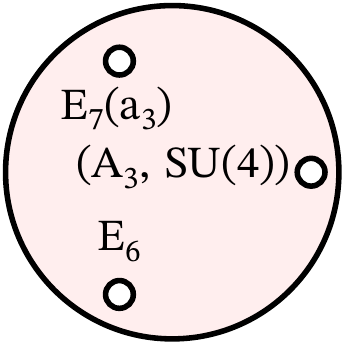}\end{matrix}$&0&empty\\
\hline
33&$\begin{matrix} \includegraphics[width=93px]{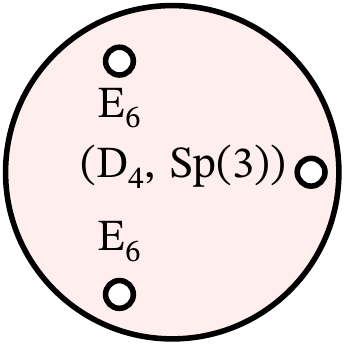}\end{matrix}$&12&$\tfrac{1}{2}(2,2,6)$\\
\hline
34&$\begin{matrix} \includegraphics[width=93px]{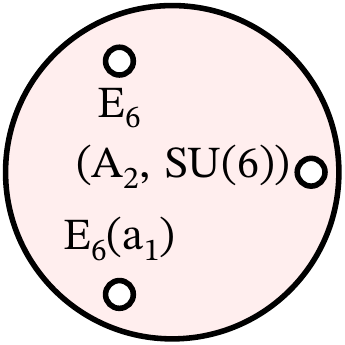}\end{matrix}$&12&$(2,6)$\\
\hline
35&$\begin{matrix} \includegraphics[width=93px]{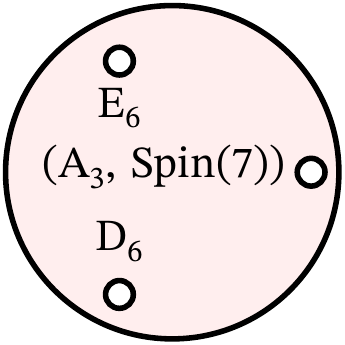}\end{matrix}$&7&$\tfrac{1}{2}(2,7)$\\
\hline
36&$\begin{matrix} \includegraphics[width=93px]{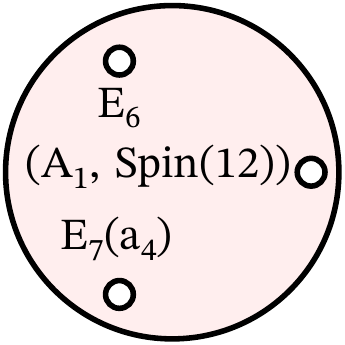}\end{matrix}$&28&$\tfrac{1}{2}(2,12)+\tfrac{1}{2}(1,32)$\\
\hline
37&$\begin{matrix} \includegraphics[width=93px]{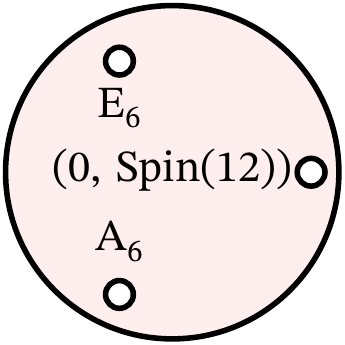}\end{matrix}$&12&$\tfrac{1}{2}(2,12)$\\
\hline
\end{longtable}

\subsection{Interacting fixtures with one irregular puncture}\label{interacting_fixtures_with_one_irregular_puncture}

There are 10 interacting fixtures involving two regular and one irregular puncture. They are all generalized Minahan-Nemeschansky theories (whose Higgs branches are (multi-)instanton moduli spaces for $E_{6,7,8}$) or products thereof, except for \#9 and \#10. They are the ${(F_4)}_{12}\times {SU(2)}_7^2$ and the ${Spin(16)}_{12}\times {SU(2)}_{8}$ theories which first appeared in \cite{Chacaltana:2011ze} as fixtures in the untwisted $D_4$ theory.

{\footnotesize
\renewcommand{\arraystretch}{2.25}

\begin{longtable}{|c|c|c|c|c|}
\hline
\#&Fixture&$(n_2,n_3,n_4,n_6,n_8,n_{10},n_{12},n_{14},n_{18})$&$(n_h,n_v)$&Theory\\
\hline
\endhead
1&$\begin{matrix} \includegraphics[width=93pt]{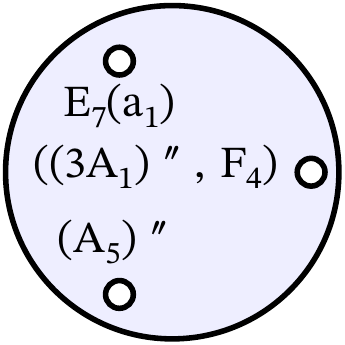}\end{matrix}$&$(0,0,0,1,0,0,0,0,0)$&$(40,11)$&${(E_8)}_{12}$ SCFT\\
\hline
2&$\begin{matrix} \includegraphics[width=93pt]{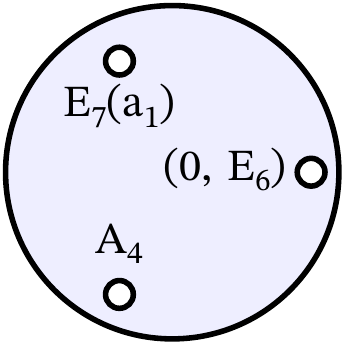}\end{matrix}$&$(0,0,0,1,0,0,0,0,0)$&$(40,11)$&${(E_8)}_{12}$ SCFT\\
\hline
3&$\begin{matrix} \includegraphics[width=93pt]{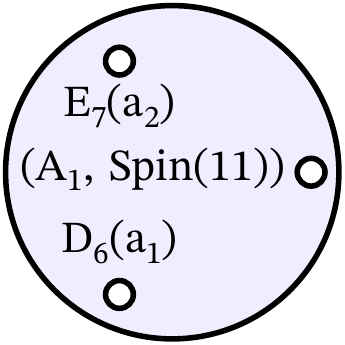}\end{matrix}$&$(0,0,1,0,0,0,0,0,0)$&$(24,7)$&${(E_7)}_{8}$ SCFT\\
\hline
4&$\begin{matrix} \includegraphics[width=93pt]{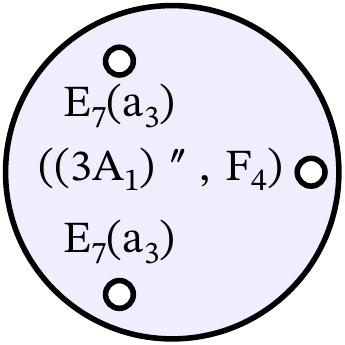}\end{matrix}$&$(0,2,0,0,0,0,0,0,0)$&$(32,10)$&${\left[{(E_6)}_{6}\, \text{SCFT}\right]}^2$\\
\hline
5&$\begin{matrix} \includegraphics[width=93pt]{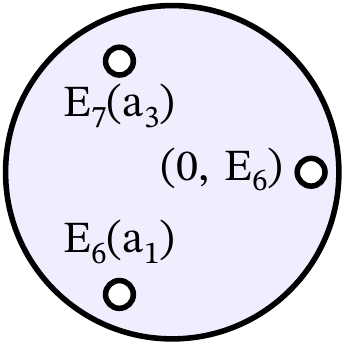}\end{matrix}$&$(0,2,0,0,0,0,0,0,0)$&$(32,10)$&${\left[{(E_6)}_{6}\, \text{SCFT}\right]}^2$\\
\hline
6&$\begin{matrix} \includegraphics[width=93pt]{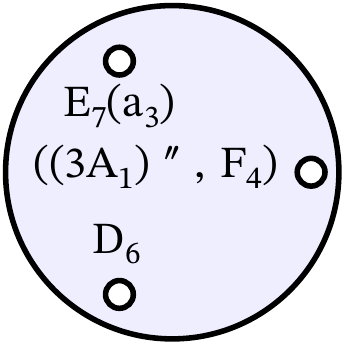}\end{matrix}$&$(0,1,0,1,0,0,0,0,0)$&$(39,16)$&${(E_6)}_{12}\times {SU(2)}_7$ SCFT\\
\hline
7&$\begin{matrix} \includegraphics[width=93pt]{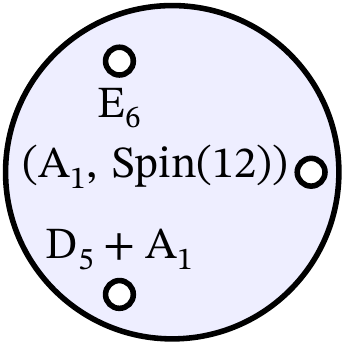}\end{matrix}$&$(0,0,0,1,0,0,0,0,0)$&$(40,11)$&${(E_8)}_{12}$ SCFT\\
\hline
8&$\begin{matrix} \includegraphics[width=93pt]{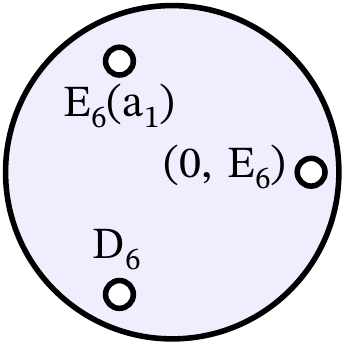}\end{matrix}$&$(0,1,0,1,0,0,0,0,0)$&$(39,16)$&${(E_6)}_{12}\times {SU(2)}_7$ SCFT\\
\hline
9&$\begin{matrix} \includegraphics[width=93pt]{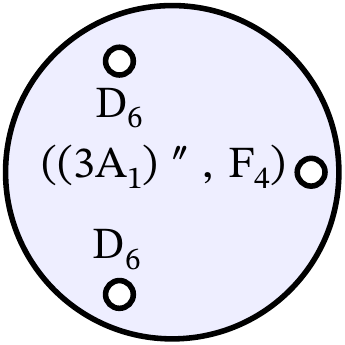}\end{matrix}$&$(0,0,0,2,0,0,0,0,0)$&$(46,22)$&${(F_4)}_{12}\times {SU(2)}_7^2$ SCFT\\
\hline
10&$\begin{matrix} \includegraphics[width=93pt]{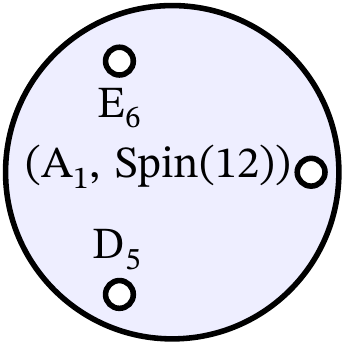}\end{matrix}$&$(0,0,1,1,0,0,0,0,0)$&$(48,18)$&${Spin(16)}_{12}\times {SU(2)}_8$ SCFT\\
\hline
\end{longtable}

}

\subsection{Mixed fixtures with one irregular puncture}\label{mixed_fixtures_with_one_irregular_puncture}

There are two mixed fixtures with two regular and one irregular puncture. The value of $n_h$ listed below is the one associated to the SCFT, \emph{after} subtracting the contribution of the free hypermultiplets.

\medskip

{\footnotesize
\renewcommand{\arraystretch}{2.25}

\begin{tabular}{|c|c|c|c|c|}
\hline
\#&Fixture&$(n_2,n_3,n_4,n_6,n_8,n_{10},n_{12},n_{14},n_{18})$&$(n_h,n_v)$&Theory\\
\hline 
1&$\begin{matrix} \includegraphics[width=93pt]{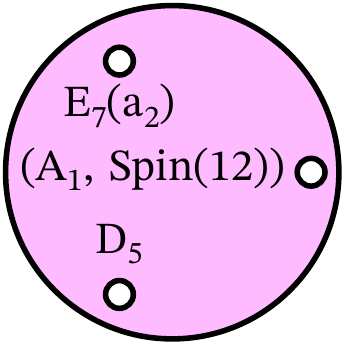}\end{matrix}$&$(0,0,1,0,0,0,0,0,0)$&$(24,7)$&$(E_7)_{8}\,\text{SCFT} + \tfrac{1}{2}(2,12)$\\
\hline 
2&$\begin{matrix} \includegraphics[width=93pt]{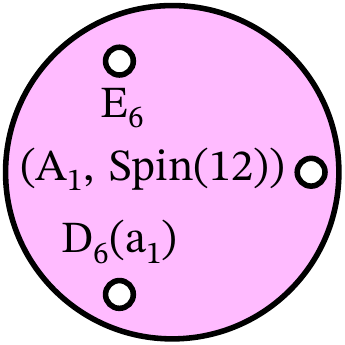}\end{matrix}$&$(0,0,1,0,0,0,0,0,0)$&$(24,7)$&$(E_7)_{8}\,\text{SCFT} + \tfrac{1}{2}(2,12)$\\
\hline
\end{tabular}
}

\subsection{Interacting and mixed fixtures}\label{interacting_and_mixed_fixtures}

There are exactly 11,000 fixtures with three regular punctures. Of these, 654 have enhanced global symmetry, 262 are mixed, and 1 is free.

Rather than listing all of these, we have created a web application where the interested reader can explore these theories for him or herself. The website, \href{https://golem.ph.utexas.edu/class-S/E7/}{https://golem.ph.utexas.edu/class-S/E7/}\, , has three sections:

\begin{itemize}
\item A compendium of the 44 regular punctures and their properties:\newline\phantom{A}\qquad\href{https://golem.ph.utexas.edu/class-S/E7/punctures/}{https://golem.ph.utexas.edu/class-S/E7/punctures/}
\item A compendium of the 11,000 3-punctured spheres:\newline\phantom{A}\qquad\href{https://golem.ph.utexas.edu/class-S/E7/fixtures/}{https://golem.ph.utexas.edu/class-S/E7/fixtures/}
\item A compendium of the 178,365 4-punctured spheres and their S-duality frames: \newline\phantom{A}\qquad\href{https://golem.ph.utexas.edu/class-S/E7/four_punctured_sphere/}{https://golem.ph.utexas.edu/class-S/E7/four\_punctured\_sphere/}
\end{itemize}

\noindent
For each S-duality frame, clicking on a fixture brings up its properties. When viewing a fixture, clicking on a puncture brings up the latter's properties.

If you find the data on the website useful in your own work, please cite \emph{this} paper instead.

\section{Applications}\label{applications}

\subsection{$E_7+3(56)$}\label{}

$E_7$ gauge theory, with three fundamental hypermultiplets, is superconformal. It is realized as the 4-punctured sphere

\begin{displaymath}
 \includegraphics[width=264pt]{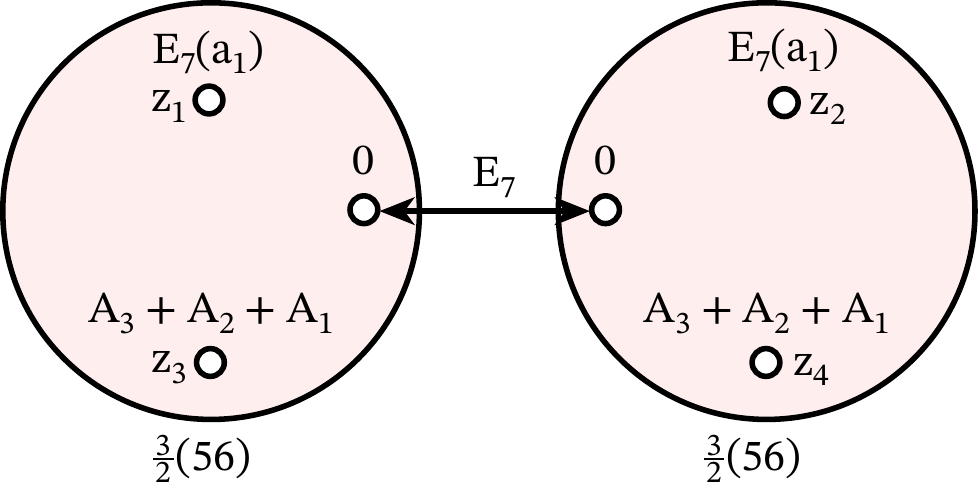}
\end{displaymath}
The S-dual theory is an $SU(2)$ gauging of the $SU(4)_{112} \times SU(2)_7$ SCFT, with an additional half-hypermultiplet in the fundamental.

\begin{displaymath}
 \includegraphics[width=264pt]{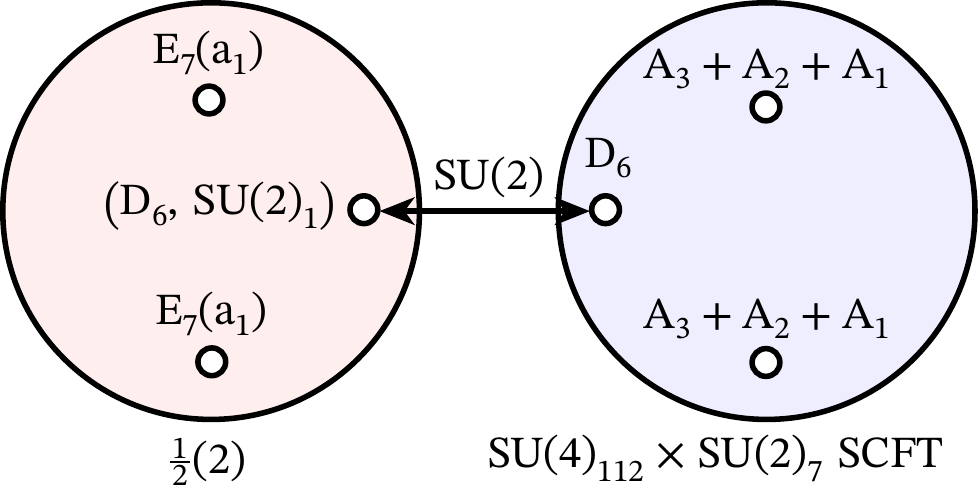}
\end{displaymath}
The $k$-differentials, which determine the Seiberg-Witten solution, are

\begin{displaymath}
\begin{split}
\phi_2(z) &= \frac{u_2\, z_{1 2} z_{3 4}\, {(d z)}^2}{ {(z-z_1)}{(z-z_2)}{(z-z_3)}{(z-z_4)} }\\
\phi_6(z) &= \frac{u_6\, {z_{1 2}^2} {z_{3 4}^4} {(d z)}^6}{{(z-z_1)^2}{(z-z_2)^2}{(z-z_3)^4}{(z-z_4)^4}} \\
\phi_8(z) &= \frac{u_8\, {z_{1 2}^2} {z_{3 4}^6} {(d z)^8}}{{(z-z_1)^2}{(z-z_2)^2}{(z-z_3)^6}{(z-z_4)^6}}\\
\phi_{10}(z) &= \frac{u_{10}\, {z_{1 2}^2} {z_{3 4}^8} {(d z)^{10}}}{{(z-z_1)^2}{(z-z_2)^2}{(z-z_3)^8}{(z-z_4)^8}}\\
\phi_{12}(z) &= \frac{u_{12}\, {z_{1 2}^3} {z_{3 4}^9} {(d z)^{12}}}{{(z-z_1)^3}{(z-z_2)^3}{(z-z_3)^9}{(z-z_4)^9}}\\
\phi_{14}(z) &= \frac{u_{14}\, {z_{1 2}^3} {z_{3 4}^{11}} {(d z)^{14}}}{{(z-z_1)^3}{(z-z_2)^3}{(z-z_3)^{11}}{(z-z_4)^{11}}}\\
\phi_{18}(z) &=  \frac{u_{18}\, {z_{1 2}^4} {z_{3 4}^{14}} {(d z)^{18}}}{{(z-z_1)^4}{(z-z_2)^4}{(z-z_3)^{14}}{(z-z_4)^{14}}}
\end{split}
\end{displaymath}

\subsection{Adding $(E_8)_{12}$ SCFTs}\label{adding__scfts}

Since the index of the 56 of $E_7$ is 12, we can start with the $E_7+3(56)$ gauge theory and trade half-hypermultiplets in the 56 for copies of the $(E_8)_{12}$ SCFT. A similar analysis was carried out for the $E_6+4(27)$ gauge theory in \cite{Chacaltana:2014jba}.

For $n$ half-hypermultiplets in the 56 and $6-n$ copies of the $(E_8)_{12}$ SCFT, the theory has flavor symmetry

\begin{displaymath}
F = SU(2)_{12}^{6-n} \times SO(n)_{k},
\end{displaymath}
where $k=112$ for $n \neq 3$, and $k=224$ for $n=3$. Each of these theories has an S-dual description as an $SU(2)$ gauging of the $SU(2)_{12}^{6-n} \times SO(n)_{k} \times SU(2)_7$ SCFT, with an additional half-hypermultiplet in the fundamental.

\subsubsection*{$n=5$}\label{_2}

With one copy of the $(E_8)_{12}$ SCFT,

\begin{displaymath}
 \includegraphics[width=264pt]{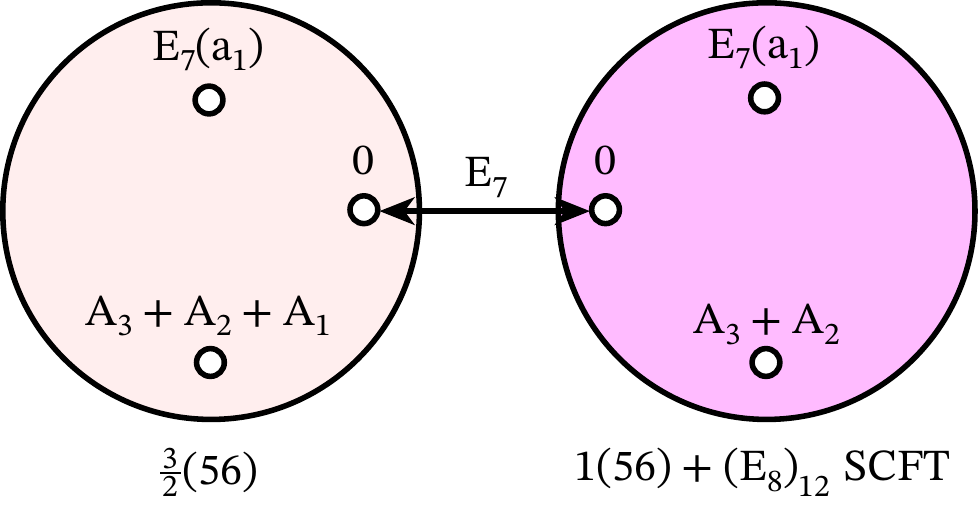}
\end{displaymath}
is dual to

\begin{displaymath}
 \includegraphics[width=311pt]{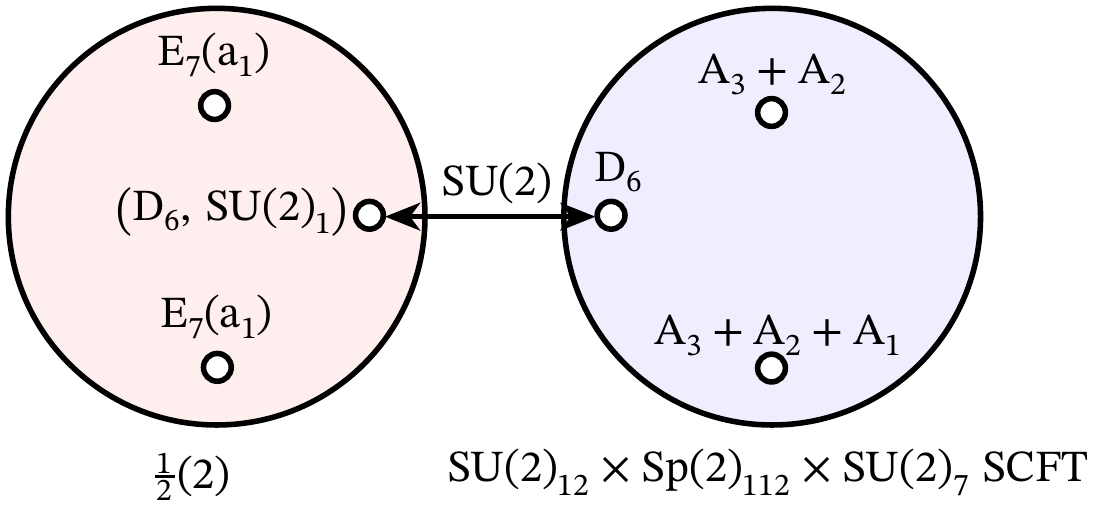}
\end{displaymath}

\subsubsection*{$n=4$}\label{_3}

With two copies of the $(E_8)_{12}$ SCFT,

\begin{displaymath}
 \includegraphics[width=264pt]{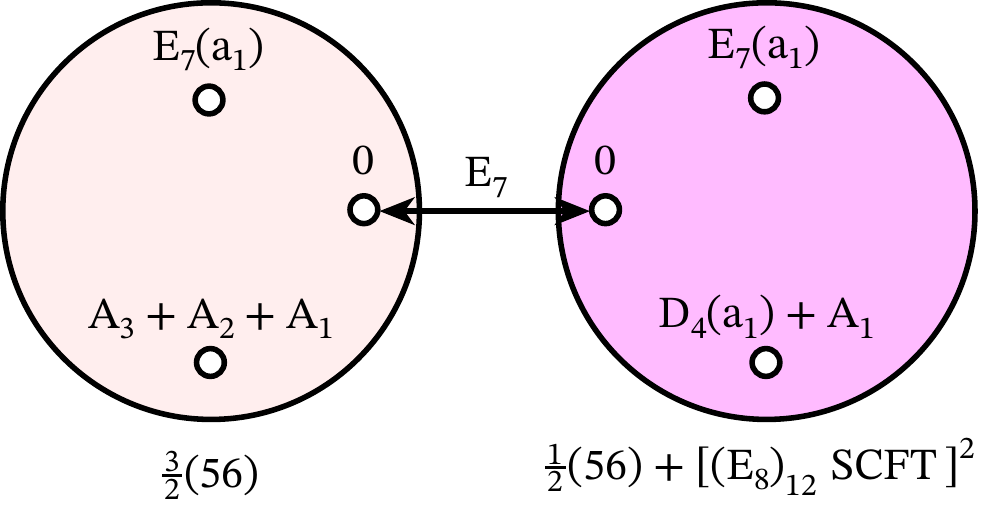}
\end{displaymath}
is dual to
\begin{displaymath}
 \includegraphics[width=311pt]{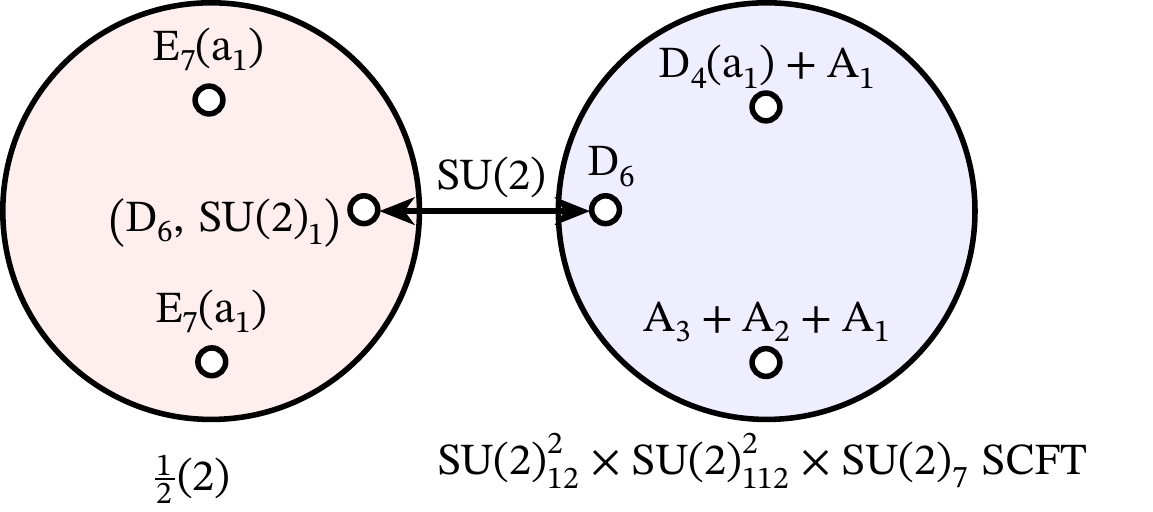}
\end{displaymath}

\subsubsection*{$n=3$}\label{_4}

With three copies of the $(E_8)_{12}$ SCFT,

\begin{displaymath}
 \includegraphics[width=264pt]{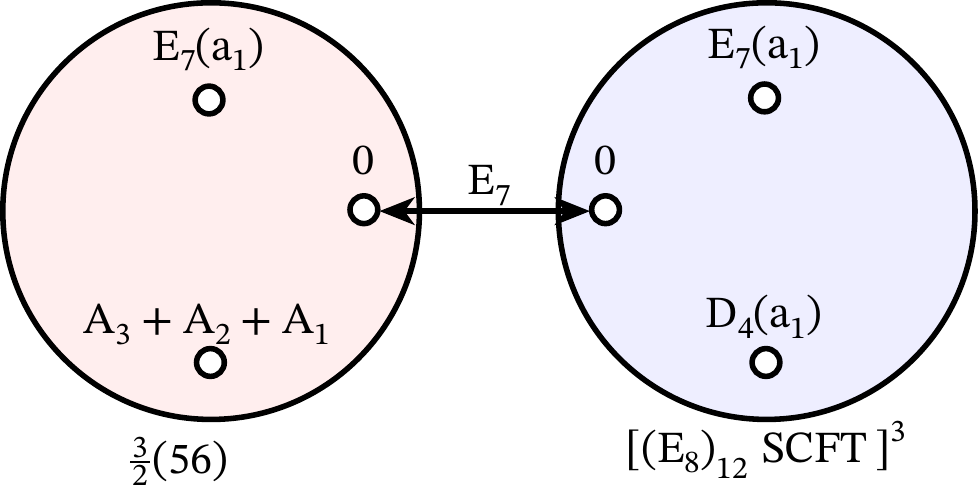}
\end{displaymath}
is dual to

\begin{displaymath}
 \includegraphics[width=311pt]{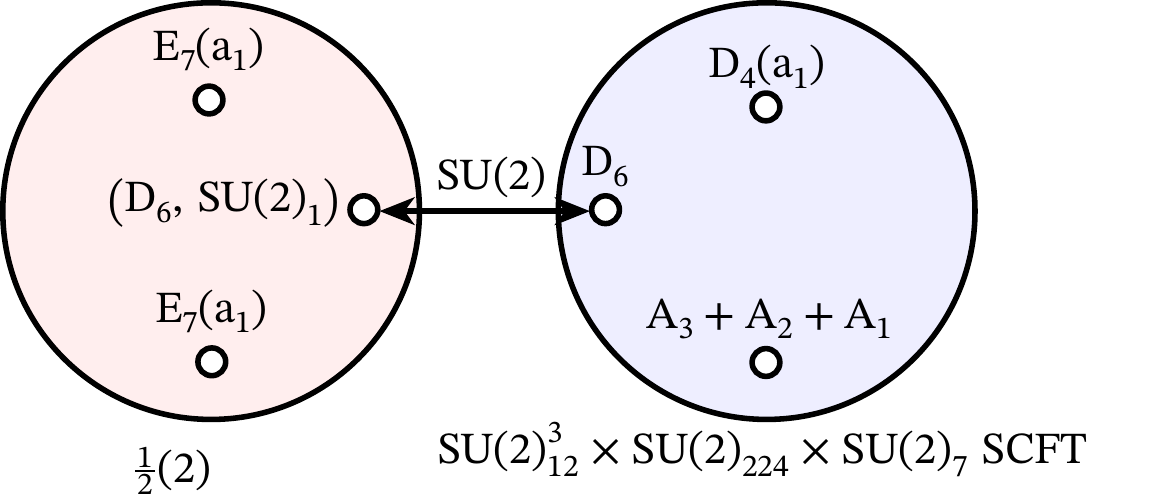}
\end{displaymath}

\subsubsection*{$n=2$}\label{_5}

With four copies of the $(E_8)_{12}$ SCFT, we have two possible realizations. Either,

\begin{displaymath}
 \includegraphics[width=264pt]{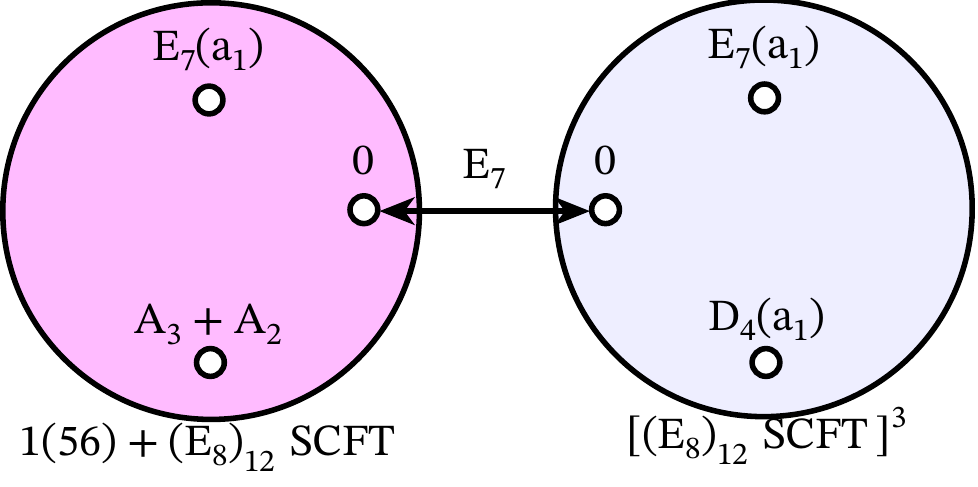}
\end{displaymath}
is dual to

\begin{displaymath}
 \includegraphics[width=310pt]{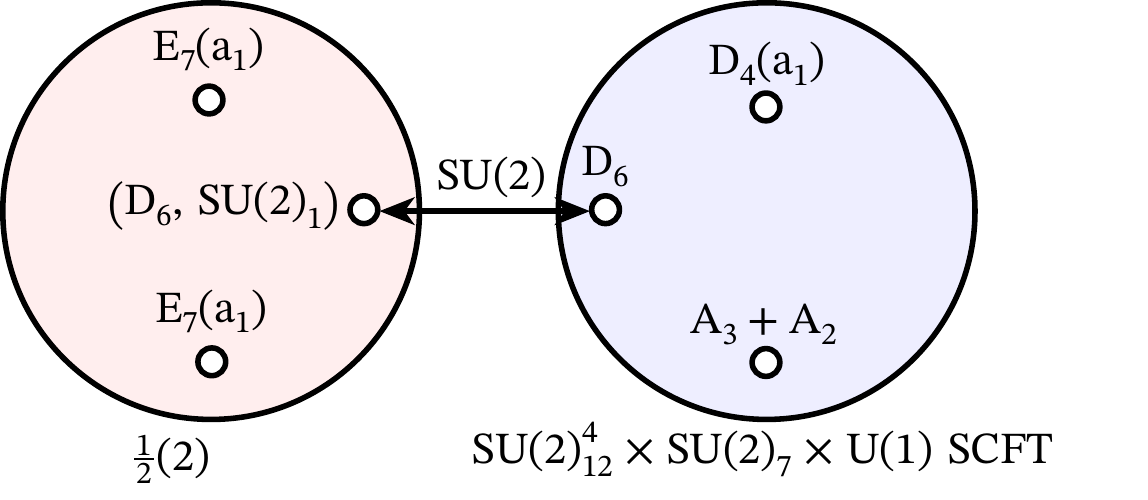}
\end{displaymath}
or

\begin{displaymath}
 \includegraphics[width=278pt]{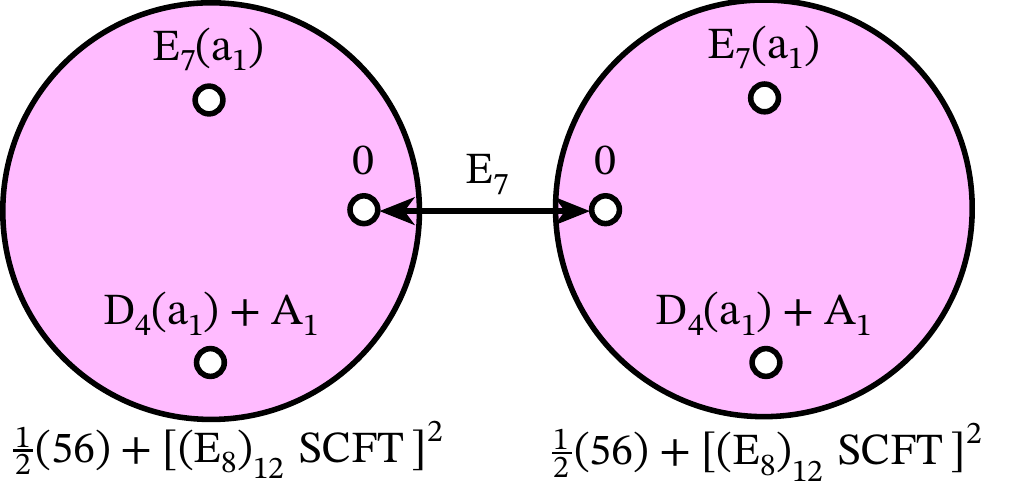}
\end{displaymath}
is dual to

\begin{displaymath}
 \includegraphics[width=310pt]{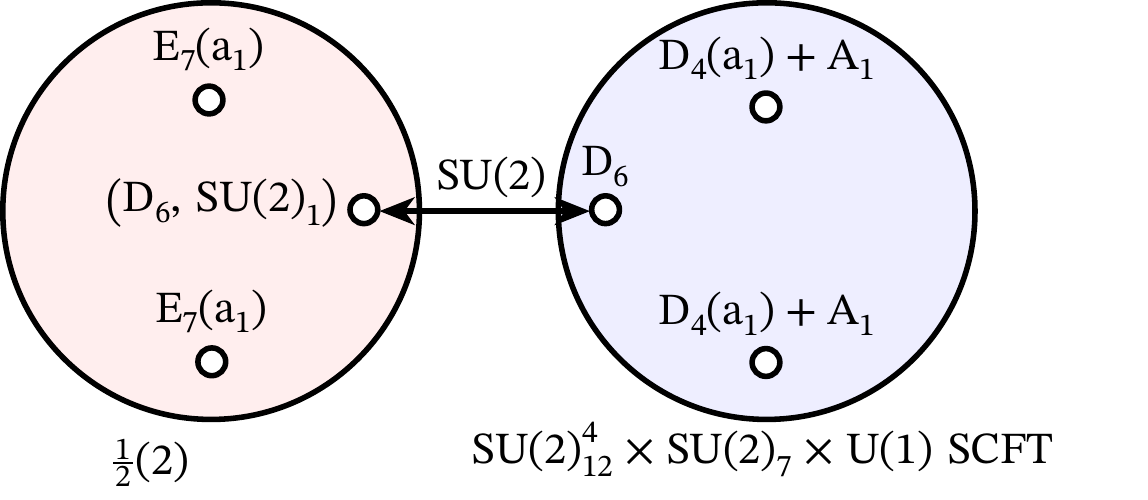}
\end{displaymath}
This gives two distinct realizations of the $SU(2)_{12}^4 \times SU(2)_7 \times U(1)$ SCFT.

\subsubsection*{$n=1$}\label{_6}

With five copies of the $(E_8)_{12}$ SCFT,

\begin{displaymath}
 \includegraphics[width=264pt]{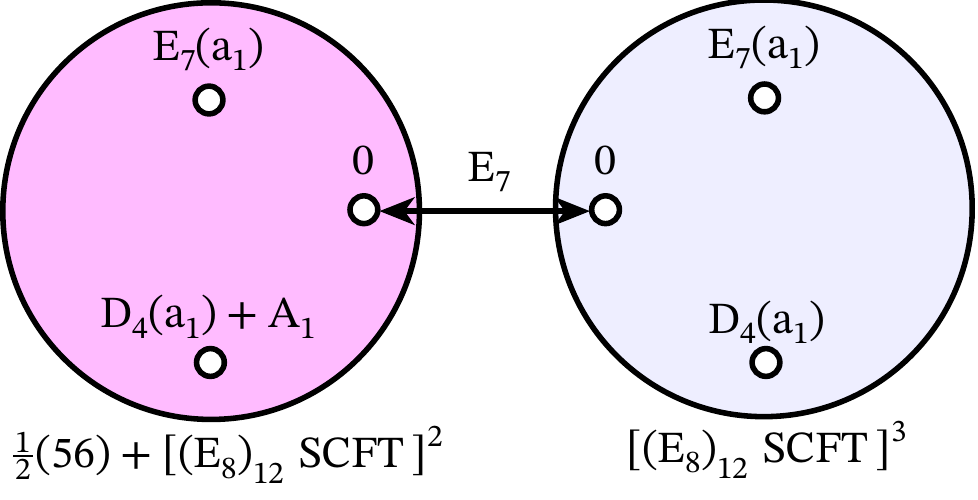}
\end{displaymath}
is dual to

\begin{displaymath}
 \includegraphics[width=310pt]{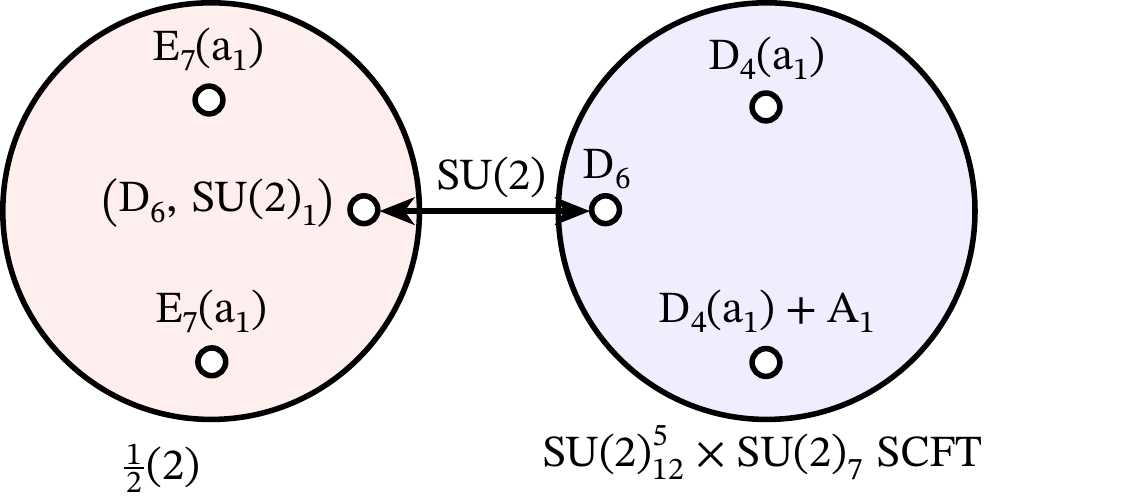}
\end{displaymath}

\subsubsection*{$n=0$}\label{_7}

Finally, the $E_7$ gauging of six copies of the $(E_8)_{12}$ SCFT,

\begin{displaymath}
 \includegraphics[width=264pt]{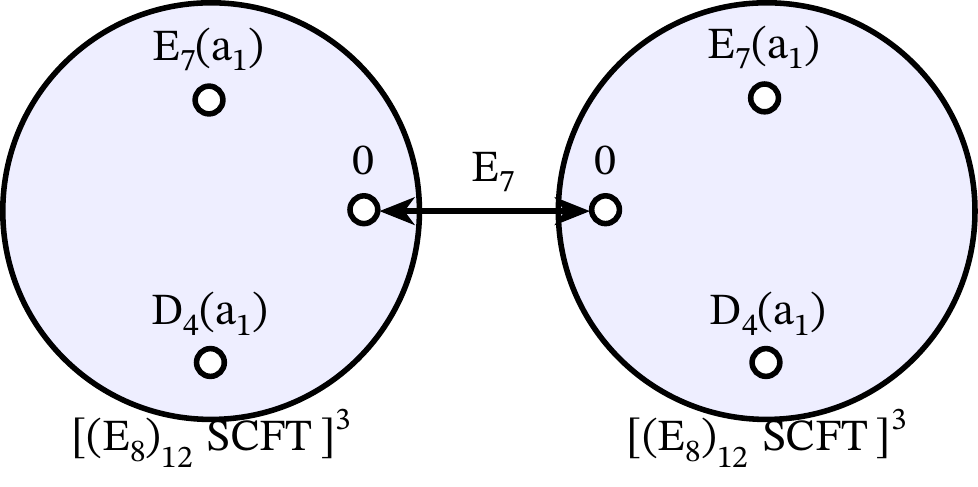}
\end{displaymath}
is dual to

\begin{displaymath}
 \includegraphics[width=273pt]{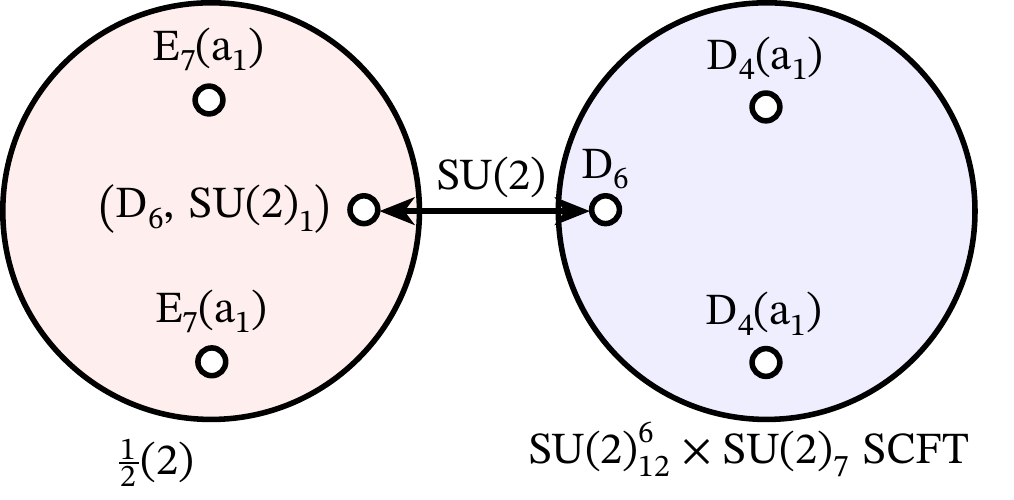}
\end{displaymath}

\subsection{New 6d realizations of SCFTs}\label{new_6d_realizations_of_scfts}

\subsubsection{Higher-rank Minahan-Nemeschansky SCFTs}\label{higher_rank_minahannemeschansky_scfts}

The rank-$n$ Minahan-Nemeschansky theories have Higgs branches which are the $n$-instanton moduli space for $E_{6,7,8}$. They are realized in F-theory as the SCFT on $n$ D3-branes probing a $\text{IV}^*$, $\text{III}^*$ or $\text{II}^*$ singularity. For small values of $n$, they appear ubiquitously among our fixtures. Here, we find our first realization, in the E-series, of the $(E_8)_{36} \times SU(2)_{38}$ SCFT, which is the theory on $n=3$ D3 branes probing a $II^*$ singularity in F-theory. This is realized on the fixture

\begin{displaymath}
 \includegraphics[width=92pt]{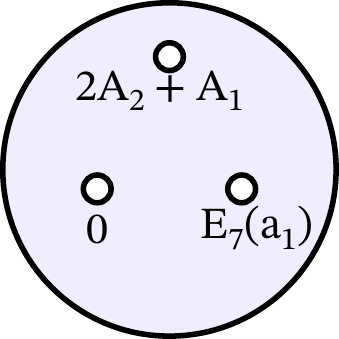}
\end{displaymath}
\subsubsection{Other low-rank SCFTs}\label{other_lowrank_scfts}

In addition to various Minahan-Nemeschansky theories, the $(F_4)_{12}\times SU(2)_7^2$ theory and the $(E_8)_{12}\times SU(2)_8$ theory (see \S\ref{interacting_fixtures_with_one_irregular_puncture}), we find two additional rank-2 SCFTs. The ${Spin(20)}_{16}$ SCFT, for which we find a new realization, appeared previously as the mixed fixture

\begin{displaymath}
 \includegraphics[width=93pt]{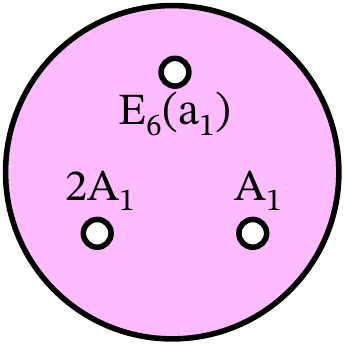}
\end{displaymath}
in the $E_6$ theory. The $Sp(6)_{11}$ theory is new.

\medskip

{\parindent=-.125in
\renewcommand{\arraystretch}{2.25}

\begin{tabular}{|c|c|c|c|c|}
\hline 
Fixture&$(n_2,n_3,n_4,n_5,n_6,n_8,n_9,n_{10},n_{12},n_{14},n_{18})$&$(n_h,n_v)$&\mbox{\shortstack{Global\\Symmetry}}&\mbox{\shortstack{Free\\Hypers}}\\
\hline 
$\begin{matrix} \includegraphics[width=93pt]{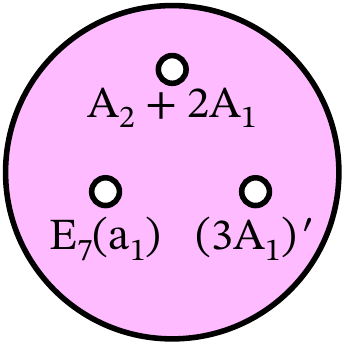}\end{matrix}$&$(0,0,0,0,1,1,0,0,0,0,0)$&$(72,26)$&${Spin(20)}_{16}$&$15$\\
\hline 
$\begin{matrix} \includegraphics[width=93pt]{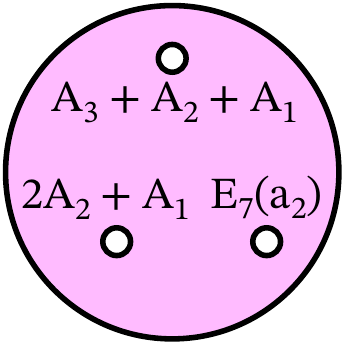}\end{matrix}$&$(0,0,1,0,0,0,0,1,0,0,0)$&$(58,26)$&${Sp(6)}_{11}$&$8$\\
\hline 
\end{tabular}
}
\medskip

We also find several new rank-3 SCFTs. The ${SU(12)}_{18}$ theory first appeared as the interacting fixture

\begin{displaymath}
 \includegraphics[width=93pt]{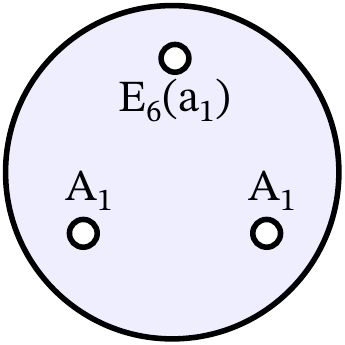}
\end{displaymath}
in the $E_6$ theory. Here, it has two distinct realizations as a mixed fixture. The ${Sp(3)}_{26}$ SCFT also appeared in the $E_6$ theory, as the mixed fixtures

\begin{displaymath}
\begin{matrix} \includegraphics[width=93pt]{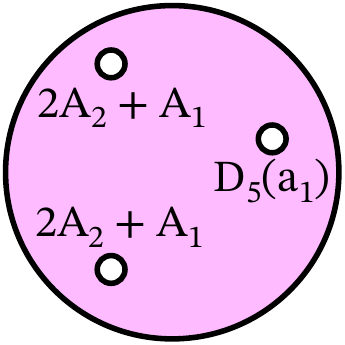}\end{matrix},\qquad
\begin{matrix} \includegraphics[width=93pt]{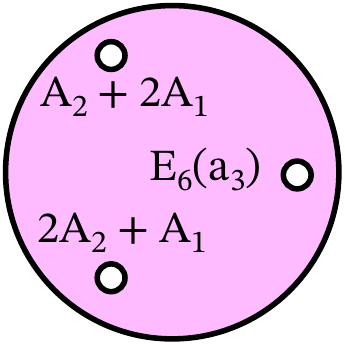}\end{matrix}
\end{displaymath}
and the ${Sp(3)}_{26}\times {SU(2)}_7$ SCFT appeared as

\begin{displaymath}
 \includegraphics[width=93pt]{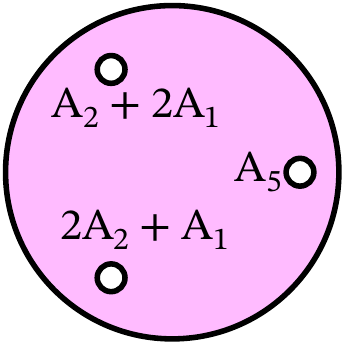}
\end{displaymath}

{\footnotesize
\setlength\LTleft{-.25in}

\begin{longtable}{|c|c|c|c|c|}\hline
Fixture&$(n_2,n_3,n_4,n_5,n_6,n_8,n_9,n_{10},n_{12},n_{14},n_{18})$&$(n_h,n_v)$&Global Symmetry&\mbox{\shortstack{Free\\Hypers}}\\
\hline
\endhead
$\begin{matrix} \includegraphics[width=93pt]{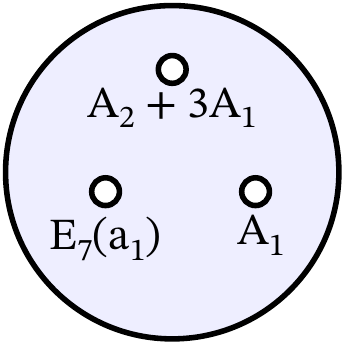}\end{matrix}$&$(0,0,0,0,1,0,0,0,1,1,0)$&$(136,61)$&${Spin(19)}_{28}$&$0$\\\hline
$\begin{matrix} \includegraphics[width=93pt]{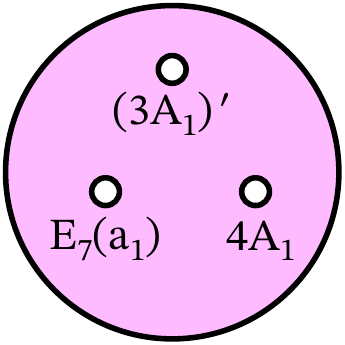}\end{matrix}$&$(0,0,0,0,1,1,0,0,0,0,1)$&$(125,61)$&${Sp(7)}_{19}$&$3$\\\hline
$\begin{matrix} \includegraphics[width=93pt]{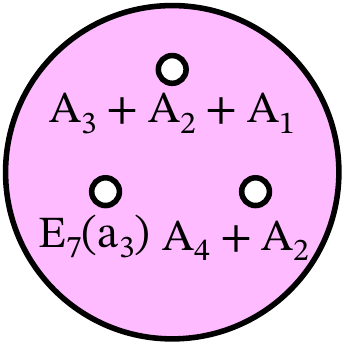}\end{matrix}$&$(0,1,0,0,0,1,0,0,1,0,0)$&$(70,43)$&${Sp(3)}_{26}$&$6$\\\hline
$\begin{matrix} \includegraphics[width=93pt]{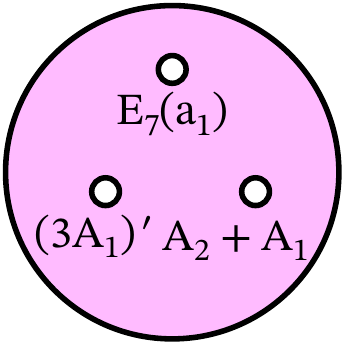}\\  \includegraphics[width=93pt]{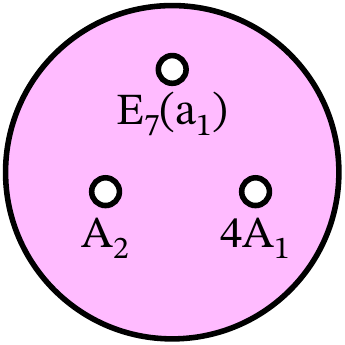}\end{matrix}$&$(0,0,0,0,1,1,1,0,0,0,0)$&$(100,43)$&${SU(12)}_{18}$&$\begin{matrix}7\\ \itexspace{10}{10}{0} \\ \itexspace{10}{10}{0} \\ 9\end{matrix}$\\\hline
$\begin{matrix} \includegraphics[width=93pt]{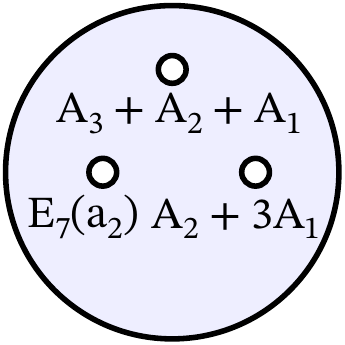}\end{matrix}$&$(0,0,1,0,0,0,0,1,0,1,0)$&$(96,53)$&${(E_6)}_{28}$&$0$\\\hline
$\begin{matrix} \includegraphics[width=93pt]{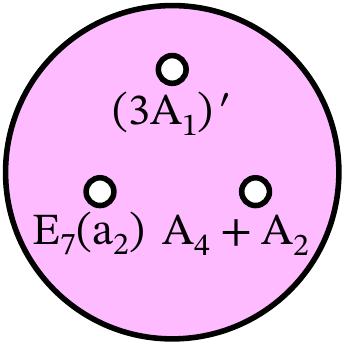}\end{matrix}$&$(0,0,1,0,0,1,0,1,0,0,0)$&$(88,41)$&$Spin(15)_{20} \times SU(2)_{16}$&$3$\\\hline
$\begin{matrix} \includegraphics[width=93pt]{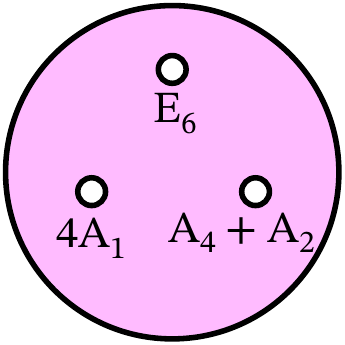}\end{matrix}$&$(0,0,1,0,1,1,0,0,0,0,0)$&$(72,33)$&${Spin(12)}_{16} \times Spin(7)_{12}$&$9$\\\hline
$\begin{matrix} \includegraphics[width=93pt]{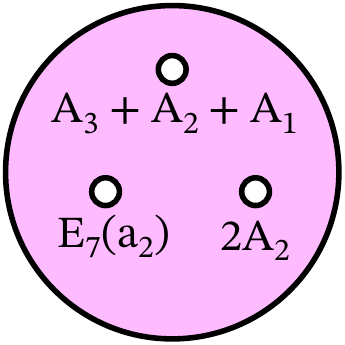}\end{matrix}$&$(0,0,1,0,0,1,0,1,0,0,0)$&$(81,41)$&${(F_4)}_{16} \times Sp(3)_{11}$&$3$\\\hline
$\begin{matrix} \includegraphics[width=93pt]{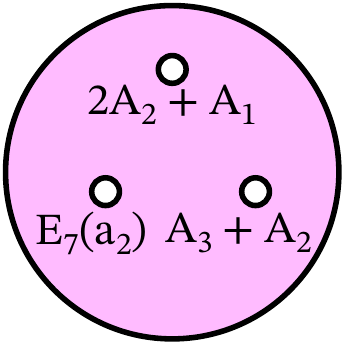}\\  \includegraphics[width=93pt]{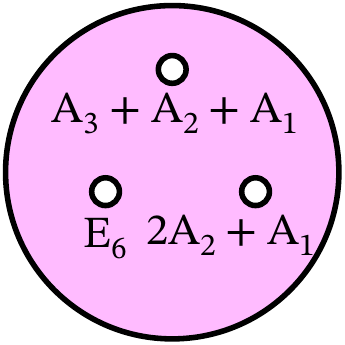}\end{matrix}$&$(0,0,1,0,1,0,0,1,0,0,0)$&$(73,37)$&$Sp(4)_{12} \times Sp(3)_{11}$&$5$\\\hline
$\begin{matrix} \includegraphics[width=93pt]{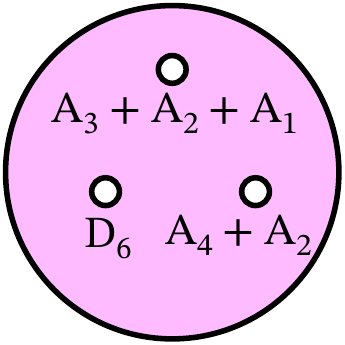}\end{matrix}$&$(0,0,0,0,1,1,0,0,1,0,0)$&$(77,49)$&$Sp(3)_{26} \times SU(2)_7$&$6$\\\hline
$\begin{matrix} \includegraphics[width=93pt]{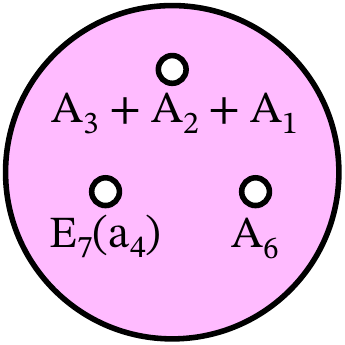}\end{matrix}$&$(0,0,0,0,1,1,0,1,0,0,0)$&$(73,45)$&$\begin{gathered}SU(2)_{128-k} \times SU(2)_k\\ \times Sp(3)_{11}\end{gathered}$&$3$\\\hline
\end{longtable}

}

\subsubsection{New SCFTs with exceptional global symmetry}\label{new_scfts_with_exceptional_global_symmetry}

In our list, we find 10 new SCFTs whose global symmetry is a simple exceptional group. One of these is the rank-3 ${(E_6)}_{28}$ example listed in \S\ref{other_lowrank_scfts}. We also find two new realizations of the ${(E_7)}_{24}$ SCFT, which was first found in \cite{Chacaltana:2014jba} as the interacting fixture

\begin{displaymath}
 \includegraphics[width=93pt]{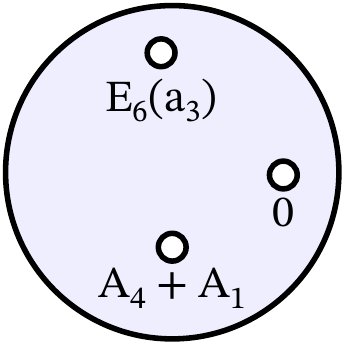}
\end{displaymath}
in the $E_6$ theory. Here, it has two realizations, both as mixed fixtures.

\begin{longtable}{|c|c|c|c|}
\hline
Fixture&$(n_2,n_3,n_4,n_5,n_6,n_8,n_9,n_{10},n_{12},n_{14},n_{18})$&$(n_h,n_v)$&Global Symmetry\\
\hline 
\endhead
$\begin{matrix} \includegraphics[width=75pt]{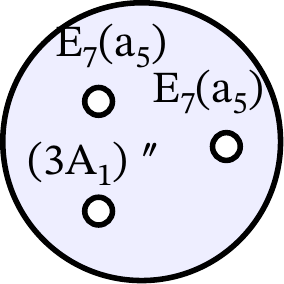}\end{matrix}$&$(0,0,4,0,2,1,0,1,3,3,4)$&$(416,374)$&${(G_2)}_{28}$\\
\hline
$\begin{matrix} \includegraphics[width=75pt]{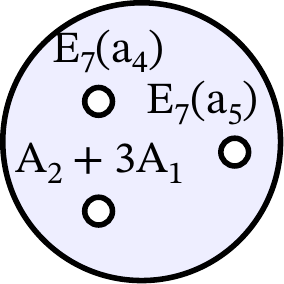}\end{matrix}$&$(0,0,2,0,2,1,0,1,2,2,2)$&$(280,240)$&${(G_2)}_{28}$\\
\hline
$\begin{matrix} \includegraphics[width=75pt]{A2A1A1A1_E7a5_E7a5}\end{matrix}$&$(0,0,4,0,2,2,0,1,4,3,5)$&$(504,447)$&${(F_4)}_{24}$\\
\hline
$\begin{matrix} \includegraphics[width=75pt]{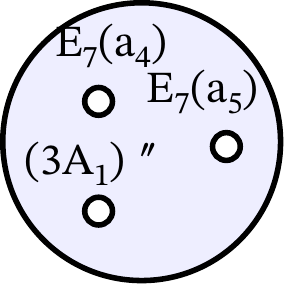}\end{matrix}$&$(0,0,2,0,2,2,0,1,3,2,3)$&$(368,313)$&${(F_4)}_{24}$\\
\hline
$\begin{matrix} \includegraphics[width=75pt]{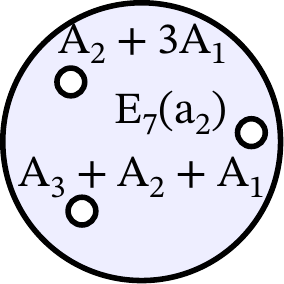}\end{matrix}$&$(0,0,1,0,0,0,0,1,0,1,0)$&$(96,53)$&${(E_6)}_{28}$\\
\hline
$\begin{matrix} \includegraphics[width=75pt]{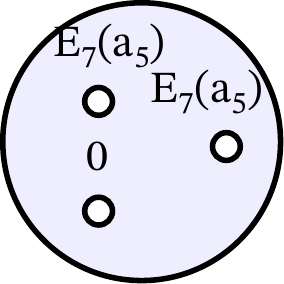}\end{matrix}$&$(0,0,4,0,2,2,0,2,4,4,6)$&$(612,528)$&${(E_7)}_{36}$\\
\hline
$\begin{matrix} \includegraphics[width=75pt]{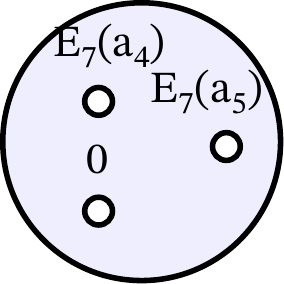}\end{matrix}$&$(0,0,2,0,2,2,0,2,3,3,4)$&$(476,394)$&${(E_7)}_{36}$\\
\hline
$\begin{matrix} \includegraphics[width=75pt]{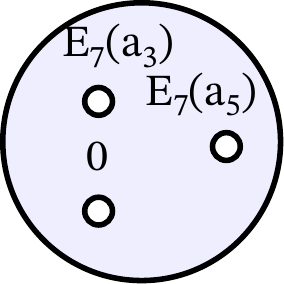}\end{matrix}$&$(0,1,2,0,1,1,0,0,2,1,2)$&$(268,188)$&${(E_7)}_{36}$\\
\hline
$\begin{matrix} \includegraphics[width=75pt]{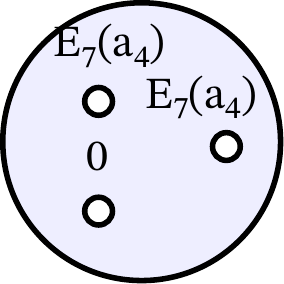}\end{matrix}$&$(0,0,0,0,2,2,0,2,2,2,2)$&$(340,260)$&${(E_7)}_{36}$\\
\hline
$\begin{matrix} \includegraphics[width=75pt]{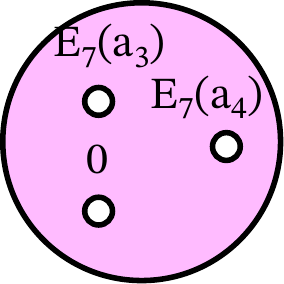}\end{matrix}$&$(0,1,0,0,1,1,0,0,1,0,0)$&$(104,54)$&${(E_7)}_{24}+\frac{1}{2}(56)$\\
\hline
$\begin{matrix} \includegraphics[width=75pt]{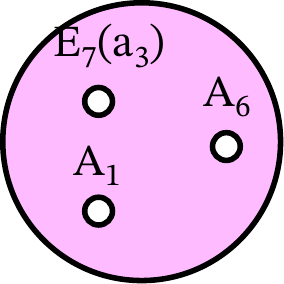}\end{matrix}$&$(0,1,0,0,1,1,0,0,1,0,0)$&$(104,54)$&${(E_7)}_{24}+\frac{1}{2}(12,2)$\\
\hline
$\begin{matrix} \includegraphics[width=75pt]{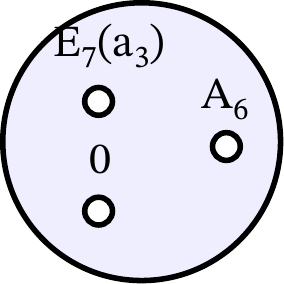}\end{matrix}$&$(0,1,0,0,1,1,0,0,1,0,1)$&$(168,89)$&${(E_8)}_{36}$\\
\hline
\end{longtable}

In the $E_6$ theory, fixtures of the form $(0,D,D)$ or $(2A_2,D,D)$, where $D$ is either $E_6(a_3)$ or $E_6(a_1)$, are all bad, so we do not get additional SCFTs with simple exceptional global symmetries in this way. However, we can construct 5 more of these in the twisted sector:

\begin{longtable}{|c|c|c|c|}\hline
Fixture&$(n_2,n_3,n_4,n_5,n_6,n_8,n_9,n_{12})$&$(n_h,n_v)$&\mbox{\shortstack{Global\\Symmetry}}\\
\hline 
\endhead
$\begin{matrix} \includegraphics[width=75pt]{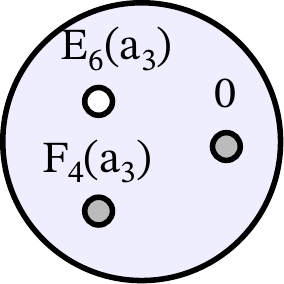}\end{matrix}$&$(0,4,0,1,1,2,2,3)$&$(208,173)$&${(F_4)}_{18}$\\\hline
$\begin{matrix} \includegraphics[width=75pt]{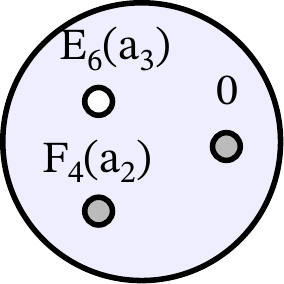}\end{matrix}$&$(0,1,1,1,1,1,1,1)$&$(120,87)$&${(F_4)}_{18}$\\\hline
$\begin{matrix} \includegraphics[width=75pt]{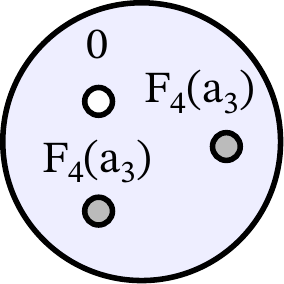}\end{matrix}$&$(0,6,0,2,2,4,4,6)$&$(384,336)$&${(E_6)}_{24}$\\\hline
$\begin{matrix} \includegraphics[width=75pt]{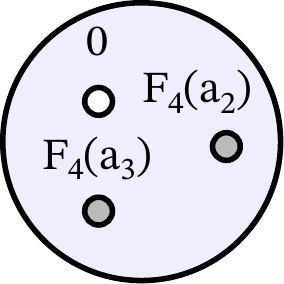}\end{matrix}$&$(0,3,1,2,2,3,3,4)$&$(296,250)$&${(E_6)}_{24}$\\\hline
$\begin{matrix} \includegraphics[width=75pt]{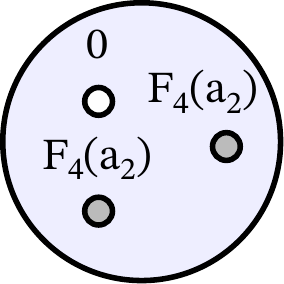}\end{matrix}$&$(0,0,2,2,2,2,2,2)$&$(208,164)$&${(E_6)}_{24}$\\\hline
\end{longtable}

\subsubsection{Enhanced global symmetries and Sommers-Achar group action on the Higgs branch}\label{enhanced_global_symmetries_the_sommersachar_group_on_the_higgs_branch}

As in section 5 of \cite{Chacaltana:2015bna}, we can consider families of fixtures where we fix two punctures and let the third vary over a special piece, $\{O\}$. We denote by $O_s$ the special puncture in this special piece and by $O_m$ the puncture with the maximal Sommers-Achar group, whose Hitchin pole is $(d(O),S_n)$ \cite{Chacaltana:2012zy}. It is often the case that, when $O=O_s$, a simple factor in the manifest global symmetry group associated to one of the two fixed punctures becomes enhanced as

\begin{displaymath}
F_{k n} \to {(F_k)}^n
\end{displaymath}
When this happens, then, for $O=O_m$, the $F_{k n}$ is unenhanced and, as $O$ varies over the special piece, the enhancement is the subgroup of ${(F_k)}^n$ which is invariant under $C(O)$ acting by permutations of the $n$ copies of $F_k$. 

We found numerous examples of this in \cite{Chacaltana:2014jba} and \cite{Chacaltana:2015bna}, and were able to verify, using various S-dualities (see, e.g., Section 4 of \cite{Chacaltana:2014jba}) that the levels of the factors of $F$ in the global symmetry behaved as predicted by this permutation action.

The $E_7$ theory provides further examples of this phenomenon. One interesting example is given by fixtures with 0 and $E_7(a_1)$ punctures and the third puncture $\color{green}O$ coming from the special piece $\{D_4(a_1), (A_3+A_1)^\prime, 2A_2+A_1\}$. The $(E_7)_{36}$ of puncture $0$ is enhanced to the subgroup of $(E_7)_{12}^3$ that is invariant under $C({\color{green}O})$. With certain $SU(2)$ groups coming from $O$, the enhanced $E_7$ groups are further enhanced to $E_8$ groups. The resulting theories are $E_8$ Minahan-Nemeschansky SCFTs of various rank $l$ whose Higgs branches are the moduli space of $l$ $E_8$ instantons, denoted by $M(E_8,l)$.

\begin{displaymath}
 \includegraphics[width=92pt]{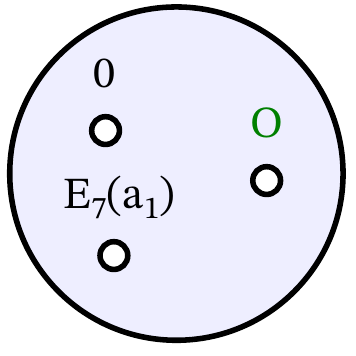}
\end{displaymath}

\medskip
{
\setlength\LTleft{-.5in}

\begin{longtable}{|c|c|c|c|c|c|}\hline
${\color{green}O}$&$C({\color{green}O})$&Theory&Higgs Branch&$\dim_{\mathbb{H}}\mathcal{H}$&$(n_h,n_v)$\\
\hline
\endhead
$D_4(a_1)$&1&$[(E_8)_{12}\, \text{ SCFT}]^3$&$M(E_8,1)^3$&$87$&$(120,33)$\\\hline
$(A_3+A_1)^\prime$&$\mathbb{Z}_2$&$\footnotesize{[(E_8)_{12}\, \text{ SCFT}] \times  [(E_8)_{24}\times SU(2)_{13}\,\text{SCFT}]}$&$\footnotesize{M(E_8,1) \times M(E_8,2)}$&$88$&$(133,45)$\\\hline
$2A_2+A_1$&$S_3$&$[(E_8)_{36}\times SU(2)_{38}\,\text{ SCFT}]$&$M(E_8,3)$&$89$&$(158,69)$\\\hline
\end{longtable}

}
\medskip

We can use this behavior to fill in some of the missing levels which cannot be determined from the information extracted from superconformal index. For example, still from the special piece $\{D_4(a_1), (A_3+A_1)^\prime, 2A_2+A_1\}$, we find another sequence, given by the fixtures with punctures $(A_6, D_5+A_1, {\color{green}O})$:

\begin{displaymath}
 \includegraphics[width=92pt]{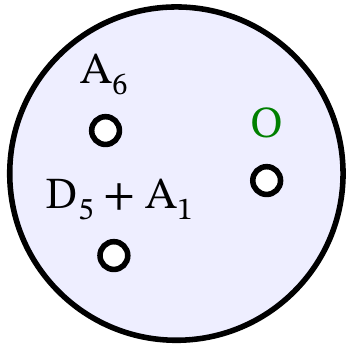}
\end{displaymath}

\medskip
\begin{center}
\begin{tabular}{|c|c|c|}\hline
${\color{green}O}$&$C({\color{green}O})$&Global symmetry\\
\hline 
$D_4(a_1)$&1&$SU(2)_{12}^4\times\color{red}SU(2)_{k_1}\times SU(2)_{k_2}\times SU(2)_{36-k_1-k_2}$\\\hline
$(A_3+A_1)^\prime$&$\mathbf{Z}_2$&$SU(2)_{13}\times SU(2)_{24}\times SU(2)_{12}^2\times\color{red}SU(2)_k\times SU(2)_{36-k}$\\\hline
$2A_2+A_1$&$S_3$&$SU(2)_{36}\times SU(2)_{38}\times SU(2)_{12}\times\color{red}SU(2)_{36}$\\\hline
\end{tabular}
\end{center}
\medskip
The $\color{red}SU(2)_{36}$ from the $A_6$ puncture is enhanced to subgroups of $SU(2)_{12}^3$. The Sommers-Achar group action tells us $k_1=k_2=k=12$.

For another example, let's look at the special piece $\{E_7(a_5), D_6(a_2), A_5+A_1\}$. Consider the fixture with punctures $(A_4+A_2, D_5+A_1, {\color{green}O})$:

\begin{displaymath}
 \includegraphics[width=92pt]{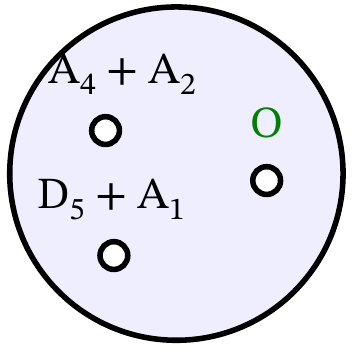}
\end{displaymath}

\medskip
\begin{center}
\begin{tabular}{|c|c|c|}\hline
${\color{green}O}$&$C({\color{green}O})$&Global symmetry\\
\hline 
$E_7(a_5)$&1&$(G_2)_{12}\times\color{red}SU(2)_{k_1}\times SU(2)_{k_2}\times SU(2)_{72-k_1-k_2}$\\\hline
$D_6(a_2)$&$\mathbf{Z}_2$&$(G_2)_{12}\times SU(2)_9\times\color{red}SU(2)_{72-k}\times SU(2)_k$\\\hline
$A_5+A_1$&$S_3$&$(G_2)_{12}\times SU(2)_{26}\times\color{red}SU(2)_{72}$\\\hline
\end{tabular}
\end{center}
\medskip

Similar to the previous examples, we can dertermine that $k_1=k_2=k=24$.

For fixture $(A_6, D_5(a_1),{\color{green}O})$ where $\color{green}O$ belongs to the special piece $\{E_6(a_3),A_5\prime\}$

\begin{displaymath}
 \includegraphics[width=90pt]{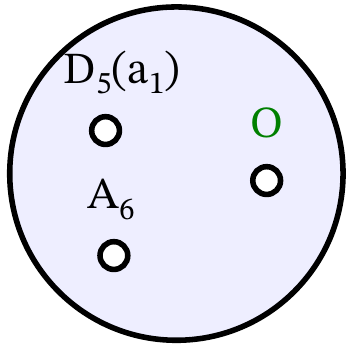}
\end{displaymath}

\medskip
\begin{center}
\begin{tabular}{|c|c|c|}
\hline
${\color{green}O}$&$C({\color{green}O})$&Global symmetry\\
\hline
$E_6(a_3)$&1&$SU(2)_{10}\times SU(4)_{20}\times U(1)\times\color{red}SU(2)_{16-k}\times SU(2)_k$\\\hline
$A_5^\prime$&$\mathbf{Z}_2$&$SU(2)_{10}\times SU(2)_9\times SU(4)_{20}\times U(1)\times\color{red}SU(2)_{16}$\\\hline
\end{tabular}
\end{center}
\medskip
from which we conclude that $k=8$.

\subsection{Connections with 6d (1,0) SCFTs on $T^2$}\label{connections_with_6d_10_scfts_on_}
Another large class of $4d$ $\mathcal{N}=2$ SCFTs arises from compactifications of $6d$ $(1,0)$ SCFTs on $T^2$. Following the recent classification of $6d$ $(1,0)$ SCFTs \cite{Heckman:2013pva,Heckman:2015bfa,Bhardwaj:2015xxa}, the study of their $T^2$ compactifications was initiated in \cite{Ohmori:2015pua,Ohmori:2015pia,DelZotto:2015rca}. In those papers, various $T^2$ compactifications of $(1,0)$ theories were found to also have class $S$ realizations. Here, we comment on the models which were conjectured to have a class $S$ realization in either the $E_7$ or $E_8$ theories.

\subsubsection{Very Higgsable theories on $T^2$}\label{very_higgsable_theories_on_}

In \cite{Ohmori:2015pua}, a subset of the $6d$ $(1,0)$ SCFTs of \cite{Heckman:2015bfa} was singled out by the authors, which they termed ``very Higgsable''. These SCFTs are those which have a Higgs branch with no tensor multiplet degrees of freedom. In their F-theory realization, these are the theories for which successive blow-downs of -1 curves in the base of the elliptically-fibered Calabi-Yau threefold removes (after a further complex structure deformation) the singularity in the base completely. They found that the central charges of the $4d$ $\mathcal{N}=2$ SCFT resulting from the $T^2$ compactification of a very Higgsable $6d$ $(1,0)$ theory are given by

\begin{equation}\label{ack}
\begin{split}
a &= 24\alpha - 12\beta - 18\gamma, \\
c &= 64\alpha - 12\beta - 8\gamma, \\
k_i &= 192\kappa_i,
\end{split}
\end{equation}
where $\alpha, \beta$, and $\kappa_i$ are the coefficients appearing in the anomaly 8-form of the $6d$ theory

\begin{displaymath}
I_8 \supset \alpha p_1(T)^2 + \beta p_1(T)c_2(R) + \gamma p_2(T) + \sum_i \kappa_i p_1(T) \text{Tr }F_i^2,
\end{displaymath}
which can be computed following \cite{Ohmori:2014pca,Ohmori:2014kda}; see also \cite{Intriligator:2014eaa}.

Using this formula, the authors argued that the minimal ``conformal matter'' theory, $\mathcal{T}(G,1)$ (the theory on a single M5-brane at a $G=ADE$-type singularity), on $T^2$ coincides with the class $S$ theory of type $G$ on a fixture with two full punctures and one minimal puncture. For $G=E_7, E_8$, these fixtures are

\medskip
{\footnotesize
\parindent=-.25in

\begin{tabular}{|c|c|c|c|c|}
\hline
G&Fixture&$(n_2,n_3,n_4,n_5,n_6,n_8,n_9,n_{10},n_{12},n_{14},n_{18},n_{20},n_{24},n_{30})$&$(n_h,n_v)$&\mbox{\shortstack{Global\\Symmetry}}\\\hline 
$E_7$&$\begin{matrix} \includegraphics[width=75pt]{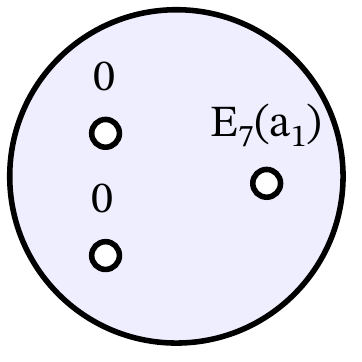}\end{matrix}$&$(0,0,0,0,1,1,0,1,2,2,3,0,0,0)$&$(384,250)$&$(E_7)_{36} \times (E_7)_{36}$\\\hline
$E_8$&$\begin{matrix} \includegraphics[width=75pt]{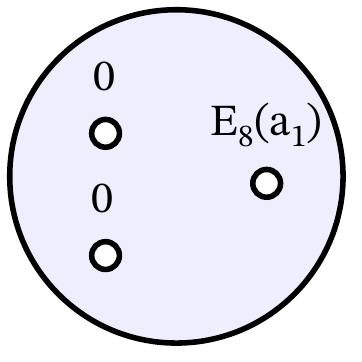}\end{matrix}$&$(0,0,0,0,1,1,0,0,2,2,3,3,4,5)$&$(1080,831)$&$(E_8)_{60} \times (E_8)_{60}$\\\hline
\end{tabular}

}
\medskip

The graded Coulomb branch dimensions for these two fixtures are in agreement with those computed from the mirror geometry of the corresponding $6d$ $(1,0)$ theories on $T^2$ in \cite{DelZotto:2015rca} and the central charges agree with those obtained in \cite{Ohmori:2015pua}.

We can also realize some of the $(G,G')$ conformal matter theories of \cite{DelZotto:2014hpa} on $E_7$ and $E_8$ fixtures. These conformal matter theories correspond to fractional M5-branes on an ALE singularity:

\medskip
\begin{center}
	\begin{tabular}{|c|c|c|}\hline
Global Symmetry&$\#$ M5&ALE type\\\hline 
$(E_7,SO(7))$&$\frac{1}{2}$&$E_7$\\\hline
$(E_8,G_2)$&$\frac{1}{3}$&$E_8$\\\hline
$(E_8,F_4)$&$\frac{1}{2}$&$E_8$\\\hline
\end{tabular}
\end{center}
\medskip

In \cite{Ohmori:2015pua}, the first of these was identified with the $E_6$ fixture

\begin{displaymath}
 \includegraphics[width=92pt]{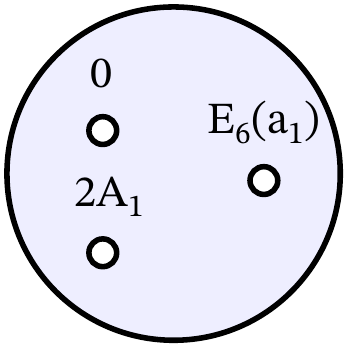}
\end{displaymath}
which appears as the third entry in table 3.4 of \cite{Chacaltana:2014jba}.

Computing the anomaly polynomial of the other two theories following \cite{Ohmori:2014pca,Ohmori:2014kda} and plugging into \eqref{ack}, we find that the $T^2$ compactified $(E_8,G_2)$ conformal matter theory has central charges

\begin{displaymath}
a=\frac{149}{6}, c=\frac{86}{3}, k_{E_8}=36, k_{G_2}=16.
\end{displaymath}
These are the central charges of the class $S$ theory realized by compactifying the $E_7$ $(2,0)$ theory on

\begin{displaymath}
 \includegraphics[width=105pt]{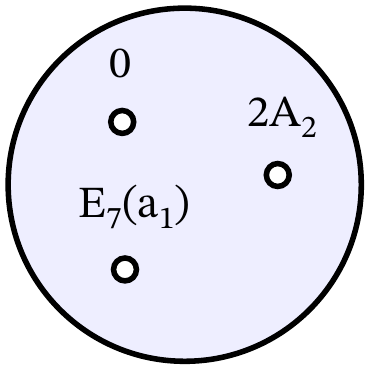}
\end{displaymath}
In this realization, only a $(E_7)_{36} \times SU(2)_{36} \times (G_2)_{16}$ subgroup of the global symmetry group is manifest. We can check the enhancement to $(E_8)_{36} \times (G_2)_{16}$ by computing the order $\tau^2$ expansion of the superconformal index, which is given by

\begin{displaymath}
\mathcal{I}=1+(\chi^{\mathbf{248}}_{E_8}+\chi^{\mathbf{14}}_{G_2})\tau^2+\dots
\end{displaymath}
where

\begin{displaymath}
\chi^{\mathbf{248}}_{E_8}=\chi^{\mathbf{133}}_{E_7}+\chi^{\mathbf{3}}_{SU(2)}+\chi^{\mathbf{56}}_{E_7}\chi^{\mathbf{2}}_{SU(2)}.
\end{displaymath}

Similarly, we find the $T^2$ compactified $(E_8,F_4)$ conformal matter theory has central charges

\begin{displaymath}
a=\frac{179}{3}, c=\frac{196}{3}, k_{E_8}=48, k_{F_4}=24
\end{displaymath}
Comparing with the $E_7$ and $E_8$ tinkertoys \cite{Future}, we do not find a direct realization in class $S$. The closest one can come\footnote{This fixture was conjectured in \cite{DelZotto:2015rca} to realize the SCFT we are seeking. We see here that it realizes, instead, a \emph{product} of the desired SCFT and the Minahan-Nemeschansky ${(E_8)}_{12}$ SCFT.}  is the fixture

\begin{displaymath}
 \includegraphics[width=196pt]{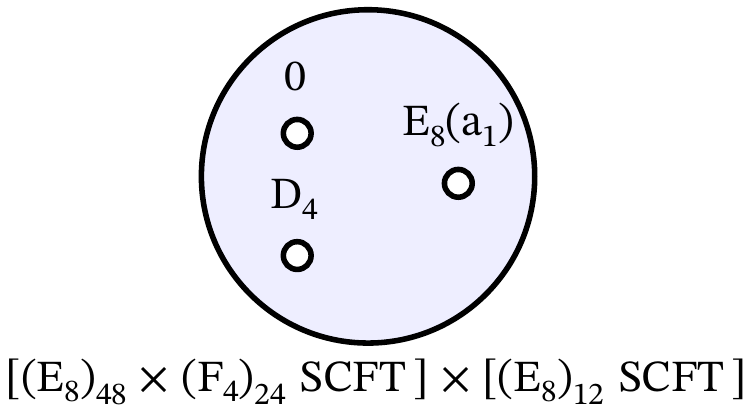}
\end{displaymath}
in the $E_8$ theory. This is a product of the desired SCFT (whose global symmetry is ${(E_8)}_{48}\times{(F_4)}_{24}$ and $(n_h,n_v)=(352,216)$) with the ${(E_8)}_{12}$ SCFT $(n_h,n_v)=(40,11)$).

Similarly, \cite{DelZotto:2015rca} also conjectured that the $T^2$-compactification of the $(E_8,G_2)$ theory is realized in Class-S as the fixture

\begin{displaymath}
 \includegraphics[width=210pt]{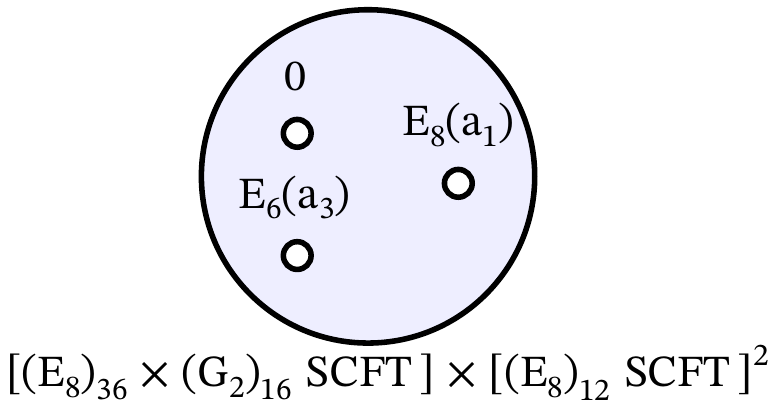}
\end{displaymath}
in the $E_8$ theory. In fact, this fixture is a product of the desired SCFT with \emph{two} copies of the ${(E_8)}_{12}$ SCFT.

The fact that these fixtures yield not the desired SCFT, but rather its product with some additional decoupled degrees of freedom, is not unheralded. Already in the case of the $T^2$ compactification of the $(E_7,SO(7))$ conformal matter theory, \cite{DelZotto:2015rca} noticed that their class-S realization, the fixture

\begin{displaymath}
 \includegraphics[width=174pt]{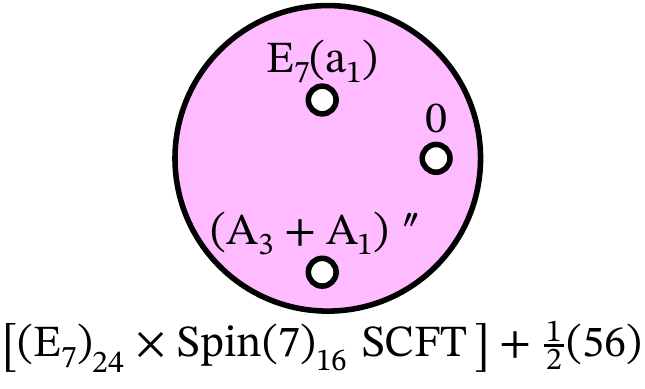}
\end{displaymath}
in the $E_7$ theory, yields not the ${(E_7)}_{24}\times Spin(7)_{16}$ SCFT, but rather the desired SCFT with additional hypermultiplets in the $\tfrac{1}{2}(56)$ of $E_7$.

\subsubsection{$N$ $M5$ branes probing an ADE singularity on $T^2$}\label{__branes_probing_an_ade_singularity_on_}

The $T^2$ compactification of the $(1,0)$ theory on the worldvolume of $N \gt 1$ M5-branes on an ALE singularity was studied in \cite{DelZotto:2015rca,Ohmori:2015pia}. In the F-theory realization of these theories, after successively blowing down all $(-1)$-curves, one reaches an endpoint which is a chain of $(-2)$-curves, intersecting as an $A_{N-1}$ Dynkin diagram. Thus, these theories are not in the class of very Higgsable SCFTs, but are instead Higgsable to a $(2,0)$ theory. The $T^2$ compactifications of such $(1,0)$ theories were systematically studied in \cite{Ohmori:2015pia}. They found that, in general, the $T^2$ compactification of a $(1,0)$ SCFT Higgsable to a $(2,0)$ SCFT of type $G$ does not give an SCFT, but rather has following structure (following the notation of \cite{Ohmori:2015pia}):

\begin{displaymath}
\mathcal{T}^{6d}\langle T^2_\tau\rangle = \frac{\mathcal{U}^{4d}\{G,H\}\times \mathcal{V}^{4d}\{H\}}{G_\tau \times H}
\end{displaymath}
where $\mathcal{U}^{4d}\{G,H\}$ is a $4d$ $\mathcal{N}=2$ SCFT with $G \times H$ global symmetry and $\mathcal{V}^{4d}\{H\}$ is a $4d$ $\mathcal{N}=2$ SCFT with $H$ global symmetry. These two SCFTs are coupled by $G \times H$ gauge fields, where the gauge coupling for $G$ is exactly marginal and can be identified with the complex structure parameter $\tau$ of the torus. The gauge coupling for $H$ is IR free.

For $N=2$ M5-branes at an ALE singularity of type $\mathfrak{g}$, for each singularity type the authors of \cite{Ohmori:2015pia} found that the theory $\mathcal{U}$ is a free hypermultiplet in the $\frac{1}{2}(3,2)$ of $SU(2)_u \times SU(2)_v$ while the theory $\mathcal{V}$ is a class $S$ theory of type $\mathfrak{g}$. \footnote{For $\mathfrak{g}=A_{k-1}$ there is an additional fundamental hypermultiplet of $SU(2)_v$.} Using our results, we can construct this theory for $\mathfrak{g}=E_7$ and $E_8$:

\medskip

{

\begin{tabular}{|c|c|c|c|c|}
\hline
\mbox{\shortstack{Singularity\\Type}}&Fixture&$\dim_{\text{Coul}}$&$(n_h,n_v)$&\mbox{\shortstack{Global\\Symmetry}}\\\hline
$E_7$&$\begin{matrix} \includegraphics[width=75pt]{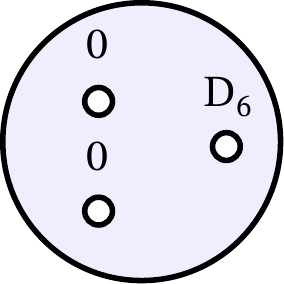}\end{matrix}$&$26$&$(767,630)$&$\begin{gathered}(E_7)_{36} \times (E_7)_{36}\\ \times SU(2)_7\end{gathered}$\\\hline
$E_8$&$\begin{matrix} \includegraphics[width=75pt]{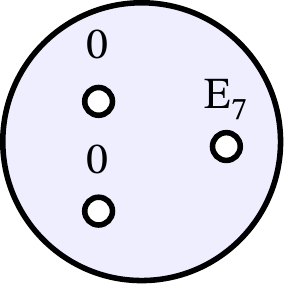}\end{matrix}$&$49$&$(2159,1907)$&$\begin{gathered}(E_8)_{60} \times (E_8)_{60}\\ \times SU(2)_7\end{gathered}$\\\hline
\end{tabular}
\medskip

}

The $SU(2)_v$ factor weakly gauges the $SU(2)_7$ flavor symmetry carried by the non-maximal puncture of each fixture listed above. Since this $SU(2)$ is coupled to an additional three fundamental half-hypermultiplets, it is infrared free.

For a general number $N$ of M5 branes probing an ALE singularity of type $\mathfrak{g}$, the theory $\mathcal{V}$ is the class $\mathcal{S}$ theory of type $\mathfrak{g}$ on a fixture with three full punctures, i.e., the $T_\mathfrak{g}$ theory. The theory $\mathcal{U}$ is given by a $4d$ SCFT $\mathcal{S}^{4d}_{(\varnothing,\mathfrak{g}),N}\{SU(N),\mathfrak{g}\}$, which is the $T^2$ compactification of the $6d$ $(1,0)$ SCFT living on $N$ M5-branes at the intersection of the Ho\v{r}ava-Witten $E_8$-wall and an ALE singularity of type $\mathfrak{g}$. It was calculated in \cite{Ohmori:2015pia} that this $4d$ SCFT has flavor symmetry $SU(N)_{4N} \times \mathfrak{g}_{2h^\vee(\mathfrak{g})+2}$. For $\mathfrak{g}=A_{k-1}$, they identified this theory as the class $S$ theory on the fixture \footnote{We quote the result here for $N>k$. The theory for $k<N$ is obtained by exchanging $k \leftrightarrow N$. For $k=N$, they identified $S^{4d}_{(\varnothing,\mathfrak{su}(k)),N}$ with the $T_N$ theory with an additional free hypermultiplet in the fundamental of $SU(N)$.}

\begin{displaymath}
 \includegraphics[width=92pt]{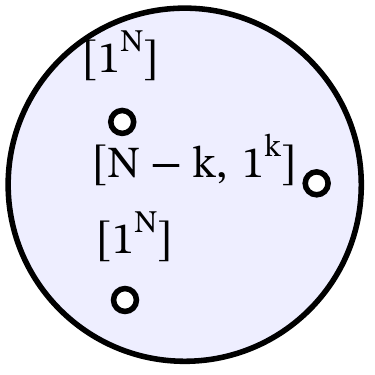}
\end{displaymath}
where the $SU(N)_{4N}$ global symmetry is realized as the diagonal subgroup of the $SU(N)_{2N}$ flavor symmetries of the two full punctures. 

For $\mathfrak{g} \neq A$, they were not able to identify this SCFT with other known $4d$ SCFTs. We also do not find this theory for any $\mathfrak{g}, N$ on any $E_7$ fixture. We have not yet checked if it appears in the list of $E_8$ fixtures, which is work in progress \cite{Future}.

Mass deforming $\mathcal{S}^{4d}_{(\varnothing,\mathfrak{g}),N}\{SU(N),\mathfrak{g}\}$ by the $SU(N)$ mass parameter, one obtains the generalized quiver tail produced by colliding $N-1$ minimal punctures on a sphere. For $\mathfrak{g}=A_{k-1}$, the class $S$ realization of this quiver tail is well-known \cite{Gaiotto:2009we}. In \cite{Ohmori:2015pia}, the authors worked out the quiver tails for $\mathfrak{g}=D_k, E_6$, and, from the structure of the $E_6$ quiver tail, conjectured the answer for $\mathfrak{g}=E_7$ and $E_8$ as well. Using our results, we can confirm their prediction for $E_7$ and $E_8$:

For $E_7$, the quiver is given by

\begin{displaymath}
 \includegraphics[width=475pt]{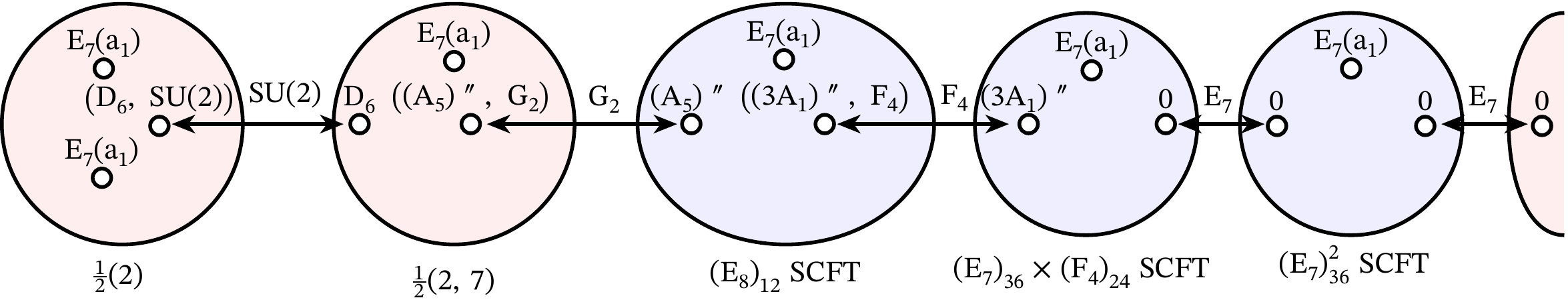}
\end{displaymath}
while for $E_8$, the quiver is

\begin{displaymath}
 \includegraphics[width=475pt]{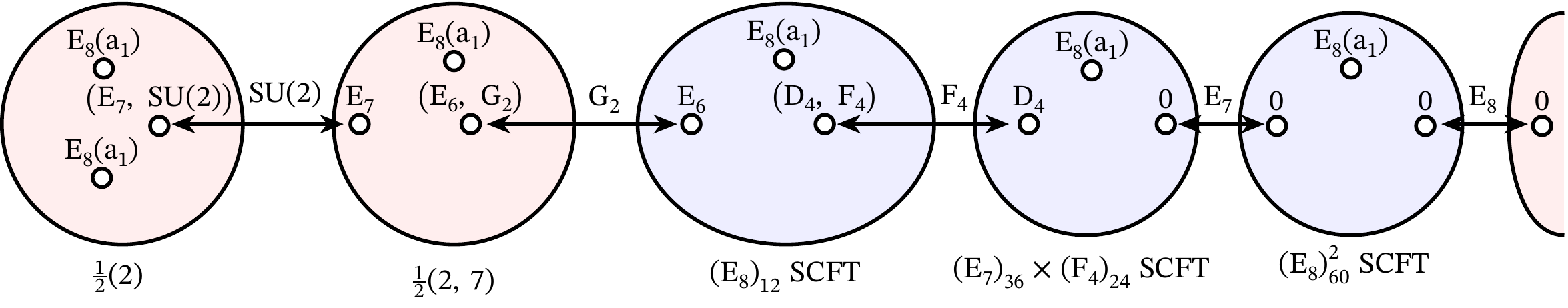}
\end{displaymath}
Here, the ${(E_7)}_{36}\times{(F_4)}_{24}$ SCFT has $(n_h,n_v)=(276,169)$ and graded Coulomb branch dimensions $(d_2,d_6,d_8,d_{10},d_{12},d_{14},d_{18})=(0,1,1,0,2,1,2)$. Colliding additional minimal punctures gives additional copies of the $(E_7)_{36}^2$ ($(E_8)^2_{60}$) SCFT (the $T^2$ compactification of the $E_7$ ($E_8$) minimal conformal matter theory), whose properties were discussed in section \ref{very_higgsable_theories_on_}.

\section*{Acknowledgements}
We would like to thank Mario Martone, Kantaro Ohmori, and Yuji Tachikawa for helpful discussions. The work of J.D. was supported in part by the National Science Foundation under Grant No. PHY-1620610. J.D. and A.T. would like to thank the Aspen Center for Physics, which is supported by National Science Foundation grant PHY-1066293,  for hospitality during the workshop ``Superconformal Field Theories in $d \geq 4$", where the final stages of this work was completed. O.C. would like to thank the Johns Hopkins University, and especially Jared Kaplan, for hospitality while this work was completed. O.C. would also like to thank the Aspen Center for Physics, supported by NSF grant PHY-1066293, and acknowledge kind support by a grant from the Simons Foundation, for hospitality during the 2016 Summer Program of the ACP, while this work was completed. O.C. also gratefully acknowledges support from the Simons Center for Geometry and Physics, Stony Brook University, for hospitality at the 2016 Simons Summer Workshop in Mathematics and Physics, where part of this work was completed.
\begin{appendices}

\section{Embeddings of $SU(2)$ in $E_7$}\label{appendix_embeddings_of__in_}

{
\setlength\LTleft{-.25in}

\begin{longtable}{|c|c|l|l|} \hline
Bala-Carter&$\mathfrak{f}$&$56$&$133$\\
\hline 
\endhead
$A_1$&$\mathfrak{so}(12)$&$(1,32)+(2,12)$&$(1,66)+(2,\overline{32})+(3,1)$\\ \hline
$2A_1$&$\mathfrak{so}(9)\times \mathfrak{su}(2)$&$\begin{gathered}(1;9,2)+(2;16,1)+\\(3;2,1)\end{gathered}$&$\begin{gathered}(1;1,3)+(1;36,1)+(2;16,2)+\\(3;1,1)+(3;9,1)\end{gathered}$\\ \hline
$(3A_1)''$&$\mathfrak{f}_4$&$(2;26)+(4;1)$&$(1;52)+(3;1)+(3;26)$\\ \hline
$(3A_1)'$&$\mathfrak{sp}(3)\times\mathfrak{su}(2)$&$\begin{gathered}(1;14',1)+(2;6,2)+\\(3;6,1)\end{gathered}$&$\begin{gathered}(1;1,3)+(1;21,1)+(2;14,2)+\\(3;1,1)+(3;14,1)+(4;1,2)\end{gathered}$\\ \hline
$A_2$&$\mathfrak{su}(6)$&$(1;20)+(3;6)+(3;\overline{6})$&$\begin{gathered}(1;35)+(3;1)+(3;15)+\\(3;\overline{15})+(5;1)\end{gathered}$\\ \hline
$4A_1$&$\mathfrak{sp}(3)$&$\begin{gathered}(1;6)+(2;14)+\\(3;6)+(4;1)\end{gathered}$&$\begin{gathered}(1;21)+(2;6)+(2;14')+\\2(3;1)+(3;14)+(4;6)\end{gathered}$\\ \hline
$A_2+A_1$&$\mathfrak{su}(4)\times\mathfrak{u}(1)$&$\begin{gathered}(1;4)_{-3/2}+(1;\overline{4})_{3/2}+\\(2;1)_1+(2;1)_{-1}+\\(2;6)_0+(3;4)_{1/2}+\\(3;\overline{4})_{-1/2}+(4;1)_1+\\(4;1)_{-1}\end{gathered}$&$\begin{gathered}(1,1)_0+(1;15)_0+(2;4)_{3/2}+\\(2;4)_{-1/2}+(2;\overline{4})_{1/2}+ (2;\overline{4})_{-3/2}+\\(3;1)_2+2(3;1)_0+(3;1)_{-2}+\\(3;6)_1+(3;6)_{-1}+(4;4)_{-1/2}+\\(4;\overline{4})_{1/2}+(5;1)_0\end{gathered}$\\ \hline
$A_2+2A_1$&${\mathfrak{su}(2)}^3$&$\begin{gathered}(1;2,3,1)+(2;1,2,4)+\\(3;2,1,3)+(4;1,2,2)\end{gathered}$&$\begin{gathered}(1;3,1,1)+(1;1,3,1)+\\(1;1,1,3)+(2;2,2,4)+\\(3;1,1,1)+(3;1,3,3)+\\(3;1,1,5)+(4;2,2,2)+\\(5;1,1,3)\end{gathered}$\\ \hline
$A_3$&$\mathfrak{so}(7)\times \mathfrak{su}(2)$&$(1;7,2)+(4;8,1)+(5;1,2)$&$\begin{gathered}(1;1,3)+(1;21,1)+(3;1,1)+\\(4;8,2)+(5;7,1)+(7;1,1)\end{gathered}$\\ \hline
$2A_2$&$\mathfrak{g}_2\times \mathfrak{su}(2)$&$(1;1,4)+(3;7,2)+(5;1,2)$&$\begin{gathered}(1;1,3)+(1;14,1)+(3;1,1)+\\(3;7,3)+(5;1,3)+(5;7,1)\end{gathered}$\\ \hline
$A_2+3A_1$&$\mathfrak{g}_2$&$(2;14)+(4;7)$&$\begin{gathered}(1;14)+(3;1)+(3;27)+(5;7)\end{gathered}$\\ \hline
$(A_3+A_1)''$&$\mathfrak{so}(7)$&$\begin{gathered}(2;7)+(4;1)+(4;8)+(6;1)\end{gathered}$&$\begin{gathered}(1;21)+2(3;1)+(3;8)+(5;7)+\\(5;8)+(7;1)\end{gathered}$\\ \hline
$2A_2+A_1$&${\mathfrak{su}(2)}^2$&$\begin{gathered}(1;4,1)+(2;2,2)+\\(3;2,3)+(4;2,2)+\\(5;2,1)\end{gathered}$&$\begin{gathered}(1;3,1)+(1;1,3)+(2;3,2)+\\(2;1,4)+2(3;1,1)+(3;3,3)+\\(4;1,2)+(4;3,2)+ (5;3,1)+\\(5;1,3)+(6;1,2)\end{gathered}$\\ \hline
$(A_3+A_1)'$&${\mathfrak{su}(2)}^3$&$\begin{gathered}(1;1,3,2)+(2;2,1,2)+\\(3;1,2,1)+(4;2,2,1)+\\(5;1,2,1)+(5;1,1,2)\end{gathered}$&$\begin{gathered}(1;3,1,1)+(1;1,3,1)+\\(1;1,1,3)+(2;2,3,1)+\\2(3;1,1,1)+(3;1,2,2)+\\(4;2,1,1)+(4;2,2,2)+\\(5;1,2,2)+(5;1,3,1)+\\(6;2,1,1)+(7;1,1,1)\end{gathered}$\\ \hline
$D_4(a_1)$&${\mathfrak{su}(2)}^3$&$\begin{gathered}(1;2,2,2)+(3;2,1,1)+\\(3;1,2,1)+(3;1,1,2)+\\(5;2,1,1)+(5;1,2,1)+\\(5;1,1,2)\end{gathered}$&$\begin{gathered}(1;3,1,1)+(1;1,3,1)+\\(1;1,1,3)+3(3;1,1,1)+\\(3;2,2,1)+(3;2,1,2)+\\(3;1,2,2)+(5;1,1,1)+\\(5;2,2,1)+(5;2,1,2)+\\(5;1,2,2)+2(7;1,1,1)\end{gathered}$\\ \hline
$A_3+2A_1$&${\mathfrak{su}(2)}^2$&$\begin{gathered}(1;2,1)+(2;1,3)+\\(3;1,2)+(3;2,1)+\\(4;1,1)+(4;2,2)+\\(5;1,2)+(6;1,1)\end{gathered}$&$\begin{gathered}(1;3,1)+(1;1,3)+(2;1,2)+\\(2;2,3)+3(3;1,1)+(3;2,2)+\\(4;2,1)+2(4;1,2)+(5;2,2)+\\(5;1,3)+(6;2,1)+(6;1,2)+\\(7;1,1)\end{gathered}$\\ \hline
$D_4$&$\mathfrak{sp}(3)$&$(1;14')+(7;6)$&$(1;21)+(3;1)+(7;14)+(11;1)$\\ \hline
$D_4(a_1)+A_1$&${\mathfrak{su}(2)}^2$&$\begin{gathered}(2;1,1)+(2;2,2)+\\(3;1,2)+(3;2,1)+\\2(4;1,1)+(5;1,2)+\\(5;2,1)+(6;1,1)\end{gathered}$&$\begin{gathered}(1;3,1)+(1;1,3)+(2;2,1)+\\(2;1,2)+4(3;1,1)+(3;2,2)+\\2(4;2,1)+2(4;1,2)+(5;1,1)+\\(5;2,2)+(6;2,1)+(6;1,2)+\\2(7;1,1)\end{gathered}$\\ \hline
$A_3+A_2$&$\mathfrak{su}(2)\times\mathfrak{u}(1)$&$\begin{gathered}(1;2)_0+(2;1)_1+\\(2;1)_{-1}+(3;2)_2+\\(3;2)_{-2}+(4;1)_3+\\(4;1)_1+(4;1)_{-1}+\\(4;1)_{-3}+(5;2)_0+\\(6;1)_1+(6;1)_{-1}\end{gathered}$&$\begin{gathered}(1;1)_0+(1;3)_0+(2;2)_1+\\(2;2)_{-1}+(3;1)_4+2(3;1)_2+\\2(3;1)_0+2(3;1)_{-2}+(3;1)_{-4}+\\(4;2)_3+(4;2)_1+(4;2)_{-1}+\\(4;2)_{-3}+(5;1)_2+2(5;1)_0+\\(5;1)_{-2}+(6;2)_1+(6;2)_{-1}+\\(7;1)_2+(7;1)_0+(7;1)_{-2}\end{gathered}$\\ \hline
$A_4$&$\mathfrak{su}(3)\times\mathfrak{u}(1)$&$\begin{gathered}(1;3)_{-5/3}+(1;\overline{3})_{5/3}+\\(3;1)_{-1}+(3;1)_1+\\(5;3)_{1/3}+(5;\overline{3})_{-1/3}+\\(7;1)_{-1}+(7;1)_1\end{gathered}$&$\begin{gathered}(1;1)_0+(1;8)_0+(3;1)_0+\\(3;3)_{-2/3}+(3;\overline{3})_{2/3}+(5;1)_2+\\(5;1)_0+(5;1)_{-2}+(5;3)_{4/3}+\\(5;\overline{3})_{-4/3}+(7;1)_0+(7;3)_{-2/3}+\\(7;\overline{3})_{2/3}+(9;1)_0\end{gathered}$\\ \hline
$A_3+A_2+A_1$&$\mathfrak{su}(2)$&$(2;5)+(4;7)+(6;3)$&$\begin{gathered}(1;3)+(3;1)+(3;5)\\+(3;9)+(5;3)+(5;7)+(7;5)\end{gathered}$\\ \hline
$(A_5)''$&$\mathfrak{g}_2$&$(4;1)+(6;7)+(10;1)$&$\begin{gathered}(1;14)+(3;1)+(5;7)+(7;1)+\\(9;7)+(11;1)\end{gathered}$\\ \hline
$D_4+A_1$&$\mathfrak{sp}(2)$&$\begin{gathered}(1;4)+(2;5)+(6;1)+\\(7;4)+(8;1)\end{gathered}$&$\begin{gathered}(1;10)+(2;4)+2(3;1)+(6;4)+\\(7;1)+(7;5)+(8;4)+(11;1)\end{gathered}$\\ \hline
$A_4+A_1$&${\mathfrak{u}(1)}^2$&$\begin{gathered}1_{-2,-5/3}+1_{2,5/3}+2_{1,-5/3}+\\2_{-1,5/3}+3_{0,-1}+3_{0,1}+\\4_{1,1/3}+4_{-1,-1/3}+5_{-2,1/3}+\\5_{2,-1/3}+6_{1,1/3}+6_{-1,-1/3}+\\7_{0,-1}+7_{0,1}\end{gathered}$&$\begin{gathered}2(1_{0,0})+2_{3,0}+2_{-3,0}+2_{1,-2/3}+\\2_{-1,2/3}+3_{0,0}+3_{0,0}+3_{2,2/3}+\\3_{-2,-2/3}+4_{-1,2/3}+4_{1,-2/3}+4_{1,4/3}+\\4_{-1,-4/3}+5_{0,0}+5_{0,2}+5_{0,-2}+\\5_{2,-4/3}+5_{-2,4/3}+6_{1,4/3}+6_{1,-2/3}+\\6_{-1,2/3}+6_{-1,-4/3}+7_{2,2/3}+7_{0,0}+\\7_{-2,-2/3}+8_{1,-2/3}+8_{-1,2/3}+9_{0,0}\end{gathered}$\\ \hline
$D_5(a_1)$&$\mathfrak{su}(2)\times\mathfrak{u}(1)$&$\begin{gathered}(1;2)_2+(1;2)_{-2}+\\(2;1)_1+(2;1)_{-1}+\\(3;2)_0+(6;1)_1+\\(6;1)_{-1}+(7;2)_0+\\(8;1)_1+(8;1)_{-1}\end{gathered}$&$\begin{gathered}(1;1)_0+(1;3)_0+(2;2)_1+\\(2;2)_{-1}+(3;1)_2+2(3;1)_0+\\(3;1)_{-2}+(5;1)_0+(6;2)_1+\\(6;2)_{-1}+(7;1)_2+(7;1)_{-2}+\\2(7;1)_0+(8;2)_1+(8;2)_{-1}+\\(9;1)_0+(11;1)_0\end{gathered}$\\ \hline
$A_4+A_2$&$\mathfrak{su}(2)$&$(3;6)+(5;2)+(7;4)$&$\begin{gathered}(1;3)+(3;1)+(3;5)+(5;3)+\\(5;7)+(7;5)+(9;3)\end{gathered}$\\ \hline
$(A_5)'$&${\mathfrak{su}(2)}^2$&$\begin{gathered}(1;1,4)+(5;1,2)+\\(6;2,2)+(9;1,2)\end{gathered}$&$\begin{gathered}(1;3,1)+(1;1,3)+(3;1,1)+\\(4;2,1)+(5;1,3)+(6;2,3)+\\(7;1,1)+(9;1,3)+(10;2,1)+\\(11;1,1)\end{gathered}$\\ \hline
$A_5+A_1$&$\mathfrak{su}(2)$&$\begin{gathered}(4;1)+(5;2)+(6;3)+\\(7;2)+(10;1)\end{gathered}$&$\begin{gathered}(1;3)+(2;4)+2(3;1)+(4;2)+\\(5;3)+(6;2)+(7;1)+(8;2)+\\(9;3)+(10;2)+(11;1)\end{gathered}$\\ \hline
$D_5(a_1)+A_1$&$\mathfrak{su}(2)$&$\begin{gathered}(2;5)+(4;1)+(6;3)+(8;3)\end{gathered}$&$\begin{gathered}(1;3)+2(3;1)+(3;5)+(5;3)+\\(7;3)+(7;5)+(9;3)+(11;1)\end{gathered}$\\ \hline
$D_6(a_2)$&$\mathfrak{su}(2)$&$\begin{gathered}2(4;1)+(5;2)+(6;1)+\\(7;2)+(8;1)+(10;1)\end{gathered}$&$\begin{gathered}(1;3)+3(3;1)+2(4;2)+(5;1)+\\(6;2)+3(7;1)+(8;2)+(9;1)+\\(10;2)+2(11;1)\end{gathered}$\\ \hline
$E_6(a_3)$&$\mathfrak{su}(2)$&$\begin{gathered}(1;4)+2(5;2)+\\(7;2)+(9;2)\end{gathered}$&$\begin{gathered}(1;3)+3(3;1)+(5;1)+2(5;3)+\\(7;1)+(7;3)+(9;1)+(9;3)+\\2(11;1)\end{gathered}$\\ \hline
$D_5$&${\mathfrak{su}(2)}^2$&$\begin{gathered}(1;2,3)+(5;1,2)+\\(9;2,1)+(11;1,2)\end{gathered}$&$\begin{gathered}(1;3,1)+(1;1,3)+(3;1,1)+\\(5;2,2)+(7;1,1)+(9;1,3)+\\(11;1,1)+(11;2,2)+(15;1,1)\end{gathered}$\\ \hline
$E_7(a_5)$&-&$3(4)+3(6)+2(8)+10$&$6(3)+4(5)+5(7)+3(9)+3(11)$\\ \hline
$A_6$&$\mathfrak{su}(2)$&$(3;2)+(7;4)+(11;2)$&$\begin{gathered}(1;3)+(3;1)+(5;3)+(7;5)+\\(9;3)+(11;1)+(13;3)\end{gathered}$\\ \hline
$D_5+A_1$&$\mathfrak{su}(2)$&$\begin{gathered}(2;3)+(5;2)+(8;1)+\\(10;1)+(11;2)\end{gathered}$&$\begin{gathered}(1;3)+2(3;1)+(4;2)+(6;2)+\\(7;1)+(9;3)+(10;2)+(11;1)+\\(12;2)+(15;1)\end{gathered}$\\ \hline
$D_6(a_1)$&$\mathfrak{su}(2)$&$\begin{gathered}(3;2)+(4;1)+(6;1)+\\(9;2)+(10;1)+(12;1)\end{gathered}$&$\begin{gathered}(1;3)+2(3;1)+(4;2)+(6;2)+\\2(7;1)+(9;1)+(10;2)+2(11;1)+\\(12;2)+(15;1)\end{gathered}$\\ \hline
$E_7(a_4)$&-&$\begin{gathered}2+2(4)+6+\\8+2(10)+12\end{gathered}$&$\begin{gathered}4(3)+2(5)+3(7)+2(9)+4(11)\\+13+15\end{gathered}$\\ \hline
$D_6$&$\mathfrak{su}(2)$&$\begin{gathered}(1;2)+(6;1)+(10;1)+\\(11;2)+(16;1)\end{gathered}$&$\begin{gathered}(1;3)+(3;1)+(6;2)+(7;1)+\\(10;2)+2(11;1)+(15;1)+\\(16;2)+(19;1)\end{gathered}$\\ \hline
$E_6(a_1)$&$\mathfrak{u}(1)$&$\begin{gathered}1_3+1_{-3}+5_1+\\5_{-1}+9_1+9_{-1}+\\13_1+13_{-1}\end{gathered}$&$\begin{gathered}1_0+3_0+5_2+5_0+5_{-2}+7_0+\\9_2+9_0+9_{-2}+2(11_0)+13_2+\\13_{-2}+15_0+17_0\end{gathered}$\\ \hline
$E_6$&$\mathfrak{su}(2)$&$(1;4)+(9;2)+(17;2)$&$\begin{gathered}(1;3)+(3;1)+(9;3)+(11;1)+\\(15;1)+(17;3)+(23;1)\end{gathered}$\\ \hline
$E_7(a_3)$&-&$2+6+2(10)+12+16$&$\begin{gathered}2(3)+5+2(7)+9+3(11)+\\2(15)+17+19\end{gathered}$\\ \hline
$E_7(a_2)$&-&$4+8+10+16+18$&$\begin{gathered}2(3)+7+9+2(11)+2(15)+17+\\19+23\end{gathered}$\\ \hline
$E_7(a_1)$&-&$6+12+16+22$&$\begin{gathered}3+7+2(11)+15+17+19+\\23+27\end{gathered}$\\ \hline
$E_7$&-&$10+18+28$&$3+11+15+19+23+27+35$\\ \hline
\end{longtable}

}

\newpage

\section{Projection matrices}\label{projection_matrices}

\begin{longtable}{|c|c|c|}
\hline
Bala-Carter&$\mathfrak{f}$&Projection Matrix\\
\hline
\endhead
$A_1$&$\mathfrak{so}(12)$&$\begin{pmatrix}0&0&0&0&0&1&0\\0&1&2&2&2&1&1\\1&0&0&0&0&0&0\\0&1&0&0&0&0&0\\0&0&1&0&0&0&0\\0&0&0&1&0&0&0\\0&0&0&0&0&0&1\end{pmatrix}$\\\hline
$2A_1$&$\mathfrak{so}(9)\times \mathfrak{su}(2)$&$\begin{pmatrix}1&1&2&2&2&1&1\\0&1&1&1&1&0&0\\0&0&0&0&0&0&1\\1&1&1&0&0&0&0\\-1&-1&0&1&0&0&0\\1&3&4&3&2&2&2\end{pmatrix}$\\\hline
$(3A_1)''$&$\mathfrak{f}_4$&$\begin{pmatrix}2&4&6&5&4&3&3\\0&0&0&0&0&0&1\\0&0&1&0&0&0&0\\0&1&0&1&0&0&0\\1&0&0&0&1&0&0\end{pmatrix}$\\\hline
$(3A_1)'$&$\mathfrak{sp}(3)\times\mathfrak{su}(2)$&$\begin{pmatrix}3&6&8&6&4&2&4\\0&0&0&0&0&1&1\\0&0&1&0&1&0&0\\0&0&0&1&0&0&0\\1&0&0&0&0&0&0\end{pmatrix}$\\\hline
$A_2$&$\mathfrak{su}(6)$&$\begin{pmatrix}4&6&8&6&4&2&4\\0&0&0&0&0&1&0\\0&0&0&0&1&0&0\\0&0&0&1&0&0&0\\0&0&1&0&0&0&0\\0&0&0&0&0&0&1\end{pmatrix}$\\\hline
$4A_1$&$\mathfrak{sp}(3)$&$\begin{pmatrix}3&6&9&7&5&3&4\\1&0&0&0&1&0&0\\0&1&0&1&0&0&0\\0&0&1&0&0&0&1\end{pmatrix}$\\\hline
$A_2+A_1$&$\mathfrak{su}(4)\times\mathfrak{u}(1)$&$\begin{pmatrix}3&7&10&8&6&3&5\\0&0&0&1&0&0&0\\0&0&1&0&0&0&1\\1&1&0&0&0&0&0\\1/2&1/2&1&1/2&0&1&0\end{pmatrix}$\\\hline
$A_2+2A_1$&${\mathfrak{su}(2)}^3$&$\begin{pmatrix}4&8&12&9&6&3&6\\0&0&0&0&0&0&1\\0&0&0&1&0&1&0\\0&2&0&-1&0&1&0\end{pmatrix}$\\\hline
$A_3$&$\mathfrak{so}(7)\times \mathfrak{su}(2)$&$\begin{pmatrix}0&1&4&0&0&0&3\\0&0&0&0&0&1&0\\0&0&0&0&1&0&0\\0&1&2&2&0&0&1\\2&3&4&3&2&1&2\end{pmatrix}$\\\hline
$2A_2$&$\mathfrak{g}_2\times \mathfrak{su}(2)$&$\begin{pmatrix}4&8&12&10&8&4&6\\1&0&2&1&0&0&1\\0&1&0&0&0&0&0\\0&0&0&1&0&1&1\end{pmatrix}$\\\hline
$A_2+3A_1$&$\mathfrak{g}_2$&$\begin{pmatrix}4&8&12&9&6&3&5\\1&0&2&2&0&1&2\\0&1&0&0&1&0&0\end{pmatrix}$\\\hline
$(A_3+A_1)''$&$\mathfrak{so}(7)$&$\begin{pmatrix}4&10&14&11&8&5&7\\0&0&0&0&0&0&1\\0&0&1&0&0&0&0\\1&1&0&1&0&0&0\end{pmatrix}$\\\hline
$2A_2+A_1$&${\mathfrak{su}(2)}^2$&$\begin{pmatrix}5&10&14&11&8&4&7\\0&0&0&1&0&1&1\\1&0&0&-1&0&0&1\end{pmatrix}$\\\hline
$(A_3+A_1)'$&${\mathfrak{su}(2)}^3$&$\begin{pmatrix}6&11&16&12&8&4&8\\0&1&0&0&0&0&0\\0&0&0&0&0&1&1\\0&0&0&1&0&0&0\end{pmatrix}$\\\hline
$D_4(a_1)$&${\mathfrak{su}(2)}^3$&$\begin{pmatrix}6&12&16&12&8&4&8\\0&0&0&0&0&0&1\\0&0&0&1&0&0&0\\0&0&0&0&0&1&0\end{pmatrix}$\\\hline
$A_3+2A_1$&${\mathfrak{su}(2)}^2$&$\begin{pmatrix}6&11&16&13&9&5&8\\0&1&1&0&0&0&0\\0&0&1&0&1&0&1\end{pmatrix}$\\\hline
$D_4$&$\mathfrak{sp}(3)$&$\begin{pmatrix}10&18&24&18&12&6&12\\0&0&0&0&0&1&1\\0&0&1&0&1&0&0\\0&0&0&1&0&0&0\end{pmatrix}$\\\hline
$D_4(a_1)+A_1$&${\mathfrak{su}(2)}^2$&$\begin{pmatrix}6&12&17&13&9&5&9\\0&0&0&0&1&0&0\\0&0&1&0&0&0&0\end{pmatrix}$\\\hline
$A_3+A_2$&$\mathfrak{su}(2)\times\mathfrak{u}(1)$&$\begin{pmatrix}6&12&18&14&10&5&9\\0&0&0&1&0&0&0\\0&2&0&0&0&1&-1\end{pmatrix}$\\\hline
$A_4$&$\mathfrak{su}(3)\times\mathfrak{u}(1)$&$\begin{pmatrix}6&14&20&16&12&6&10\\0&0&0&1&0&0&0\\0&0&1&0&0&0&0\\0&0&2/3&1/3&0&1&1\end{pmatrix}$\\\hline
$A_3+A_2+A_1$&$\mathfrak{su}(2)$&$\begin{pmatrix}6&12&18&15&10&5&9\\4&6&6&0&2&2&4\end{pmatrix}$\\\hline
$(A_5)''$&$\mathfrak{g}_2$&$\begin{pmatrix}10&18&26&21&16&9&13\\0&1&0&1&0&0&1\\0&0&1&0&0&0&0\end{pmatrix}$\\\hline
$D_4+A_1$&$\mathfrak{sp}(2)$&$\begin{pmatrix}10&17&25&19&13&7&13\\0&1&1&0&1&0&0\\0&0&0&1&0&0&0\end{pmatrix}$\\\hline
$A_4+A_1$&${\mathfrak{u}(1)}^2$&$\begin{pmatrix}8&15&22&17&12&6&11\\0&1&0&1&0&0&1\\0&-2/3&0&1/3&0&1&1/3\end{pmatrix}$\\\hline
$D_5(a_1)$&$\mathfrak{su}(2)\times\mathfrak{u}(1)$&$\begin{pmatrix}10&18&26&20&14&7&13\\0&0&0&1&0&0&0\\0&0&0&0&0&1&1\end{pmatrix}$\\\hline
$A_4+A_2$&$\mathfrak{su}(2)$&$\begin{pmatrix}8&16&24&18&12&6&12\\0&2&0&3&4&3&1\end{pmatrix}$\\\hline
$(A_5)'$&${\mathfrak{su}(2)}^2$&$\begin{pmatrix}10&19&28&22&16&8&14\\0&1&0&0&0&0&0\\0&0&0&1&0&1&1\end{pmatrix}$\\\hline
$A_5+A_1$&$\mathfrak{su}(2)$&$\begin{pmatrix}10&19&28&22&16&9&14\\0&1&0&1&0&0&1\end{pmatrix}$\\\hline
$D_5(a_1)+A_1$&$\mathfrak{su}(2)$&$\begin{pmatrix}10&18&26&21&14&7&13\\0&0&2&0&2&2&2\end{pmatrix}$\\\hline
$D_6(a_2)$&$\mathfrak{su}(2)$&$\begin{pmatrix}10&20&29&23&16&9&15\\0&0&1&0&0&0&0\end{pmatrix}$\\\hline
$E_6(a_3)$&$\mathfrak{su}(2)$&$\begin{pmatrix}10&20&28&22&16&8&14\\0&0&0&1&0&1&1\end{pmatrix}$\\\hline
$D_5$&${\mathfrak{su}(2)}^2$&$\begin{pmatrix}14&24&36&28&20&10&18\\0&0&0&1&0&0&0\\0&0&0&0&0&1&1\end{pmatrix}$\\\hline
$E_7(a_5)$&-&$\begin{pmatrix}10&20&30&23&16&9&15\end{pmatrix}$\\\hline
$A_6$&$\mathfrak{su}(2)$&$\begin{pmatrix}12&24&36&28&20&10&18\\2&0&0&1&0&1&1\end{pmatrix}$\\\hline
$D_5+A_1$&$\mathfrak{su}(2)$&$\begin{pmatrix}11&25&37&29&20&10&19\\1&1&1&0&0&1&0\end{pmatrix}$\\\hline
$D_6(a_1)$&$\mathfrak{su}(2)$&$\begin{pmatrix}14&26&37&29&20&11&19\\0&0&1&0&0&0&0\end{pmatrix}$\\\hline
$E_7(a_4)$&-&$\begin{pmatrix}14&26&38&29&20&11&19\end{pmatrix}$\\\hline
$D_6$&$\mathfrak{su}(2)$&$\begin{pmatrix}18&33&48&39&28&15&23\\0&1&0&0&0&0&0\end{pmatrix}$\\\hline
$E_6(a_1)$&$\mathfrak{u}(1)$&$\begin{pmatrix}16&30&44&34&24&12&22\\0&0&0&1&0&1&1\end{pmatrix}$\\\hline
$E_6$&$\mathfrak{su}(2)$&$\begin{pmatrix}22&42&60&46&32&16&30\\0&0&0&1&0&1&1\end{pmatrix}$\\\hline
$E_7(a_3)$&-&$\begin{pmatrix}18&34&50&39&28&15&25\end{pmatrix}$\\\hline
$E_7(a_2)$&-&$\begin{pmatrix}22&42&60&47&32&17&31\end{pmatrix}$\\\hline
$E_7(a_1)$&-&$\begin{pmatrix}26&50&72&57&40&21&37\end{pmatrix}$\\\hline
$E_7$&-&$\begin{pmatrix}34&66&96&75&52&27&49\end{pmatrix}$\\\hline
\end{longtable}

\end{appendices}

\bibliographystyle{utphys}

\bibliography{ref}

\end{document}